\documentclass{aa}  
\usepackage{graphicx}
\usepackage{txfonts}
\usepackage[abs]{overpic}
\usepackage{float}
\usepackage{color}

\begin{document} 

\title{Observation of aerodynamic instability in the flow of a particle stream in a dilute gas}

\author{Holly L. Capelo \inst{1} \fnmsep\thanks{currently: Physikalisches Institut, Universit\"{a}t Bern, Sidlerstrasse, 5, CH-3012 Bern, Switzerland}
          \and Jan Mol\'{a}\v{c}ek \inst{1}
          \and Michiel Lambrechts\inst{2}
          \and John Lawson \inst{1}
          \and Anders Johansen \inst{2}
          \and J\"{u}rgen Blum \inst{3}
          \and \\Eberhard Bodenschatz \inst{1,4,5}
          \and Haitao Xu \inst {1}\fnmsep\thanks{currently: Center for Combustion Energy and School of Aerospace Engineering, Tsinghua University, Beijing 100084, China}
          }        
   \institute{Max Planck Institute for Dynamics and Self-Organization (MPIDS),
              Am Fa$\beta$berg 17, D-37077 G\"{o}ttingen, Germany\\
              \email{holly.capelo@space.unibe.ch}
         \and Lund Observatory, Department of Astronomy and Theoretical Physics, Lund University, 22100 Lund, Sweden
          \and Institut f\"ur Geophysik und extraterrestrische Physik, Technische Universit\"at Braunschweig, Mendelssohnstr. 3, D-38106 Braunschweig, Germany
          \and Institute for Nonlinear Dynamics, University of G\"{o}ttingen, D-37073 G\"{o}ttingen, Germany
          \and Laboratory of Atomic and Solid-State Physics and Sibley School of Mechanical and Aerospace Engineering, Cornell University, Ithaca, New York 14853, USA
             }

   \date{Received ; accepted }

\abstract{Forming macroscopic solid bodies in circumstellar discs requires local dust concentration levels significantly higher than the mean. Interactions of the dust particles with the gas must serve to augment local particle densities, and facilitate growth past barriers in the metre size range. Amongst a number of mechanisms that can amplify the local density of solids, aerodynamic streaming instability (SI) is one of the most promising. This work tests the physical assumptions of models that lead to SI in protoplanetary disks (PPDs). We conduct laboratory experiments in which we track the three-dimensional motion of spherical solid particles fluidized in a low-pressure, laminar, incompressible, gas stream. The particle sizes span the Stokes--Epstein drag regime transition and the overall dust-to-gas mass density ratio, $\epsilon$, is close to unity. \cite{lambrechts} established the similarity of the laboratory flow to a simplified PPD model flow. We study velocity statistics and perform time-series analysis of the advected flow to obtain experimental results suggesting an instability due to particle-gas interaction: i) there exist variations in particle concentration in the direction of the mean relative motion between the gas and the particles, i.e. the direction of the mean drag forces; ii) the particles have a tendency to `catch up' to one another when they are in proximity; iii) particle clumping occurs on very small scales, which implies local enhancements above the background $\epsilon$ by factors of several tens; v) the presence of these density enhancements occurs for a mean $\epsilon$ approaching or greater than 1; v) we find evidence for collective particle drag reduction when the local particle number density becomes high and when the background gas pressure is high so that the drag is in the continuum regime. The experiments presented here are precedent-setting for observing SI under controlled conditions and may lead to a deeper understanding of how it operates in nature.}

   \maketitle
%

\section{Introduction} \label{intro}

Planet formation theories that are consistent with the architecture, elemental composition, and collisional record in our Solar System require a population of $\sim$km-sized precursor bodies, termed planetesimals. Due to primarily gravitational interactions with one another and with the surrounding gas, dust, and pebbles, planetesimals, which undergo repeated collisions, grow into planetary cores and ultimately mature planets possessing atmospheres \citep{Morbidellietal:2012}. A similar process must be responsible for the planets known to exist in extra-solar systems \citep{seager_exoplanets}. 

How first-generation planetesimals form is under debate, with fundamental agreement that the only available material to compose them are the solids (`particles') observed in gaseous protoplanetary discs (PPDs; \citealt{Testi:2014}) and likely to have been present in the early Solar PPD. Studies employing isochrone age-determination techniques of star-cluster sources with near-infrared excess indicate that PPDs have statistically short lifetimes of $\sim$2--4~Myr \citep{HaischEtal:2001,wyatt2008,bell}, providing upper limits on the time-scale within which macroscopic solids can form from the disc material. Sub-mm observations place even tighter constraints of $\sim$1--2~Myr for the survival of the large-grain population \citep{LeeEtal:2011,Ansdell2017}. The dust and ice grains of $\mu$m sizes represent $\sim1\%$ of the total PPD mass, the rest of which is mainly hydrogen gas. Therefore, first-order PPD models are ideal-gas flows subject to orbital dynamics, with primitive state variables that depend upon distance from the star and height above the disc midplane \citep{Weidenschilling:1977b, HayashiEtal:1985, Armitage_book}. 

Particle--gas interactions determine the movement of solids -- with respect to each other and to the gas -- at virtually all particle size scales \citep{Turner_2014}. Depending upon local circumstances, the flow may be laminar or else turbulence may arise from a host of proposed fluid instabilities in the context of planetesimal formation \citep{Johansen_2014}. Although hydrodynamic turbulence is known to create pressure and velocity gradients that sequester particles and facilitate collisions, it can, for the same reasons, serve to diffuse high particle-concentration regions or lead to high-energy particle collisions that are ultimately destructive (\citealt{Blum_wurm2008,you_leshouches}).

The statistical outcomes of dust-aggregate collisions are well mapped from microgravity experiments conducted under vacuum conditions \citep{DominikTielens:1997,BlumEtal:1998,BlumEtal:1999,BlumWurm:2000, GuettlerEtal:PandG}. While low-energy collisions result in fractal growth, aggregates accrue momentum with increasing mass, such that further collisions may result in erosion or fragmentation, instead of continued bottom-up growth \citep{Zsom2010, GuettlerEtal:2010, mohtashim}. 

Drag dissipation from pressure-supported sub-Keplerian gas motions on particles with Keplerian velocities causes particle orbital decay \citep{Whipple:1972}. The drift rate of particles possessing typical solid densities peaks for particle diameters 10--100 cm, with associated depletion time-scales in the 100s of orbits \citep{Weidenschilling:1977a}.  Growth by collisional coagulation is thus stifled by the short survival times of the pebbles that disappear due to systematic drift \citep{BrauerEtal:2007,Birnstiel:2012}. Mechanisms to bypass the meter-size drift barrier must therefore be rapid, making strict collisional growth inefficacious. 

A complementary approach to coagulation models is to follow the dust evolution as if it is a fluid without internal pressure (e.g. \citealt{Nakagawa}), where both the gas and solid phases obey continuum mass and momentum balance equations. Taking into account the collective drag-force coupling between the two phases, \cite{You_good2005} found that such a flow possesses an oscillatory instability, referred to as the streaming instability (SI) . The critical control parameters for the onset of instability are the spatially averaged dust-to-gas mass density ratio $\epsilon=\rho_{\rm p}/\rho_{\rm g}$ and the Stokes number $\tau_{f}=T_{\rm f}\Omega$, defined in this context as the viscous coupling time $T_{\rm f}$ normalised by the particles' Keplerian frequency $\Omega$. For a particle with mass $m_{\rm {p}}$ and velocity $|\vec{v}|$, the viscous coupling time derives from comparing its momentum to the drag force, $F_{\rm d}$, that it experiences: $T_{\rm f}=\frac{m_{\rm{p}}|\vec{v}|}{F_{\rm d}}$. The growth of SI is strongest when particles are weakly decoupled from the gas ($\tau_{f}>\sim 0.1$) and when $\epsilon$ is close to unity.  When SI becomes fully developed, particles in regions with high particle density feel a reduced collective drag and therefore the mass density can become further enhanced by isolated particles that catch up to the slower drifting cluster.
The amplitude of the resultant density wave could become sufficient for gravitational instability to occur and form planetesimals by direct collapse \citep{You_jo2007j,Jo_you2007j}. 
 
 This effect has become integral to a number of subsequent theoretical studies on disc evolution and planet formation (e.g. \citealt{Jo_you_mac_2009, Bai_2010a, Bai_2010b, Bai_2010c, Miniati_2010,  Johansen2012AA,Kowalik_2013,blum:2014,Yang_2014,Carrera_2015, simon,schaefer:2017}). The paradigm revisits the sequence originally proposed by \cite{Safronov:1969} and \cite{goldreich_ward}, whereby dust grains are first concentrated by vertical settling and collisional coagulation, followed by gravitational fragmentation of the resultant dust sub-disc. 
 
 A common finding in the aforementioned theoretical works is that the onset rate of the instability is appreciable for $\epsilon \gtrsim 1$. This is consistent with the consensus from studies of dispersed two-phase flow, where two-way drag coupling is significant for $\epsilon=1$ and four-way coupling\footnote{Fluid drag and particle back reaction, plus collisional momentum exchange.} is significant when volumetric filling factors $\phi$ increase \citep{multiphase10, collective_particle}. The work of \cite{Drazkowska_2014} shows that the collisional coagulation process can produce agglomerates large enough to seed SI, with especially favourable conditions beyond the radial distance at which ices can form \citep{Bitsch_2015, Drazkowska_2016,schoonenberg_2017}.
 
Recent studies have sought to generalise the SI and understand its model dependencies. \cite{jacquet_2011} demonstrated that a laminar two-fluid shearing sheet is linearly stable to SI when $\Omega = 0$. \cite{lambrechts} numerically evolved the same model equations, i.e. incompressible two-fluid drag-coupled mass and momentum balance, with terms relating to Keplerian rotation absent, and found the system to be unstable to non-linear perturbations. Together with the lack of requirement of rotation, the presence of an instability on small scales led the authors to suggest that the effect may occur in a range of circumstances where particles experience drag from a dilute gas, such as in the vertical sedimentation phase of dust-subdisc evolution, the dust in the atmosphere of an embedded giant planet, or in the formation of chondrules.
Additional works find dust-drag induced fluid instabilities to occur generically and in various stages of PPD formation and evolution (e.g. \citealt{Bate:2017,Squire_Hopkins:2018a,Squire_Hopkins:2018b,youdin_singlefluid}).

The SI is a robust effect that arises in a laminar disc model but also persists in the presence of turbulence. Independent of how the viscous evolution of PPDs proceeds, SI is considered a promising way to help overcome the long-standing meter-barrier problem. However, small-scale dust concentration is not yet possible to observe directly with astronomical observations, due to the optically thick nature of discs at short wavelengths, because spontaneous concentration of particles due to fluid instabilities is likely occurring over extremely rapid time-scales in very young discs, and because the length scales involved are too small to resolve. 

The topic of angular-momentum transport in PPDs by anomalous turbulent viscosity has been addressed experimentally with Taylor--Couette devices rotating at Keplerian frequencies \citep{Dubrulle2005,Hersant2005}. The study of dust-grain drag-induced aerodynamic instabilities, which could eventually lead to turbulence, in sufficiently dilute scaled-down laboratory flow has never been previously attempted and is the unique goal of this work.  

We devise a very reduced system, starting with the premise that the instability arises principally from the differential motion between the gas (`carrier') and solid (`disperse') phases, and that this is simply analogous to the case where particles, initially homogeneously mixed within a fluid, settle against a pressure gradient\footnote{A minor difference between the setup in \cite{lambrechts} and the present one is that they assumed a hydrostatic pressure gradient against which particles could settle and unmix, whereas the change in hydrostatic column density is negligible in our experimental volume and we instead drive an upward flow by applying a pressure differential. In the reference frame between the disperse and carrier phases, however, the effect of there being a net velocity difference is the same.
} under constant gravitational acceleration $g$ and relax to their terminal velocity $u_{\rm t}$. As will be shown, the particle-tracking data analysis exhibits particle clustering and modified particle dynamics on small scales, which we interpret as a collective particle-drag induced instability. 

Particle suspensions in the Stokes regime (continuum drag assumption) have been studied for centuries (see, e.g.  \citealt{darcy_early,Batchelor_jfm} and references therein and thereto), but broaching into the Epstein-drag regime\footnote{In general, solids that participate in planetesimal formation can be in either regime.} (\citealt{epstein}; free molecular flow or relaxed surface slip conditions) is a novelty of this work. Similarly, the use of a rarefied gas as the carrier phase, where the value of $\epsilon \sim1$, results in volumetric filling factors $\phi$ several orders of magnitude (at least 3--4) lower than what has ever been studied before in the literature of fluidized beds and suspension chambers (see reviews such as \citealt{multiphase10}, or \citealt{Guaz_Hinch11}). 

We describe the apparatus and our experimental techniques in Section~\ref{techniques}. We then outline the parameters of the experiments in Section~\ref{character}. The results of the particle tracking are shown in Section~\ref{results}. We discuss our findings in Section~\ref{discussion} and suggest ideas for follow-up in Section~\ref{soa}. We summarize and conclude in Section~\ref{summary}.

\section{Experimental Techniques}\label{techniques}

\subsection{Apparatus} \label{apparatus}

\begin{figure}
\includegraphics[scale=0.9]{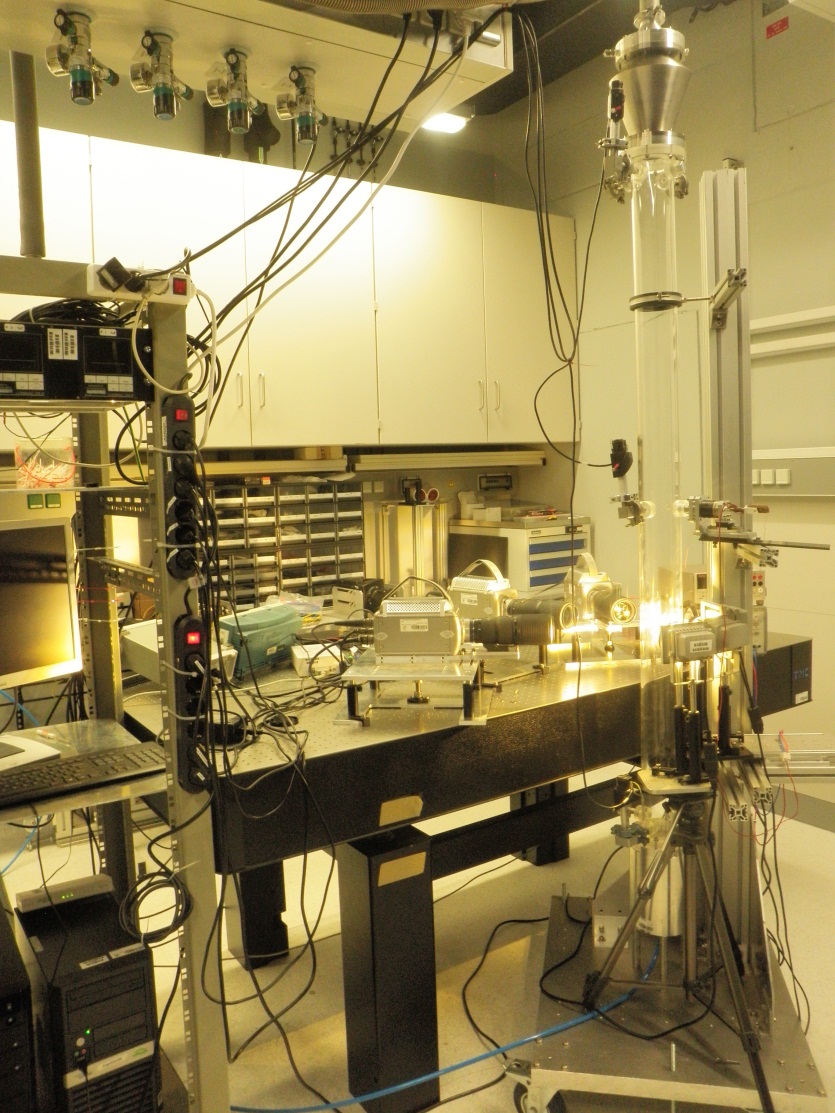}
\caption{\label{apparatus_image} The particle fluidisation/sedimentation chamber, shown towards the right of the image, is supported on a heavy aluminium plate and fixed to an extruded aluminium profile. A gas line leads into the pressure-reducing filter at the bottom and vacuum tubing leading to a vacuum pump extends from the hollow gas expander at the top. Cables extending from the device connect pressure transducers to a data-logging system. On the optical table are three high-speed phantom V10 cameras, positioned to have overlapping field of view inside of the apparatus, and the signal generator used to synchronise the three cameras via external triggering. LED spotlights are mounted near the apparatus for use as backlighting. }
\end{figure}

The experimental facility is shown in Figure~\ref{apparatus_image}. In a cylindrical plexiglas chamber, solid particles are transported by an upward stream of dilute gas under low pressure. 
The top of the chamber is connected to a vacuum pump and the base of the apparatus contains a pressure-reducing filter composed of porous material. The filter creates a pressure drop $\Delta P=9.9\times 10^{5}$ Pa from the outside of the apparatus to the inside. The pressure differential across the whole system drives a continuous flow. 

The facility is equipped to supply different gases to the apparatus, but all the experiments described here were conducted with dry air as the carrier phase. We choose spherical steel particles particles for their high mass density and high electrical conductivity to eliminate the effect of electrostatic forces. 

During each experiment, the steady-state pressure and particle-seeding density are controlled and we record the particle positions from three viewing angles in order to reconstruct their trajectories. 

\begin{figure*}[!h]
\includegraphics[scale=.1]{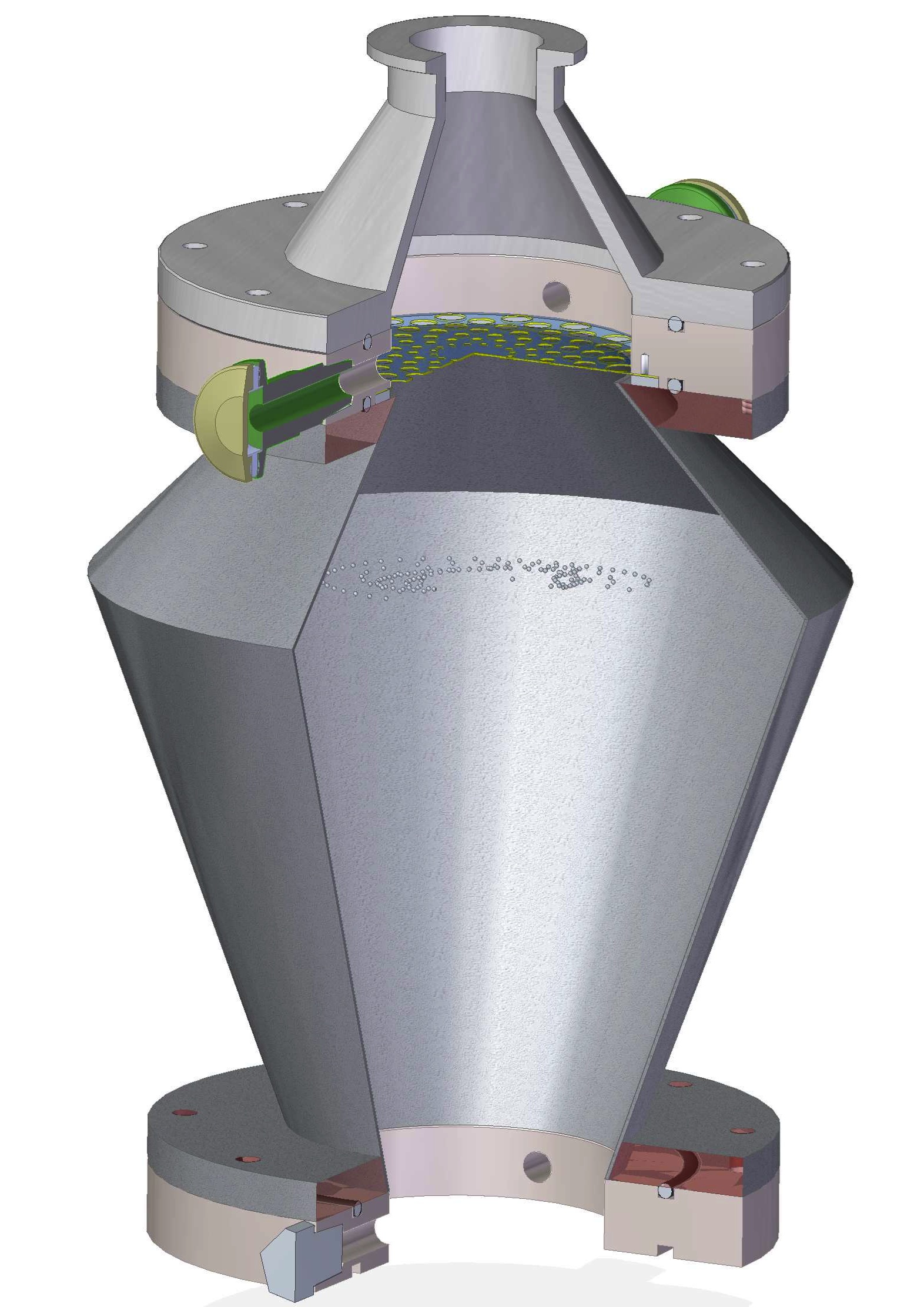}
\hspace{50mm}
\includegraphics[scale=.3]{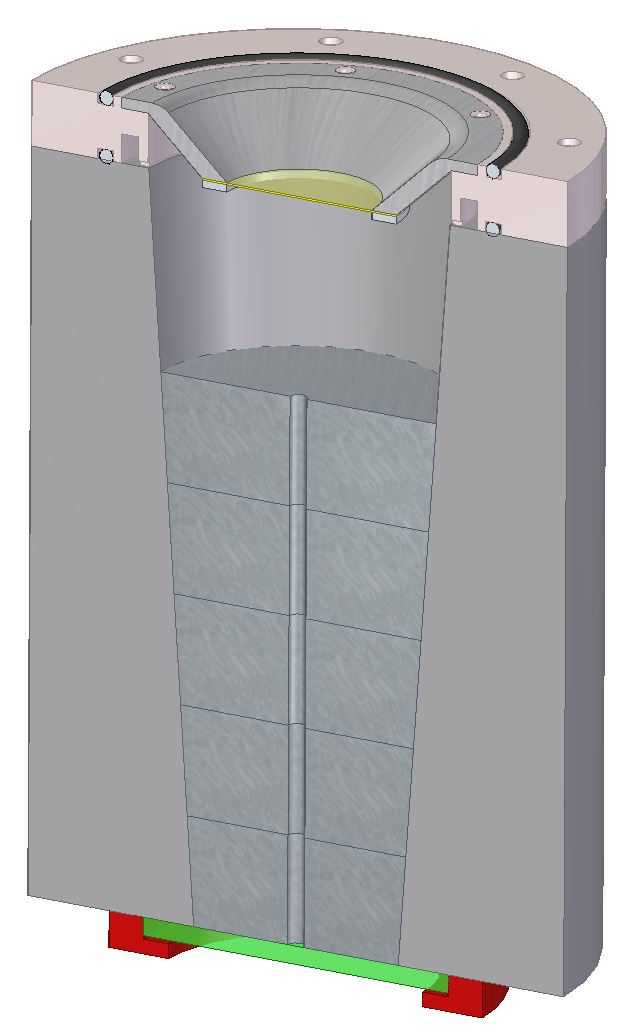}

\caption{\label{seeding} Cut-away views of components used to control the gas pressure and two-phase flow velocity. Left: Hollow gas expansion unit where particles enter and slow when they reach the top of the sedimentation chamber. Right: Apparatus base with pressure-reducing filter composed of sintered metal plates of porosity coefficient $\alpha=10^{-12} $ m$^{2}$.  The funnel insert holds a fine wire mesh that is used as a particle staging platform; low-pressure air streaming through the mesh fluidises the particles.}
\end{figure*}

The particles are pre-loaded onto a particle seeding platform, which is composed of a fine wire mesh, through which air can pass - but particles cannot (see right-hand panel of Figure~\ref{seeding}). Particles that reach the top of the container enter a hollow expansion where the gas slows down (due to mass conservation), preventing particles from travelling further upstream (see left-hand panel of Figure~\ref{seeding} for a view of the expansion chamber). Because the gas velocity close to the wall is smaller, particles that migrate towards the walls fall down, then they funnel towards the particle-seeding platform, where they re-enter in the flow and circulate.
The result of the particle circulation is that the centre-line of the flow has a persistent upward-directed particle stream that is constantly replenished with particles. The particle stream persists indefinitely, therefore we wait until transients in the pressure disappear and take recordings of the particle motions during steady-state pressure conditions.

The particle positions are tracked in time using three-dimensional Lagrangian particle tracking (LPT), described in Section~\ref{lpt}. The stereoscopic camera arrangement is shown in Figure~\ref{apparatus_image}, and we used LED backlighting to image the solid particles using shadowgraphy.

The measurements presented in this section will demonstrate that, before adding particles, the underlying gas flow is incompressible and is laminar. We made time-resolved pressure (temperature) measurements with Pirani pressure heads (thermocouples) secured at three heights in the measurement volume.  We measured the radial profile of the upward gas velocity using particle image velocimetry (PIV), described in Section~\ref{flowprofile}.  Many more details of the apparatus and its tests can be found in \cite{capelo:2018}; see also the summary section of \cite{lambrechts}. Below we describe our measurement techniques.

\subsection{Flow profile}\label{flowprofile}

The carrier-phase flow was calibrated using PIV, with smoke aerosols as the seeding material. This technique involves correlating the positions of tracer particles in subsequent image pairs to extract the instantaneous vector flow field. An average of time-series PIV data can then be used to determine the mean field and its fluctuations. Due to the axisymmetry of the flow in question, we determine the radial flow profile from two-dimensional measurements of a single slice through the centre-line of the apparatus. Particles with short viscous coupling time, $T_{\rm f}$, are desirable to approach tracer particles and, due to the high contrast in density between the gas and solid particles, it is necessary to use the smallest possible tracers. We therefore generated $\mu$m-sized smoke particles by combustion at atmospheric pressure outside the apparatus and entrained the fluidised aerosols into the bottom of the sedimentation chamber, where the flow was maintained at a steady-state operational pressure of $\sim$10 mbar.  

The measurements were made using one high-speed camera placed perpendicularly to a laser sheet\footnote{This camera and illumination scheme is different than the one shown in Figure~\ref{apparatus_image}, since the two measurement techniques, PIV and LPT, require a different setup. The tracer (smoke, 1~$\mu$m) particles used in the carrier phase flow calibration should not be confused with the inertial (steel, 25--65~$\mu$m) particles used in the LPT experiments.} from a single cavity Q-switched Nd:YAG laser at 532 nm (IB Laser Chronos 400 MM IC SHG). The laser was operating at a repetition rate of 3.3 kHz, with a pulse duration of 140 ns and pulse energy of 6.4 mJ.  A signal generator was used to synchronise the rising edge of the laser pulses with the camera. The image pairs consisted of 20~$\mu$s exposures recorded at a temporal separation of 300 $\mu$s.  We measured for a 6s duration, with an image pair acquired every 450 $\mu$s. 

We used the commercially available DaViS 7.3 software \footnote{https://www.lavision.de/en/products/davis-software/} to perform the image pre-processing and processing. The pre-processing consisted of the following steps:  i) mean background image subtraction; ii) local non-linear subtraction; iii) sliding-minimum filter over a scale of 64 pixels; iv) linear 3x3 pixel Gaussian smoothing filter; v) intensity renormalisation filter. The processing involved interrogating pre-processed image pairs with a 3-pass PIV scheme, with the first pass using 1.5$\times$1.5 mm windows at 50\% overlap and subsequent passes using 0.75$\times$0.75 mm windows at 50\% overlap. 

\begin{figure}[!h]
\includegraphics[scale=.57]{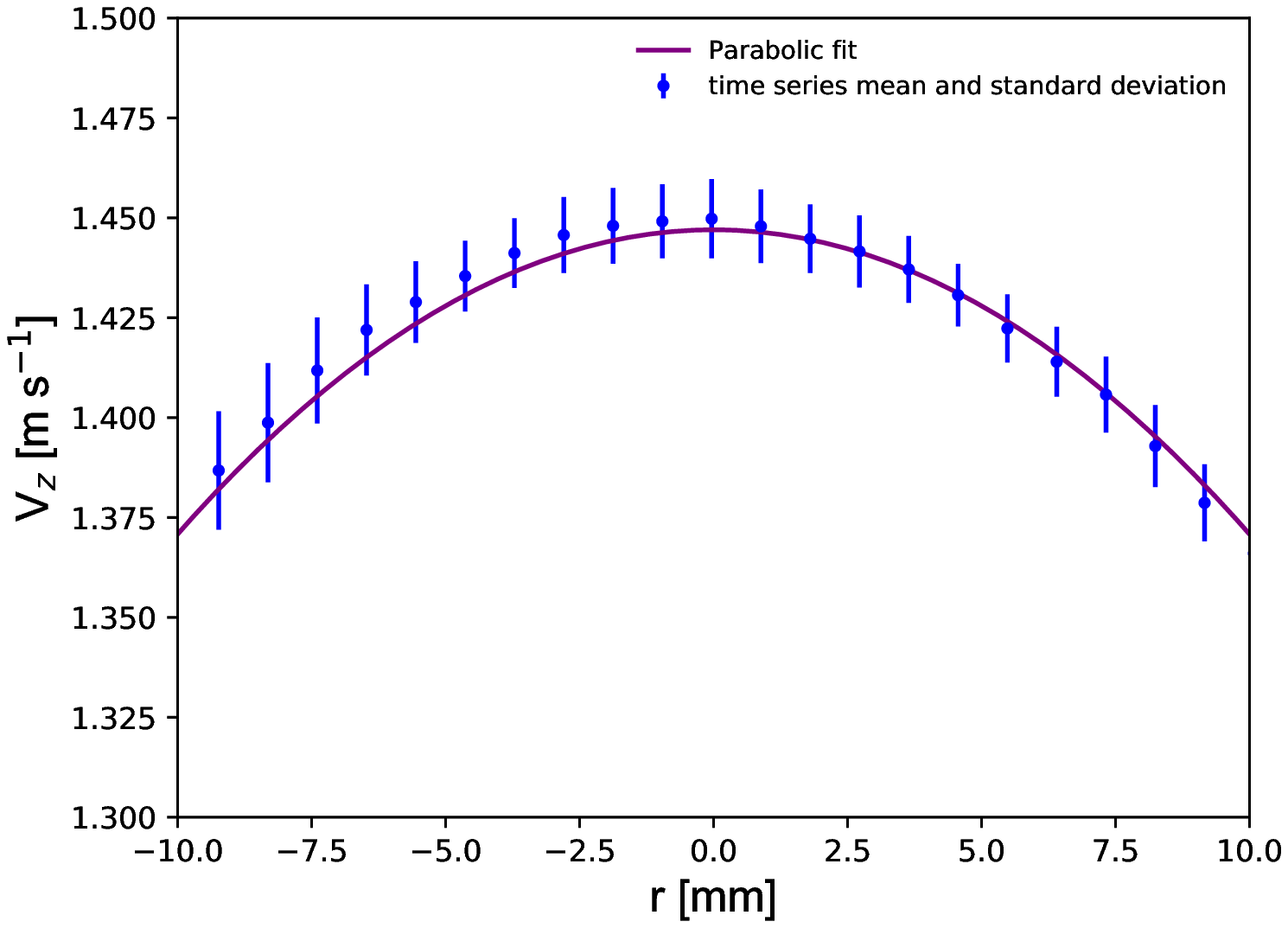}\\
\includegraphics[scale=.57]{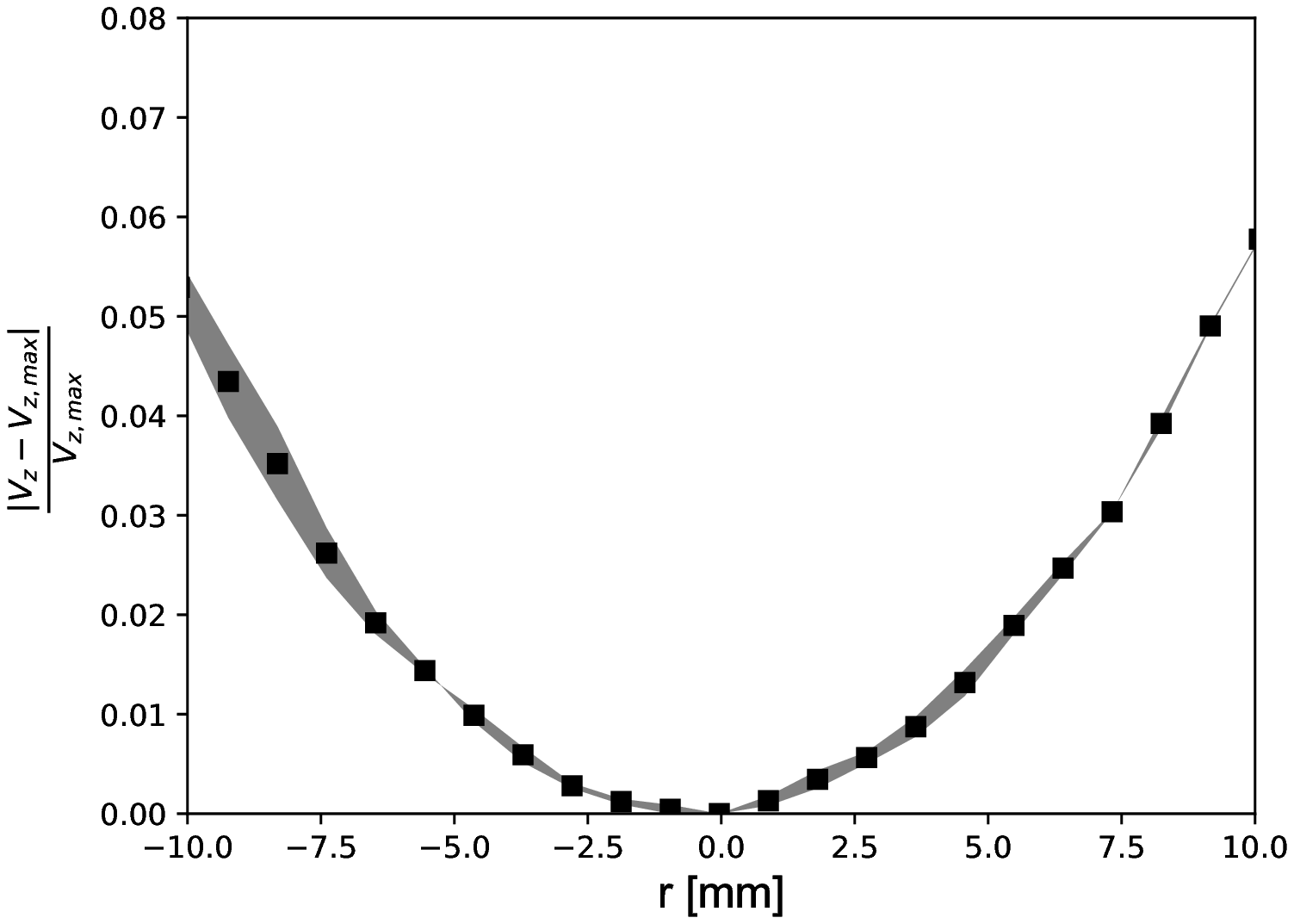}\\
\caption{\label{central_profile} Flow profile calibration measurement using PIV, limited to the region of possible overlap with the LPT measurements, which are conducted within a $\sim$1 cm$^{3}$ region centred within the vessel with an accuracy of $\sim0.5$cm. Top: Measured mean velocity (blue dots), and parabolic fit (purple line), limited to the central two cm of the flow vessel. Error bars represent the standard deviation of the time-series PIV data. Bottom: calculation of the relative difference in velocity from the maximum centre-line velocity. The shaded region represents the uncertainty range due to error bars in the top figure. }
\end{figure}

 Line-profiles of the two-dimensional vector fields in the PIV time-series data demonstrate that the gas flow has a cylindrical Pouseille profile, with a centre-line velocity of 1.45 ms$^{-1}$. Figure~\ref{central_profile} zooms into the central two cm of the gas velocity profile. The measured mean profile is shown with discrete points, indicating the spatial resolution of the vector field. The error bars represent the fluctuations (standard deviation) on the mean flow. A parabolic fit to the mean field, shown as a solid purple line in the figure, goes to zero at radii of $\pm45$mm (not shown), as expected for a laminar flow with small container-scale Knudsen number \citep{Sharipov}.  Although there is radial variation in the flow velocity, it has a difference of no more than 5\% over the central two cm, which is about twice the size of the region where we conduct our LPT measurements.  

\subsection{Inertial particle motion}\label{lpt}

In the past decades, traditional Eulerian flow-measurement techniques have been complemented by new methods to study individual flow elements in their Lagrangian frame. Using multiple synchronised high-speed cameras with overlapping fields of view, fluid motions of tracer particles have been used to further understand the energy transfer across scales in turbulent flows \citep{LaPorta:2000ka, Ouellette_2006, xu2006,Xu2007, Xu2008b}, and the behaviour of inertial particles in turbulent fluids has been studied as well \citep{Bourgoin2006,Xu2008,Gibert:2010cu,ewe-wei, ewe-weiII}. Although we have devised a laminar gas flow, one expects that the presence of the disperse particle phase will result in complex turbulent-like motions, if an instability develops. We therefore employ the same particle-tracking technique used in previous studies of turbulent flow, with customisation for the present setup. 

We acquire simultaneous shadowgraph images with three high-speed Phantom 10 cameras at a frame rate of 2000 Hz and frame size of 768 $\times$ 1024 pixels. The spatial scale of the recordings is $\sim12$  $\mu$m/pixel. After the image acquisition has taken place, the particle trajectory reconstruction consists of the following steps in the order stated: i) particle image finding; ii) stereoscopic position reconstruction; iii) tracking in time.  

In step (i), we perform a background image subtraction and then determine the sensor position of each particle in all movie frames by performing a Gaussian fit to pixel intensity maxima. In step (ii), we apply a least-square minimisation to the triangulation errors on a mapping between sensor position and real-world coordinates.\footnote{The laboratory-frame coordinates are known via a calibration procedure in which the three cameras shown in Figure~\ref{apparatus_image} were focussed on a grid-pattern calibration mask inserted in the centre of the tube. The mask was translated in the laboratory frame, with a calibration image being obtained at 2 mm separations.} 

\begin{figure}
\includegraphics[width=\columnwidth]{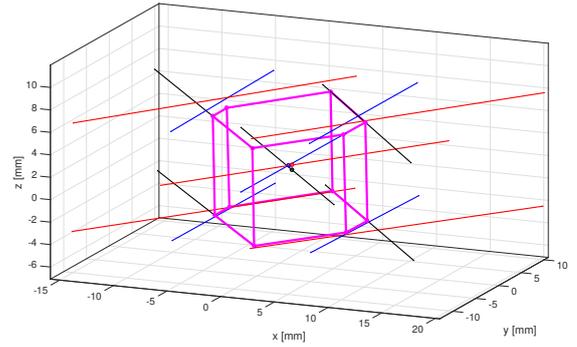}
\caption{The neighbourhood of the origin of the LPT measurements, where the black, red, and blue lines correspond to the lines of sight and centre point for cameras 0,1, and 2, respectively. The thick purple lines delineate the measurement volume, $V_{\rm{meas}}$, defined by the intersecting of lines of sight from all three camera sensors.}
\label{FigMeasVolume}
\end{figure}

It is common to construct the sensor-to-real-world mapping using a  `pinhole' model, which assumes that all points in the volume can be traced back to the imaging plane via a mutual intersection point. However, we must also account for aberrations in projected particle position due to the cylindrical walls of the experimental vessel, and we therefore implemented a generalised pinhole model that assumes that the lines of sight can pass through two mutually perpendicular slits. We also correct for smaller-scale distortions due to variations in the thickness of the experimental vessel wall by fitting a distortion profile along the horizontal axis. We verified the two-slit camera model goodness-of-fit by performing the reconstruction on the calibration images. We subtract the real-world coordinates from the reconstructed positions and find a standard deviation in offset of particle position of $\sim 2~\mu$m, which we take to be the error on particle position. 

In step (iii), the remaining task is to link the positions of particles in subsequent frames, thus producing trajectories in time. In each frame, likely particle positions are estimated based upon previous particle positions and a maximum velocity in the previous frame. Note that the particles' displacement, due to their velocity, is hundreds of $\mu$m, which is much greater than the noise due to the position reconstruction, stated above. At each frame in which a particle appears, we use the minimum variance linear extrapolation from the previous trajectory positions to determine which trajectory the particle belongs to. 

From the three-dimensional particle trajectories, the instantaneous velocity and acceleration at all times is calculated by applying a finite difference scheme, smoothed with a Gaussian filter \citep{mordant:2004}. The filter lengths for the velocity and acceleration were 10 and 40 frames, respectively.  We also calculate the moments and distributions of the velocity and acceleration, in addition to other explicit quantities that will be reported in the following sections. 

\section{Properties of the data} \label{character}

\subsection{Parameters and their definitions}\label{definitions}

We conducted experiments at fixed gas pressures in the range 200--800 Pa (2-8 mbar). Each experiment consists of a 6-s LPT recording, processed in the manner previously stated. A single data set includes several repetitions of the experiment ($\mathcal{O}(10)$) at a given gas pressure. To perform a statistical analysis, we combine the three-dimensional trajectories from all experiments of the data set. Table \ref{exp_table} summarises the parameters of the experiments, including the pressure at which the facility was operating and the total number of trajectories, $N_{\rm{traj}}$, comprising the data set. 

\begin{table*}
\caption{ Parameters of four data sets collected at different gas pressure (mbar), listed in the second column. The Reynolds numbers for both the container and particle diameters correspond to laminar gas conditions. The values of the volumetric filling factor, $\phi$, and spatially averaged solid-gas ratio, $\epsilon$, depend directly on the mean particle number, $n$, and/or the value of $V_{\rm{meas}}$, given in the text. The momentum diffusion time, $t_{\rm d}$, is less than the friction time, $T_{\rm f}$, in all cases, and both are in units of s. The particle diameter $d_{\rm p, St-Ep}$ is obtained using Equation~\eqref{draglaws}, where the prescript Ep or St indicates whether the particles belong to either the Epstein or Stokes drag regime. The particle diameter $d_{\rm p}$ derives from Equation~\eqref{cunningham}, and is the value assumed for the calculation of $Re$, $\phi$, $\epsilon$, $T_{\rm f}$, and $t_{\rm d}$. The variance on the mean settling velocity for all experiments in a data set is $\sigma_{\langle u_{\rm z}\rangle}$ and is used to estimate the range in mean particle size d$_{\rm p}$. The last column reports the number of trajectories in a given data set, which is the total sample size for the mean quantities displayed in the data analysis.}
\label{exp_table}
\centering
\footnotesize
\begin{tabular}{c c c c c c c c c c c c c}  
\hline\hline  
Data Set& P &$Re_{\rm C}$ & $Re_{\rm p}$&$n$  &  $\phi$& $\epsilon$ &
$T_{\rm f}$&$ t_{\rm d}$ &$d_{\rm p,St-Ep}$ [$\mu$m]&$d_{\rm p}$ [$\mu$m]&$\sigma_{\langle u_{\rm z}\rangle}$&
$N_{\rm{traj}}$\\
\hline
\texttt{kn1\_2mb [DS1]} &2.71 & 15 & 0.0006 &1 $\pm$1 &2.46e-08&  0.06 &0.107&
0.00128 &$_{\rm Ep}40.03\pm 1.8$&$40.53\pm 1.8$&0.06 &5758\\
\texttt{kn1\_3mb [DS2]} & 3.63 & 20& 0.0008&15$\pm$5 &7.05e-07 &  1.32& 0.113&0.0177 &$_{\rm Ep}56.39\pm 0.2$&$48.28 \pm 0.2$&0.005 &65470\\
\texttt{kng1\_5mb[DS4]} & 5.45 &  31  &0.003& 7$\pm$8 &3.83e-07   &0.47 &0.097 & 0.0137 &$_{\rm St}62.65 \pm 3.0$&$49.83 \pm 3.0 $& 0.09 & 99816\\
\texttt{kng1\_8mb[DS3]}&8.00 & 45 & 0.005 & 21$\pm$3 &1.40e-06 & 1.18& 0.099 &0.0653&$_{\rm St}63.35 \pm 1.1$&$54.54\pm1.2$& 0.03& 172615\\
\hline
\end{tabular}
\end{table*}

Flows with vastly different length-scales, $L$, will exhibit statistically similar velocity fields, if their Reynolds numbers, $Re=uL/\eta$, are the same. In the numerator of $Re$ is the flow's characteristic velocity $u$ and in the denominator is the kinematic viscosity: the dynamic viscosity $\mu$ divided by the fluid density $\eta=\mu/\rho_{\rm g}$. When the Reynolds number is high, the flow will become unstable and even turbulent. Conversely, in low-$Re$ flows the viscous damping is significant and the flows remain laminar. For gases with a normal temperature range, the dynamic viscosity is roughly constant, especially when the pressure is low. Therefore, low gas density $\rho_{g}$ means high $\eta$, consequently low $Re$, which implies a laminar flow. The dynamic viscosity of dry air at room-temperature in our vacuum system is $\mu_{\rm air} = 1.8 \times 10^{-5}$ kg m$^{-1}$ s $^{-1}$, which gives a kinematic viscosity of $\eta_{\rm air} =$ 2.0--7.9 $\times 10^{-3}$ m$^2$s$^{-1}$ for the pressure range $200 - 800$~Pa. Therefore for the pipe flow on the container scale, $Re_{\rm C} = D_{\rm pipe} \overline{u} / \eta \approx$ 15--45, and for flow at the scale of an individual particle with diameter $d_{\rm p}$, $Re_{\rm p}=d_{\rm p} v_{\rm rms,z} / \eta \ll 1$, where $D_{\rm pipe}$ is the pipe diameter, $\overline{u}=0.95$ms$^{-1}$ is the average velocity in the pipe, $d_{\rm p}$ is the particle diameter, and $v_{\rm rms,z}$ is the root-mean-square (rms) velocity of the particles in the $z$-direction. These values are listed in Table \ref{exp_table}. 

The mean number of particles in the measurement volume over all experiments in a data set is $n$, where the error reported is the standard deviation in $n$ across individual iterations of the experiment.
From $n$, we derive a mean volumetric filling factor, 
\begin{equation} \label{filling}
\phi =\frac{\pi}{6} n d^3_{\rm p} / V_{\rm meas}, 
\end{equation}
where $V_{\rm meas}$ is the volume of the measurement region, and spatially averaged mass loading 
\begin{equation} \label{loading}
\epsilon = \rho_{\rm p} / \rho_{\rm g} = \phi \rho_{\rm m,p} / \rho_{\rm g}, 
\end{equation}
where the material density of the stainless steel particles used in the experiments is $\rho_{\rm m,p} =8050$ kg m$^{-3}$. We also calculate the momentum diffusion time across the mean inter-particle distance, $t_{\rm d}=n^{-2/3}/\eta$, to demonstrate that the experiments are in a tight particle-coupling regime, as indicated by this time scale being shorter than the viscous relaxation time (see also equation~4 of \citealt{lambrechts}). 

\begin{figure*}[!h]
\includegraphics[scale=1]{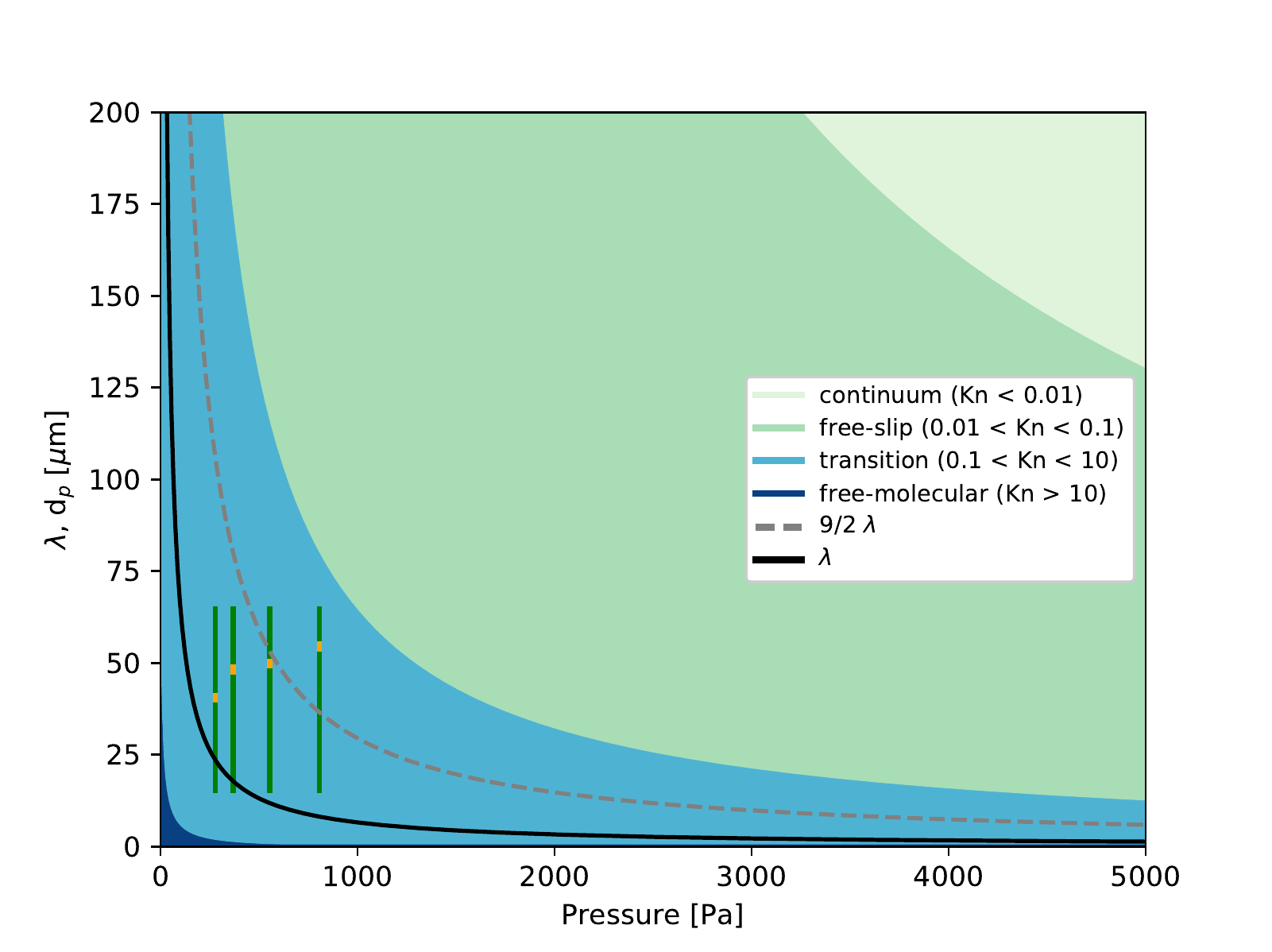}
\caption{\label{mfp} Flow conditions at the scale of particle size in the experiments. Black curve: the mean free path of gas molecules as a function of pressure $\lambda(P)$. Coloured swaths denote $Kn$-dependent drag regimes. Grey dashed curve: Stokes--Epstein transition. The four vertical green bars represent each of the four data sets; the position of the bars denotes the pressure, their height denotes the spread in particle size for all experiments, and their width denotes the offset error in the pressure measurement. The orange subsegments indicate the particle sizes that have the steady state velocity equaling the measured particle velocity in the experiment:$u_{\rm p,ss} = \langle u_{\rm pz} \rangle$. All the data sets can be considered transition flow, with the two at higher pressure further towards continuum and those at lower pressure approaching free molecular flow. 
} 
\end{figure*}

All of the experiments listed in Table \ref{exp_table} were conducted with the same distribution of particle diameters: 15--65 $\mu$m initially seeded into the apparatus. On the other hand, because the experiments are conducted in a flow regime where the velocity of particles depends upon $\rho_{\rm g}$, for each experiment under a given gas pressure, the flow naturally selects only particles with diameters in a subrange of this size distribution by settling and levitation. 

We determined the particle diameters by using the fact that at steady state, the measured relative velocity between the particles and the gas stream should equal the terminal velocity of a particle of size $d_{\rm p}$. It is common to assume a piece-wise drag-law transition for $Re_{\rm p}<1$,
\begin{equation} \label{draglaws}
F_{\rm D} = \frac{\pi}{6} d_{\rm p}^3 \rho_{\rm m,p} g =
  \begin{cases}
   3 \pi d_{\rm p}\mu_{\rm air} \delta u & \text{Stokes}, \\
   \frac{\pi}{3}\rho_{\rm g} d_{\rm p}^{2} v_{\rm th} \delta u      & \text{Epstein},  
  \end{cases}
\end{equation}
where $\delta u = u_{\rm g} - u_{\rm pz} $ is the difference between the $z$-component particle velocity, $u_{\rm pz}$, and the gas velocity, $u_{\rm g}=1.45$ ms$^{-1}$, $v_{\rm th} = \sqrt{8 R_{\rm g} T / \pi M}$ is the thermal speed of molecules of molecular weight $M$, and $R_{\rm g}$ is the universal gas constant.

The resistance that a sphere experiences is thus broadly classified into whether continuum viscous forcing, or rather impingement from individual molecules, is responsible for the drag. Between these two extremes, the boundary conditions of the flow around the particle will be characterised by increasing slip on particle boundaries. The appropriate drag regime \citep{Whipple:1972, Weidenschilling:1977a} is given in general by the Knudsen number: 
\begin{equation}
Kn \equiv \lambda/d_{\rm p}.
\end{equation}

Further sub-divisions in flow regime use the following as approximate guidelines (see, e.g. \citealt{Allen_Raabe:1985,Karniadakis_etal:2005}):
\begin{eqnarray}
   Kn < 0.01 && \text{continuum (Stokes)}, \nonumber\\
   0.01< Kn <0.1 && \text{continuum (with free-slip correction)}, \nonumber\\
   0.1< Kn <10 && \text{transition}, \nonumber\\
   10 < Kn  && \text{free molecular (Epstein)}\nonumber.
\end{eqnarray}

Figure~\ref{mfp} shows $\lambda$ as a function of pressure, with the $Kn$-dependent regimes color-coded for reference. Particles with sizes used in our experiments are represented by vertical green bars spanning $15 <d_{\rm p} < 65~\mu$m, with orange sub-segments indicating $d_{\rm p}$ selected by the flow. 

Equating the expressions for the Stokes and Epstein drag forces gives an algebraic dividing point for particle diameter of $(9/2) \lambda$, which is shown as a dashed line and divides our data sets in Figure~\ref{mfp}. 

While our data is representative of the two different drag regimes delineated by Equation~\eqref{draglaws}, Figure~\ref{mfp} also demonstrates that even the data collected at higher pressure can be considered transitional flow and that a purely continuum assumption for the gas flow would require significantly higher pressure or much larger particles. In observance of this issue, we further constrain d$_{\rm p}$, as is appropriate for transitional flow, using the empirically derived $Kn$-dependent Cunningham correction factor \citep{cunningham} to Stokes drag:
\begin{equation} \label{cunningham}
F_{D}=F_{D,\rm Stokes}/C_{\rm Kn}
\end{equation} 
with $C_{\rm Kn}=1.0+2Kn\left(\alpha+\beta \exp{(-\gamma/2Kn)}\right)$ . We adopt the values from \cite{Allen_Raabe:1985}$:\alpha=1.142$, $\beta=0.558$ and $\gamma=0.99$. The values of $d_{\rm p}$ obtained using this correction are shown in Table \ref{exp_table} and are those used to calculate the other listed parameters\footnote{Assuming a drag law with abrupt transition might lead to slightly over-estimating the particle size and hence deriving somewhat different values for the other parameters that depend on $d_{\rm p}$. However, the two approaches to calculating $d_{\rm p}$ do not yield qualitatively different results in terms of the flow or mass-loading regimes that result.}. For comparison, we list $d_{\rm p,St-Ep}$ obtained using Equation~\eqref{draglaws}. In both cases, the errors are calculated by propagating the variance on the mean settling velocity in the $z$-direction, denoted $\sigma_{\langle u_{\rm z} \rangle}$ and also listed in table \ref{exp_table}, and demonstrate narrow dispersion in particle size. The quantity $\sigma_{\langle u_{\rm z}\rangle}$ is distinct from and much smaller than the fluctuation velocities reported in table \ref{rms_table} below.

\begin{figure}
\includegraphics[scale=.57]{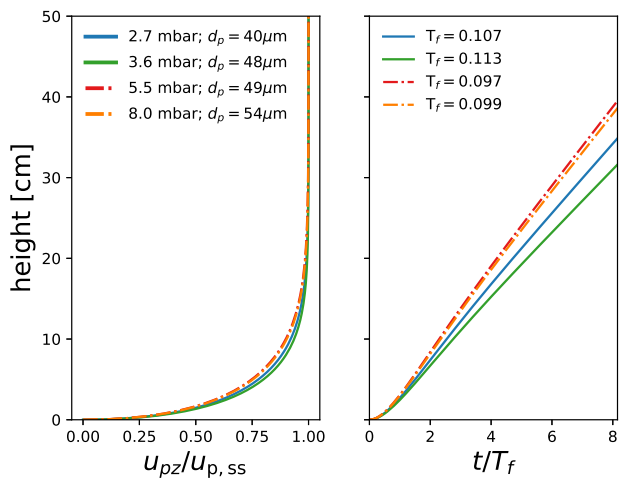}
\caption{\label{coupling_distance} Left: height as a function of the ratio of particle velocity to the steady state velocity $u_{\rm pz}/u_{\rm p,ss}$.  Right: height as a function of time in units of the viscous relaxation time scale $T_{\rm f}$.  Particles reach a steady state well before reaching the measurement window at $\sim 50$ cm.}
\end{figure}

We verify that the particles have had time to reach their terminal velocity before reaching the measurement window, which is h$_{\rm meas}=$ $5\times D_{\rm pipe} \sim 50$ cm downstream the bottom mesh where the particles settle when there is no flow, by solving the dynamic equation for particle velocity,

\begin{equation}
\frac{d u_{\rm pz}}{dt}=\frac{u_{\rm g}}{T_{\rm f}}-\frac{u_{\rm pz}}{T_{\rm f}}+g, \nonumber\\
\end{equation}
which has the solution
\begin{equation} \label{terminalvelocity}
u_{\rm pz} = (u_{\rm t} -u_{\rm g})\left(e^{\frac{-t}{T_{\rm f}}}-1\right) = (g T_{\rm f} - u_{\rm g})\left(e^{\frac{-t}{T_{\rm f}}}-1\right),
\end{equation}
where $u_{\rm t} = g T_{\rm f}$ is the terminal velocity. At long times, i.e. when $t \gg T_{\rm f}$, $u_{\rm pz}$ reaches its steady state value $u_{\rm p,ss} = u_{\rm g} - u_{\rm t}$. Equation~\eqref{terminalvelocity} is used to obtain the relation between the height $z_{\rm p} = \int_0^t u_{\rm pz} dt$ above the bottom mesh and the normalized particle velocity $u_{\rm pz}/u_{\rm p,ss}$, and between $z_{\rm p}$ and the normalized time $t/T_{\rm f}$. The results are plotted in Figure~\ref{coupling_distance}, which shows that particles closely approach steady-state velocity in no more than 20 cm above the bottom mesh, which corresponds to a time approximately 4--6 $T_{\rm f}$ and that is indeed justified for the time-dependent term to be neglected according to Equation~\eqref{terminalvelocity}. 

\subsection{Summary statistics}\label{summary_stats}

The particle velocities, in each of the data sets listed in Table \ref{exp_table}, are by far dominant in the vertical-component, with $0.3~\text{m s}^{-1} \lesssim \langle u_{\rm pz} \rangle \lesssim 0.5~\text{m s}^{-1}$. 
By comparison, the horizontal components are $0.001 \lesssim  \langle u_{\rm px,y} \rangle \lesssim 0.005 \text{ ms}^{-1}$.  
Note that in the laboratory frame, the particles travel vertically up during the experiments, against gravity. However, because the mean gas velocity is much larger, the particles are actually falling downward relative to the gas, with a relative velocity $\delta u$ in the range of $\sim$0.9--1.2~ms$^{-1}$. 

We also find comparably larger velocity spread in the vertical than in the horizontal components, as quantified by the rms velocity for each component $i$, defined as $u_{rms,i} \equiv \langle u_i^2 \rangle^{1/2}$, which we also refer to as the velocity dispersion or fluctuation. Table \ref{rms_table} shows $u_{rms,i}$ at the gas pressures we used in the experiments (data set). 
\begin{table}
\caption{The component-wise rms velocities in units of ms$^{-1}$. Horizontal velocity fluctuations, $u_{{\rm rms},x}$ and $u_{{\rm rms},y}$, are typically 1-2 orders of magnitude lower than those in the vertical direction, $u_{{\rm rms},z}$. The velocity dispersion generally decreases with lower gas pressure.}
\label{rms_table}
\centering
\footnotesize
\begin{tabular}{c c c c }  

Data Set &$u_{{\rm rms},x}$&$u_{{\rm rms},y}$&$u_{{\rm rms},z}$\\
\hline
\texttt{kn1\_2mb [DS1]} & $ 4.99\times 10^{-3}$& $ 5.66\times 10^{-3}$&
  $ 8.18\times 10^{-2}$\\
\texttt{kn1\_3mb [DS2]} & $4.37\times 10^{-3}$& $4.68\times 10^{-3}$&
  $6.76\times 10^{-2}$\\
\texttt{kng1\_5mb [DS4]} & $7.29\times 10^{-3} $& $ 6.94\times 10^{-3}$&
  $ 1.45\times 10^{-1}$\\
\texttt{kng1\_8mb [DS3]} & $2.68\times 10^{-2} $& $2.93\times 10^{-2} $&
  $1.81\times 10^{-1} $\\
 
\hline

\end{tabular}
\end{table}

We measured gas-velocity fluctuations at the flow centre-line of 0.0098 ms$^{-1}$ when the pressure is at 10 mbar, corresponding to $Re_{\rm C}=56$, which is larger than the Reynolds number at any of the lower pressures at which our experiments were conducted (see Figure~\ref{central_profile} and Table \ref{exp_table}). Such fluctuations in the gas phase velocity may contribute to the dispersion in vertical particle velocity. Its effect is expected to decrease at lower gas pressures as the Reynolds number is smaller, which is consistent with the observed smaller particle velocity fluctuations at lower gas pressures. Given that the fluctuations in the gas are smaller than those in the particle phase, the trend of increasing particle velocity fluctuations with increasing gas pressure is most likely related to the different dynamics associated with the drag regime to which the particles belong, which will become evident due to the results in Section~\ref{cont-drag}. Note that the dispersion in particle size can also contribute to the rms velocities, but, the fluctuations are much too large to be accounted for by this effect. While we have found that there is a small variance on the mean settling velocity, this does not effect the rms values in Table \ref{rms_table} because the global mean of the data set is not subtracted as an ensemble. In the appendix, we plot the probability density functions (PDFs) of instantaneous particle velocity and acceleration, which show tails much wider than Gaussian.

The possibility of a fluid instability complicates arguments based on simple terminal velocity assumptions of the particles, since one would expect a complex velocity field. However, the low level of gas velocity fluctuations compared to the mean gas flow (i.e. see shaded region in bottom panel of Figure~\ref{central_profile}) and compared to the values of $u_{\rm{rms,z}}$, together with the fact that the effect of the mean shear in the gas flow is negligible, indicates that the gas flow itself is not the direct reason to cause inhomogeneities in the particle phase, if there are any.

In the following, we therefore present our analysis of the particle motions with the focus on particle--pair relative velocity statistics. We will use $\vec{u}$ and $u_{i}$ to denote particle velocity and its components and drop the subscript ``$_{\rm p}$'' for simplicity, since we henceforth only discuss particle velocity and its statistics.

\section{Results}\label{results}

In this section, we present the results of the experiments conducted at the conditions given in Table \ref{exp_table}. We present an overview of the magnitudes of the relative motion between particles in Section~\ref{sec:magnitude}, which shows that the relative motion between particles is not determined by their terminal velocities caused by their size differences, nor by the gas velocity profile. In Section~\ref{sec:relvel}, we consider the components of the relative velocity, which demonstrates the occurrence of particle aggregation and clustering on small scales.

\subsection{Magnitude of the relative motion between particles}\label{sec:magnitude}

If particles of the same size in a uniform gas stream are moving at their terminal velocities relative to the gas as if they are isolated, then the relative velocity among particles would be zero. Even if we consider the dispersion of particle size (and hence their terminal velocity), the statistics of the relative velocity between particles would not depend on the distance between particles, provided that particles with different sizes are uniformly distributed in space. In this section, we first study the variation of the magnitude of the relative velocity between particles with the distance between particles.
  
At any instance $t$, for two particles located at positions $\vec{x}$ and $\vec{x}+\vec{l}$, with velocities $\vec{u}(\vec{x}, t)$ and $\vec{u}(\vec{x}+\vec{l}, t)$, respectively, their relative velocity is $\delta \vec{u} = \vec{u} (\vec{x} + \vec{l}) - \vec{u}(\vec{x})$.
In order to see how the statistics of $\delta \vec{u}$ depend on $\vec{l}$, we consider the following quantity:
\begin{equation}
D_{jk}(\vec{l})  \equiv \left\langle [u_{j}(\vec{x}+\vec{l}, t) - u_{j}(\vec{x},t)] [u_{k}(\vec{x}+\vec{l},t) - u_{k}(\vec{x},t)] \right\rangle ,
\label{structure_function}
\end{equation}
where ``$\langle \cdot \rangle$'' means ensemble average, i.e. average over all particle pairs that are separated by $\vec{l}$, for all $\vec{x}$ and $t$. 
If the particles are fluid tracers, then the quantity $D_{jk}$ defined by Equation~\eqref{structure_function} is the so-called Eulerian second-order velocity structure function of the flow field. For heavy particles, like those we study here, it gives us a direct measure of the magnitude of the relative velocities between particles. As we are more interested in how two particles approach or separate from each other, we consider only the second moment of the relative velocity component in the direction along their separation vector, i.e.
\begin{equation}
D_{ll}(l) \equiv \left\langle \left[( \delta \vec{u}) \cdot \frac{\vec{l}}{l} \right]^2 \right\rangle .
\label{eq:Dll}
\end{equation}

\begin{figure}[!h]
\includegraphics[scale=.57]{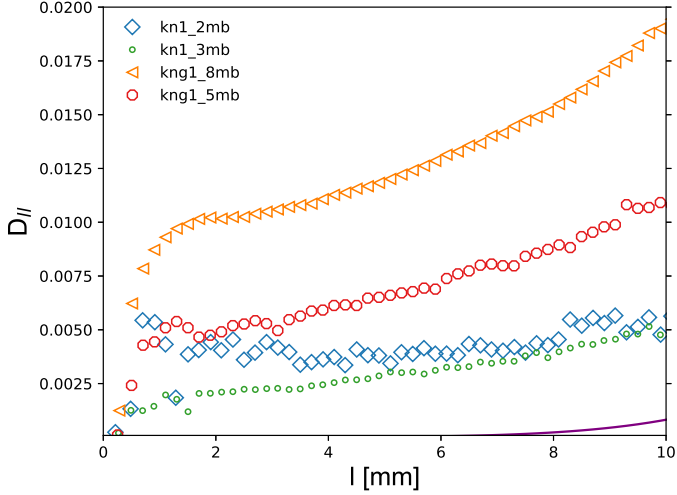}
\caption{\label{struct_a} The second-order moment of the longitudinal relative velocity component, $D_{ll}(l)$, versus $l$ for all data sets.
The solid curve corresponds to the magnitude of the velocity difference of the gas flow profile as a function of radial distance from the flow centre-line. The contribution to $D_{ll}$ from the variation in shear velocity in the gas phase is thus negligible.}
\end{figure}

In Figure~\ref{struct_a}, $D_{ll}(l)$ for all data sets are shown as open symbols. The dependence on scale $l$ is clearly visible.
A solid-line curve is plotted to show the difference in gas velocity between a point on the vessel’s centre-line and a point at radial distance l, due to the Poiseuille flow profile. If particles simply adjust to the spatially varying gas velocity, this would be the maximum possible velocity differences they could have. The curve is well below all measured values of $D_{ll}$, which means that the variation of $D_{ll}$ with $l$ cannot be attributed to the variation in the gas flow profile. This is another evidence that the mean gas flow does not contribute much to the particle velocity variations.

For all four data sets, $D_{ll}$ are rather small at separations less than $\sim 1$ mm, indicating that particle velocities may be correlated at very small separations. 
In all cases, $D_{ll}$ increases with $l$, showing that statistically the relative velocity increases with separation. The result from data set kn1\_2mb [ds1] is noisy, which is due to the low number density in the experiment. As shown in Table \ref{exp_table}, there is on average 1 particle in the measurement volume at any given time, which results in poor statistics of particle relative velocity. Similarly, data set kng1\_5mb also has relatively low number density and the statistics suffers from convergence problem as well, at least at small separations.

The second moment of the relative velocity shows how intense the relative motion is. While some hints of collective motion can be implied from that statistics, it is more informative to study directly the average relative velocity when checking for particle aggregation, as we will do next.
Since kn1\_2mb [DS1] is not a large enough sample for relative velocity analysis, we do not consider this data set further. Of the other three data sets which exhibit the potential for collective behaviour, we select kn1\_3mb [DS2] as representative of the Kn$\sim$1 regime and kng1\_8mb [DS3] as representative of the Kn$<1$ regime. We only show kng1\_5mb [DS4] when the results are not obscured by the wide range of particle number density in the experiments comprising the data set; note from Table \ref{exp_table} that the fluctuation in the number density $n$ is greater than the mean, hence although there are some high number density experiments, the mass loading is on average low. 

\subsection{Particle aggregation revealed by the relative velocity}\label{sec:relvel}
 
Now we explore the details of the relative motion between particles. In the last section, we presented statistics only as a function of distance, without concerning the directional variations. The particle motion is actually rather axisymmetric, as indicated by the fluctuation velocities shown in Table \ref{rms_table}. Therefore, we present here the average of relative motion in cylindrical coordinates, where the $z$-axis is the same as in the laboratory frame and the $\vec{r}$-coordinate is in the horizontal plane. The origin, however, is not fixed, but set on any given particle, i.e. we study the relative velocity
\begin{equation}\label{delta_uz}
\delta u_i (\delta \vec{r}, \delta z) \equiv u_i (\vec{r} + \delta \vec{r}, z + \delta z) - u_i (\vec{r}, z),
\end{equation} 
which is a function of $\delta \vec{r}$ and $\delta z$ but does not depend on $\vec{r}$ and $z$ for homogeneous flow fields, which is an accepted approximation for our cases as we perform the measurements in the central region of the apparatus, where the motion of particles is hardly influenced by the wall. Moreover, because of the symmetry of our setup, the relative velocity does not depend on the direction of $\delta \vec{r}$. Thus, the relative velocity varies only with the magnitudes of $\delta r$ and $\delta z$.

\begin{figure}[ht]
\begin{overpic}[scale=0.45]{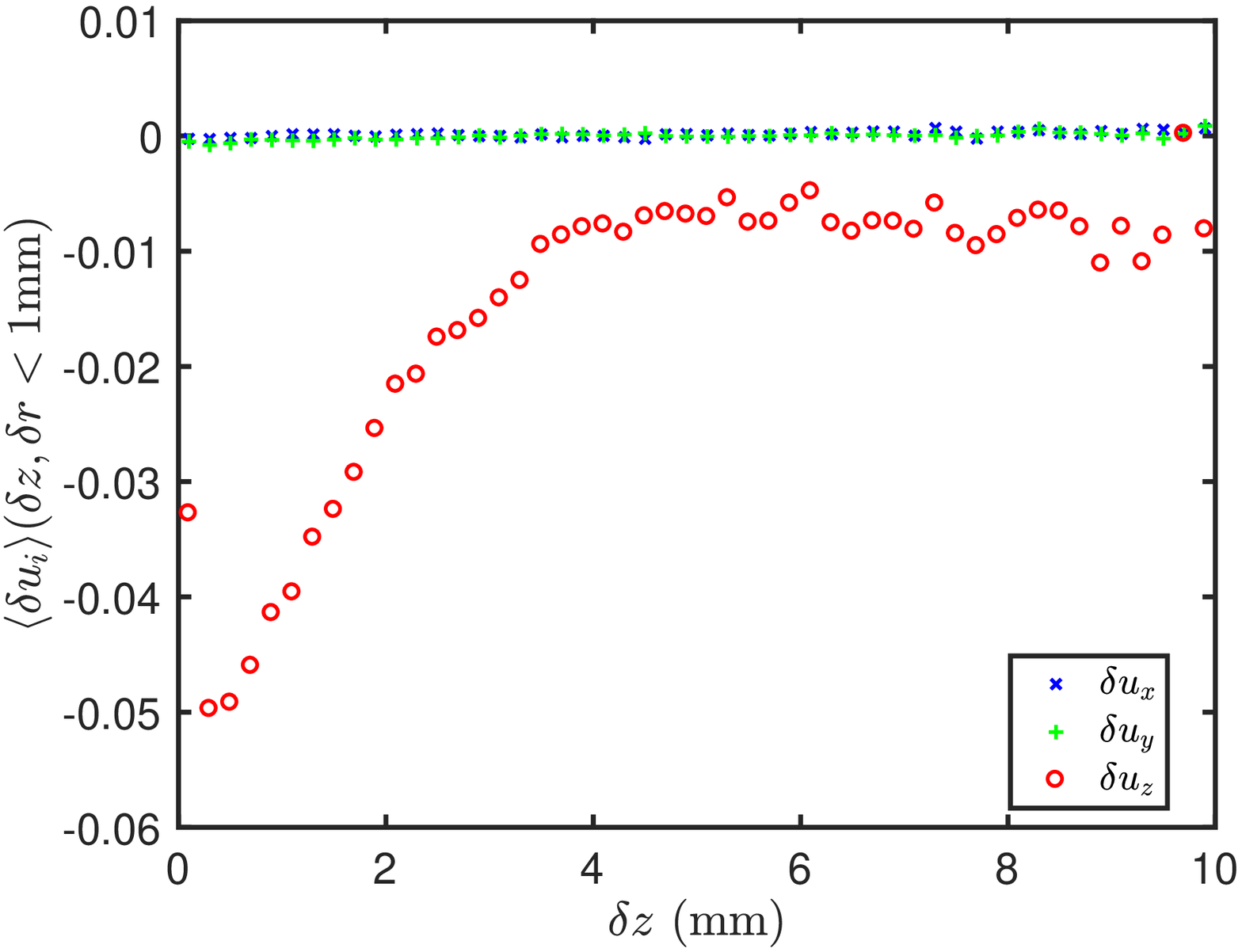}
\put (100,180) {$Kn\sim1$}
\end{overpic}\\

\begin{overpic}[scale=0.45]{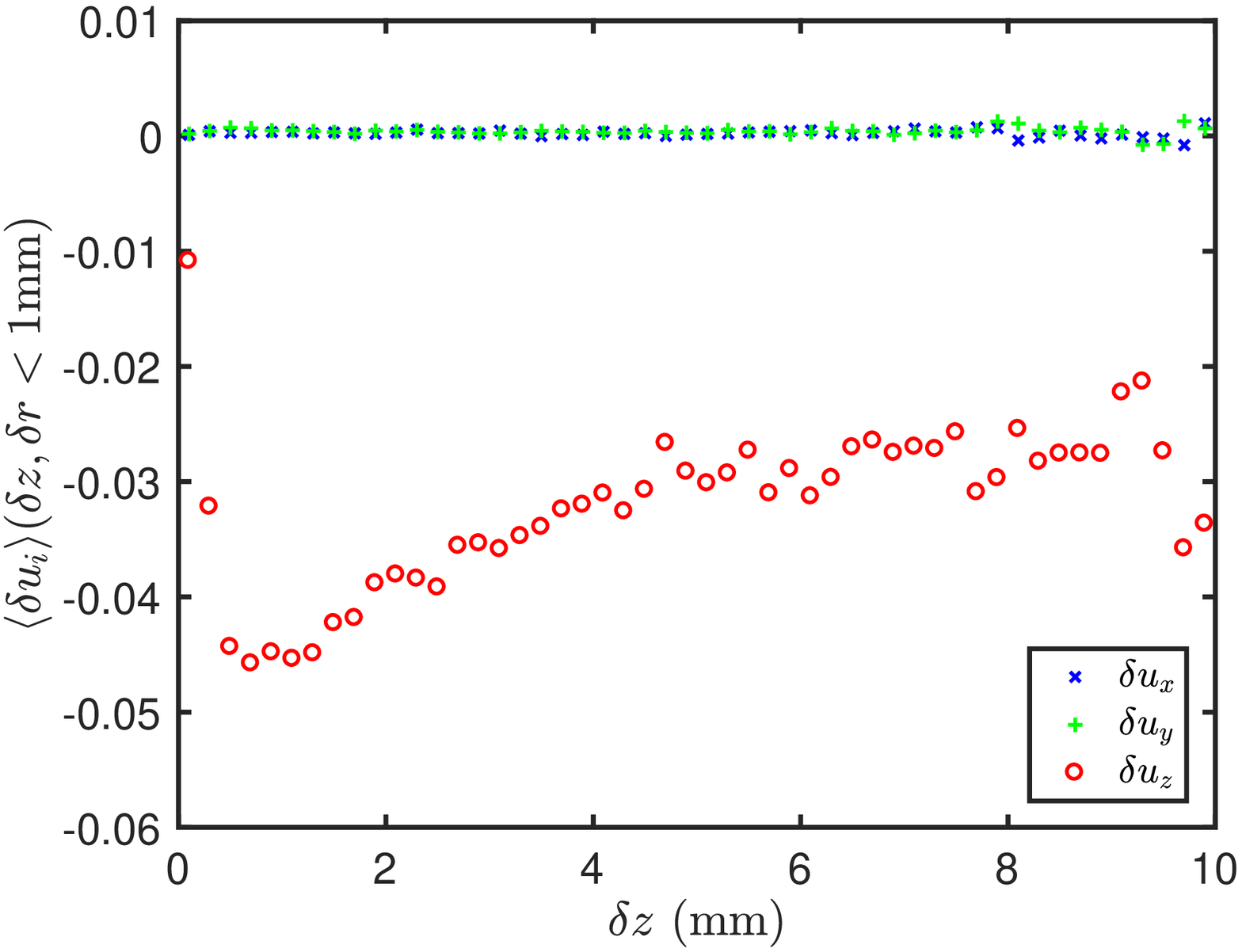}
\put (100,180) {$Kn<1$}
\end{overpic}
\caption{\label{cyllinder_a} Mean relative velocity components between particle pairs in units of m~s$^{-1}$, as a function of vertical separation. The $Kn \sim 1$ data is shown on the top and the $Kn<1$ data on the bottom. Particles aggregate for separations of 4~mm or less in both data sets. }
\end{figure}

For the two data sets of $Kn \sim 1$ and $Kn <1$, the mean of $\delta u_i (\delta r < 1 {\rm mm}, \delta z)$, as a function of $\delta z$ is shown in Figure~\ref{cyllinder_a}, in which the statistics of $\delta u_i$ is limited to the cases with two particles radially very close, $\delta r < 1 {\rm mm}$, i.e. the two particles are nearly one on top of each other, but with a vertical distance $\delta z$.

The means of $\delta u_x$ and $\delta u_y$ are nearly the same, thus supporting our axisymmetry arguments. Both $\langle \delta u_x \rangle$ and $\langle \delta u_y \rangle$ are negligibly small, which is consistent with the very small fluctuation velocities in the horizontal plane (see Table \ref{rms_table}).

The relative velocity in the vertical direction, however, is interesting. Here a negative $\langle \delta u_z \rangle$ means that on average the particle that is below is moving faster and hence would ``catch up'' the one that is above, which implies aggregation. Particles clearly aggregate towards one another at an increasing rate for diminishing separation in $\delta z$. This behaviour occurs for both the $Kn \sim1$ and $Kn<1$ cases. The separation at which this effect is most pronounced is for $\delta z \lesssim 4$~mm. In neither data set does the relative velocity return to zero for increasing separation, discussed below.

\begin{figure}[ht]
\begin{overpic}[scale=0.45]{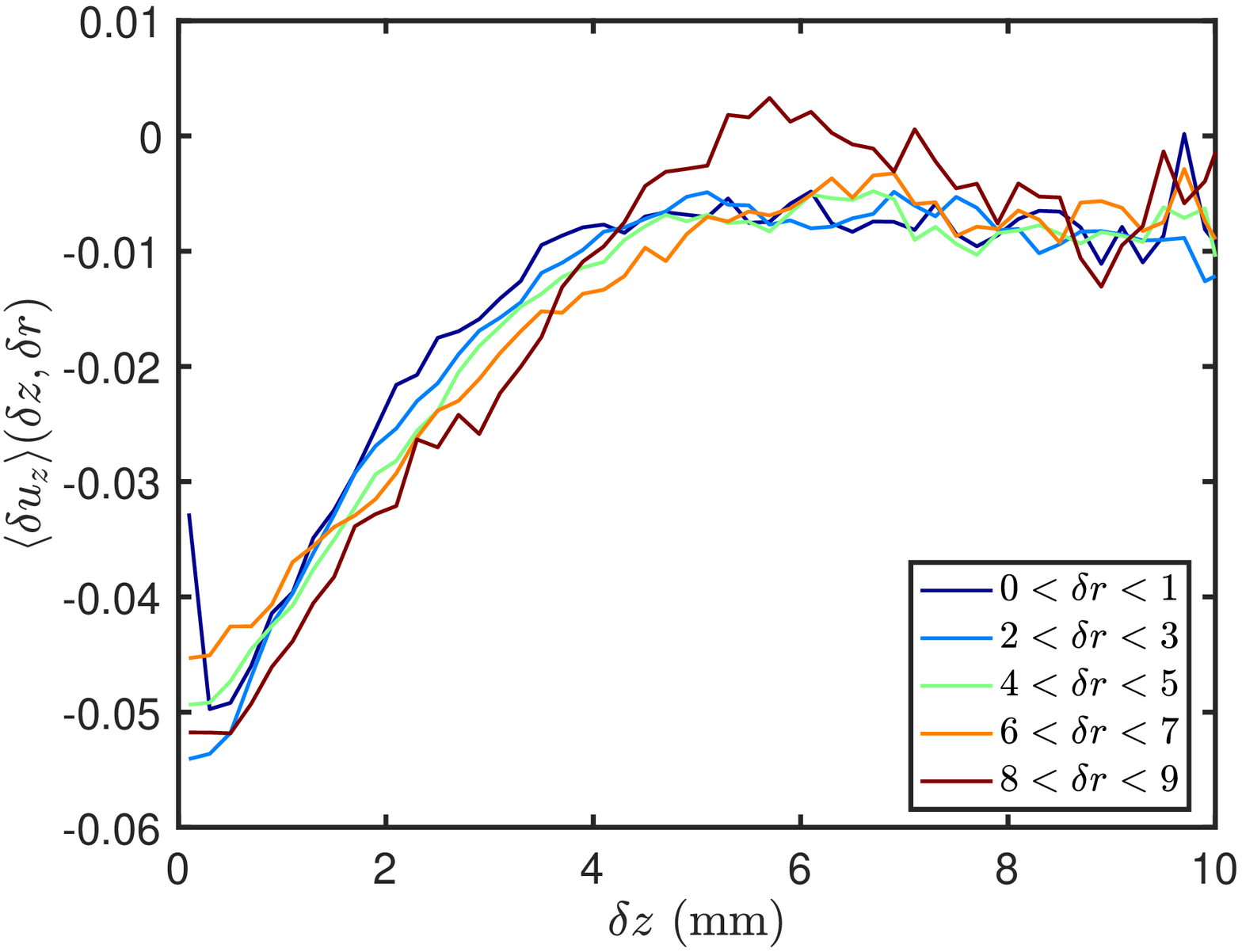}
\put (100,180) {$Kn\sim1$}
\end{overpic}\\

\begin{overpic}[scale=0.45]{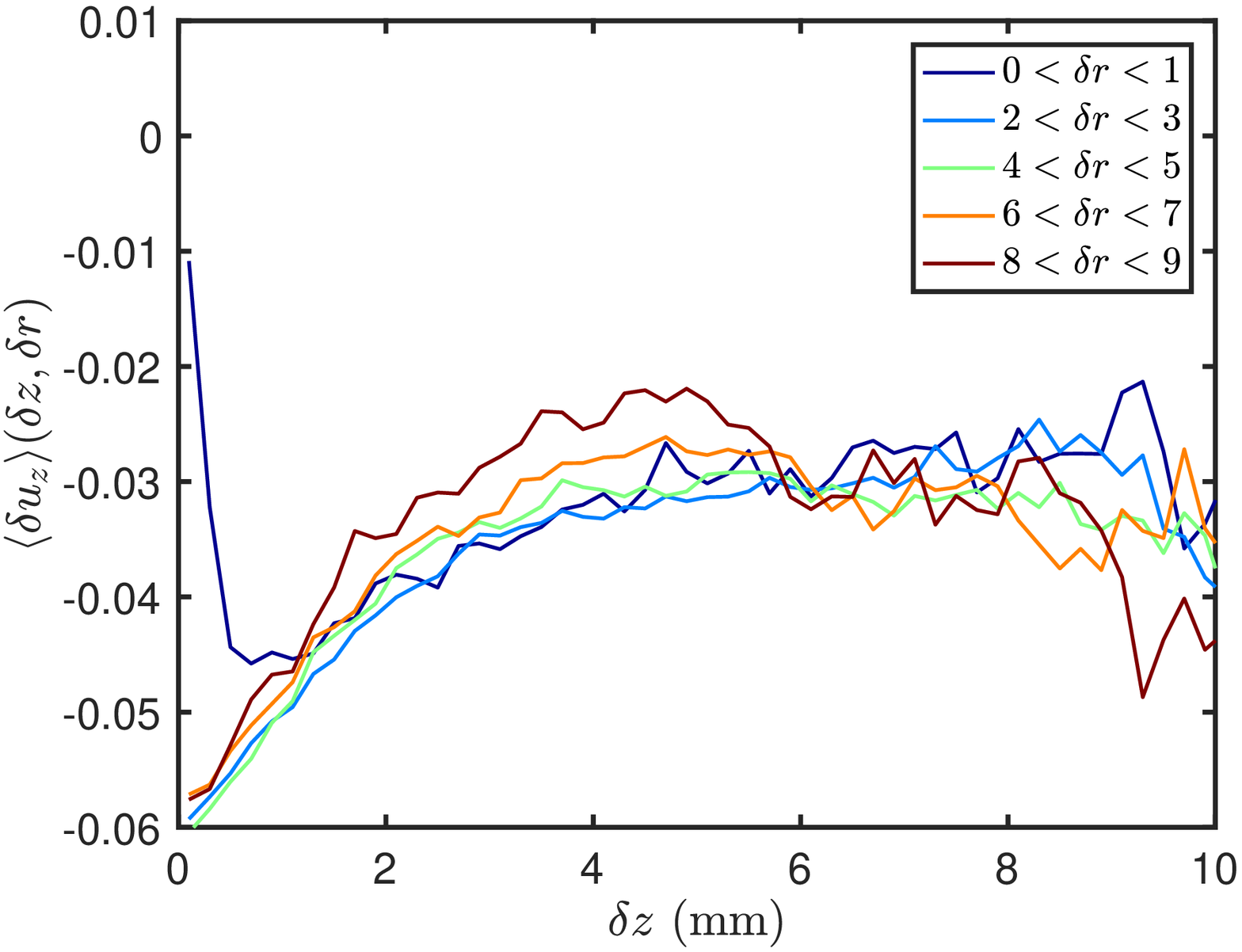}
\put (100,180) {$Kn<1$}
\end{overpic}

\caption{\label{cyllinder_b} Mean vertical relative velocity between particle pairs in units of m s$^{-1}$, as a function of vertical separation. The $Kn \sim1$ data is shown on the top and the $Kn<1$ data on the bottom. The different curves refer to increasingly large values in $\delta r$. The aggregation dynamics is the same at all radial (horizontal) separations, suggesting the formation of a particle layer.}
\end{figure}

In Figure~\ref{cyllinder_b}, we plot $\langle \delta u_z (\delta z) \rangle$ for other ranges of $\delta r$, i.e. the second particle lies in a cylindrical shell with different radius around the first particle. The statistics of the horizontal velocity components $\langle \delta u_x \rangle$ and $\langle \delta u_y \rangle$ is not shown, since they are nearly vanishing in all cases. Results in Figure~\ref{cyllinder_b} show very weak dependence on $\delta r$, which implies that the aggregation occurs in the vertical direction, at least at the scale of the measurement volume. If this is due to the instabilities in the motion of the particle phase, the wavelength of the unstable modes should be at least of the size of the measurement volume.
Theoretical calculations presented in \cite{lambrechts} predict that the smallest length scale, $\lambda_{\text{knee}}$, of the particle flow should be around $\lambda_{\text{knee}}=0.08 L_{\rm f}$. Using values of $T_{\rm f}$ from Table \ref{exp_table}, $\lambda_{\text{knee}} \approx 1$~cm for the two data sets discussed here, which is consistent with our observation.

Figures~\ref{cyllinder_a} and \ref{cyllinder_b} show that $\delta u_{z}$ does not vanish even at the largest measured separation $\delta z \approx 10$~mm, consistent with our estimate above that the smallest length scale of the unstable modes in particle motion is $\sim 1$~cm, i.e. our entire measurement volume is always in one of such ``contraction regions'' when we observe particles. Please see also details of the higher-order behaviour of the same quantity in Appendix \ref{appendix:moments}. We acknowledge that the apparent limiting behaviour may be consistent with a developing phase lag, of the type proposed in \cite{youdin_singlefluid}. Further identification of the mechanism requires simultaneous measurements of the local gas speed or pressure, together with inertial particles, which is a very challenging task and beyond the capability of our apparatus.

\subsection{Clustering of particles seen from concentration variation}\label{clustering}

The statistics of the relative velocity between particles indicates that particles tend to aggregate, which implies that the spatial distribution of particles cannot be uniform, nor stationary. Now we explore how the local concentration of particles varies with time. To that end, we divide the measurement volume into horizontal slabs, each of height 1 mm, and count the number of particles in each slab, which is then a function of time, i.e. we measure $n^\prime (t, z)$, where $n^\prime$ is the number density in the 1 mm slab and $z$ is in steps of 1 mm. 
In general, the spatial and temporal variation of $n^\prime(t,z)$ can be studied by the two-position, two-time correlation function
\begin{equation}\label{crossc}
C_{nn} ( \tau, \delta z) \equiv \frac{\langle n^\prime (t, z) n^\prime (t+\tau, z+\delta z) \rangle}{\sigma^{2}_n(z) \sigma^{2}_n(z+\delta z)},
\end{equation}
where $\sigma^{2}_{n}(z)$ and $\sigma^{2}_{n}(z+\delta z)$ are the variances of $n'$ at position $z$ and $z+\delta z$, respectively.

Because the mean particle velocity in the laboratory frame $\langle u_{\rm pz} \rangle$, which is roughly 0.4 m~s$^{-1}$ for all cases, is much larger compared to the fluctuation velocities, the change in particle relative position is very small when particles fly pass the measurement volume. Hence the particle configuration might be regarded as ``frozen'' in the measurement and the number concentration $n^\prime$ measured at fixed $z$ but different time could be used as a surrogate of the instantaneous number concentration in the vertical direction.\footnote{This is a standard approach in turbulence research when performing fixed-probe measurement in a wind tunnel, where it is called ``Taylor's frozen turbulence hypothesis'' and it is valid as long as the fluctuation velocity is much smaller than the mean \citep{tennekes:1972,pope:2000}.} Under this approximation, the spatial variation of particle concentration can be probed by the single-location, two-time autocorrelation function of $n^\prime$ at times $t$ and $t+\tau$
\begin{equation}\label{autoc}
C_{nn} ( \tau, 0) \equiv \frac{\langle n^\prime (t, z) n^\prime (t+\tau, z) \rangle}{\sigma_n^2(z)}.
\end{equation}
While $C_{nn}(\tau, 0)$ can provide information on the average size of the local concentration variation, the more revealing statistics is actually $C_{nn}$ conditioned on the local values of $n^\prime (z)$. We perform two conditional statistics, one for the local number density being greater than 1.5 times the mean number density, i.e. $C_{nn}(\tau, 0) \vert n^\prime(t,z)>1.5\bar{n}(t;z)$, and the other for it being less than 0.5 times the mean number density, i.e. $C_{nn}(\tau, 0) \vert n^\prime(t,z)<0.5\bar{n}(t;z)$.
We refer to the former conditioning as the \emph{dense} case and the latter as the \emph{dilute} case.

\begin{figure}[!h]
\includegraphics[scale=0.4]{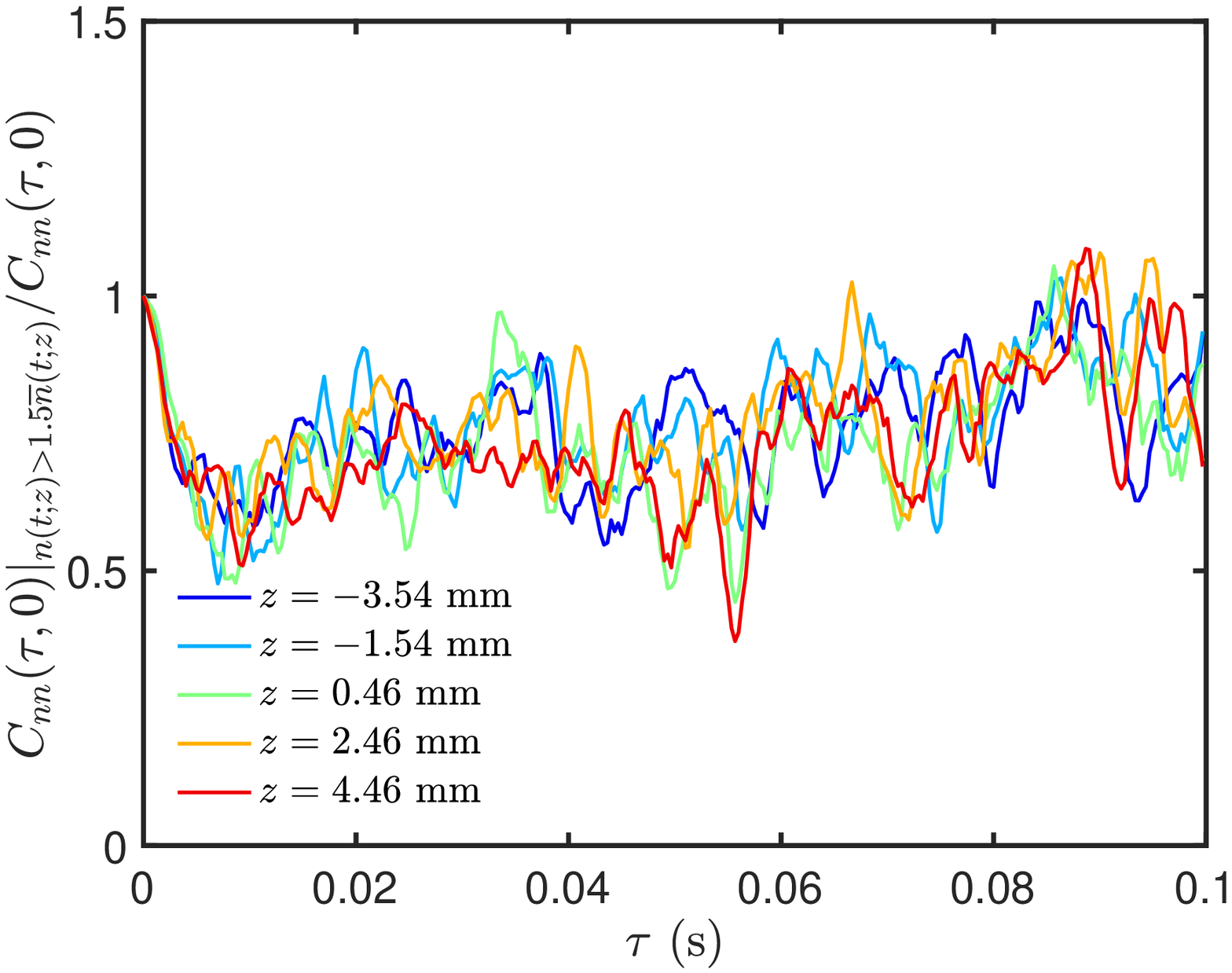}\\
\includegraphics[scale=0.4]{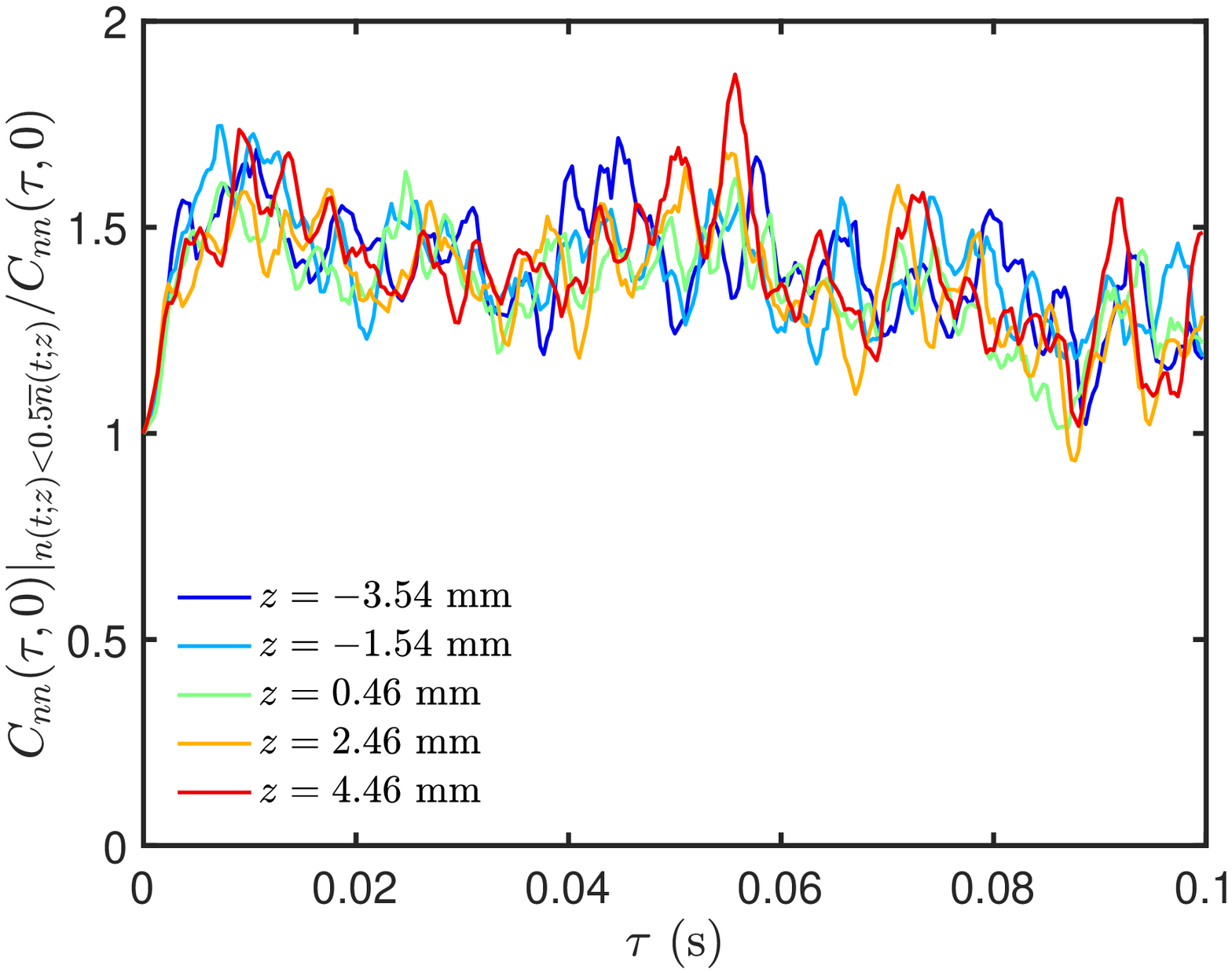}\\
\caption{\label{autocorrb_ds2} Conditional autocorrelation function of particle number
  density, normalized by the unconditional correlation, for the $Kn \sim 1$ case. 
  Top: conditional on the dense case; bottom: conditional on the dilute case. Heterogeneity in the particle density is evident from the ratios deviating from unity. The time scale for the ratio to return to unity corresponds to a characteristic length scale of the clustering. }
\end{figure}

In Figure~\ref{autocorrb_ds2}, we plot the conditional number density concentration correlation normalized by the unconditional ones, i.e. $C_{nn}(\tau,0) \vert n^\prime(t,z)>1.5\bar{n}(t,z) / C_{nn}(\tau, 0)$ and $C_{nn}(\tau,0) \vert n^\prime(t,z) < 0.5\bar{n}(t,z) / C_{nn}(\tau, 0)$, for the $Kn \sim 1$ case.

The dense-case ratio,  $C_{nn}(\tau,0)|n^\prime(t,z)>1.5\bar{n}(t,z)/C_{nn}(\tau,0)$, is always less than one for all $z$.  The converse is true for the dilute-case ratio, $C_{nn}(\tau,0)|n^\prime(t,z)<0.5\bar{n}(t,z)/C_{nn}(\tau,0)$. To gain intuition for this result, we consider a limiting case: if the particles were arranged in a perfectly homogeneous fashion with the inter-particle separation much smaller than the volume under consideration, then the correlation would remain close to one. Consequently, there would be no distinction between the dense and dilute cases and so the ratio represented by Figure~\ref{autocorrb_ds2} would also remain constant at one. 
On the other hand, if the particles are non-uniformly distributed, when we start the statistics from a locally dense region, then the correlation would decrease as we move away from the dense region. Hence, the normalized conditional correlation is smaller than one, and similarly, the normalized correlation conditioned on the dilute case is larger than one.
Furthermore, the extent of the deviation from one is an indication of the size of the inhomogeneity in number density. One expects for the local number density to de-correlate faster, the more localised that inhomogeneities of the mixture are. 
At larger distances, the effect of local concentration should vanish and the ratio of either the dense or the dilute case to the unconditional value of $C_{nn}$ should eventually return to 1, which is indeed what we observe in Figure~\ref{autocorrb_ds2}. 

\begin{figure}[!h]
\includegraphics[scale=0.4]{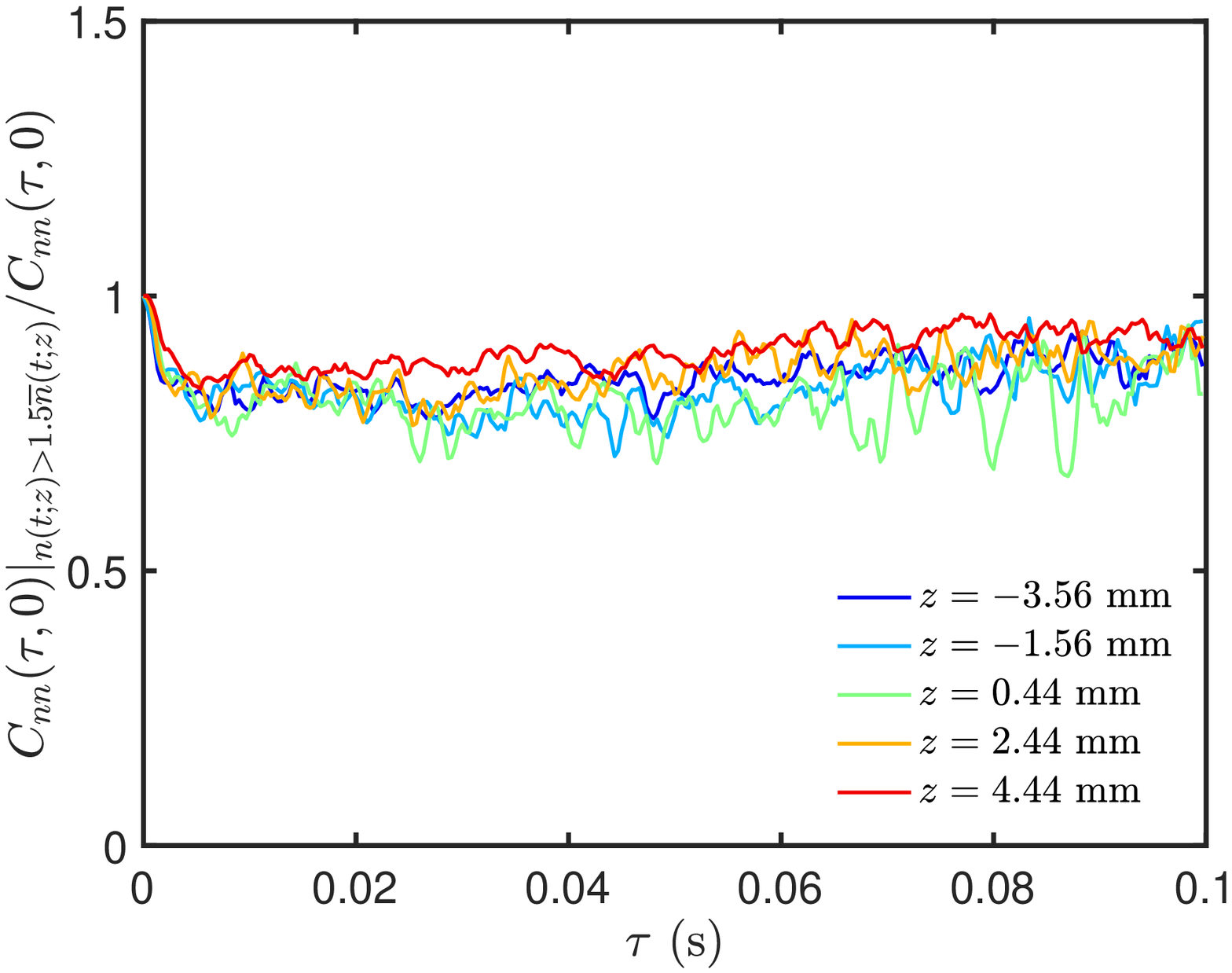}\\
\includegraphics[scale=0.4]{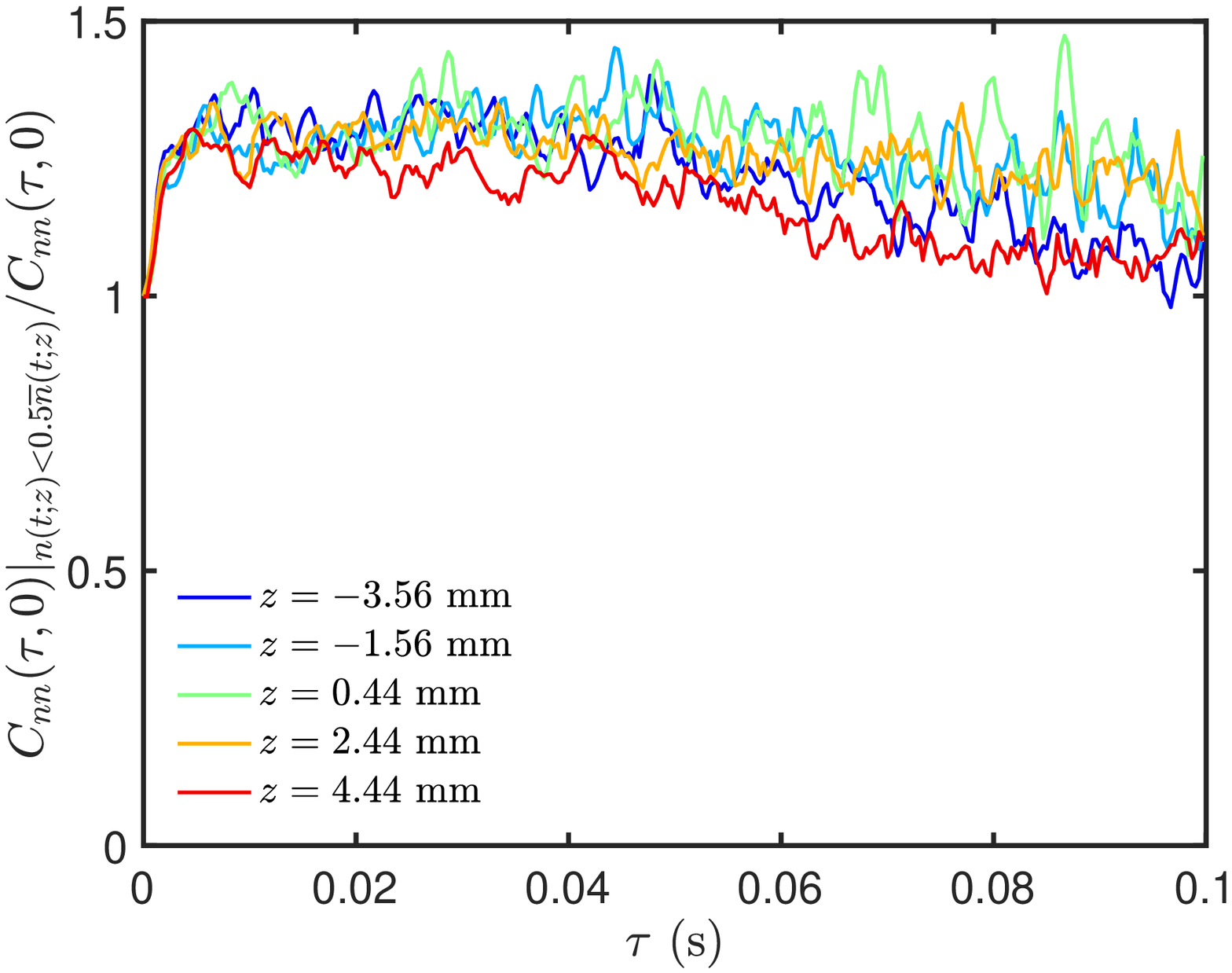}\\
\caption{\label{autocorrb_ds3} Same as Figure~\ref{autocorrb_ds2}, for the $Kn < 1$ case, showing generally similar clustering behaviour.}

\end{figure}
In Figure~\ref{autocorrb_ds3}, we plot the conditional number density concentration correlation normalized by the unconditional ones for the $Kn < 1$ case, which are basically the same as for the $Kn \sim 1$ case, except that the magnitude of the ratios is slightly smaller, which could be due to the relatively dense overall seeding.

From Figures~\ref{autocorrb_ds2} and \ref{autocorrb_ds3} we see that the inhomogeneities in particle number density de-correlate in approximately 0.03--0.05~s, which can be converted to approximately 1--2~cm when we use the mean particle velocity in the laboratory frame $\langle u_{\rm pz} \rangle \approx 0.4$~m~s$^{-1}$. This length scale is actually larger than our measurement volume size and thus cannot be directly observed. The frozen hypothesis, on the other hand, provides a tool for us to probe this spatial inhomogeneity beyond the size of the measurement volume.
We also note that this lengthscale is consistent with the wavelength of the most unstable modes that we estimated in Section~\ref{sec:relvel}.

\begin{figure}
\includegraphics[scale=0.45]{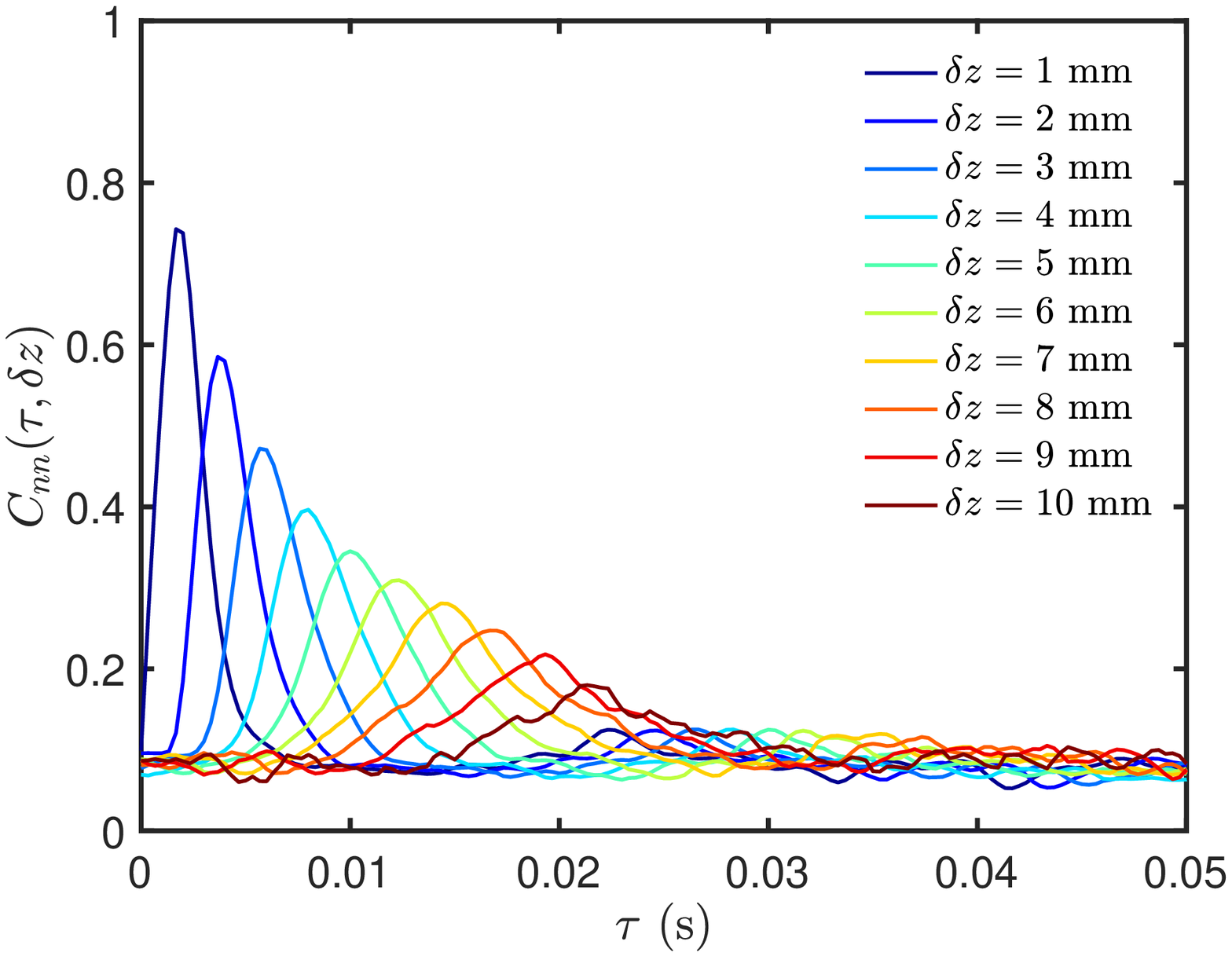}\\
\includegraphics[scale=0.45]{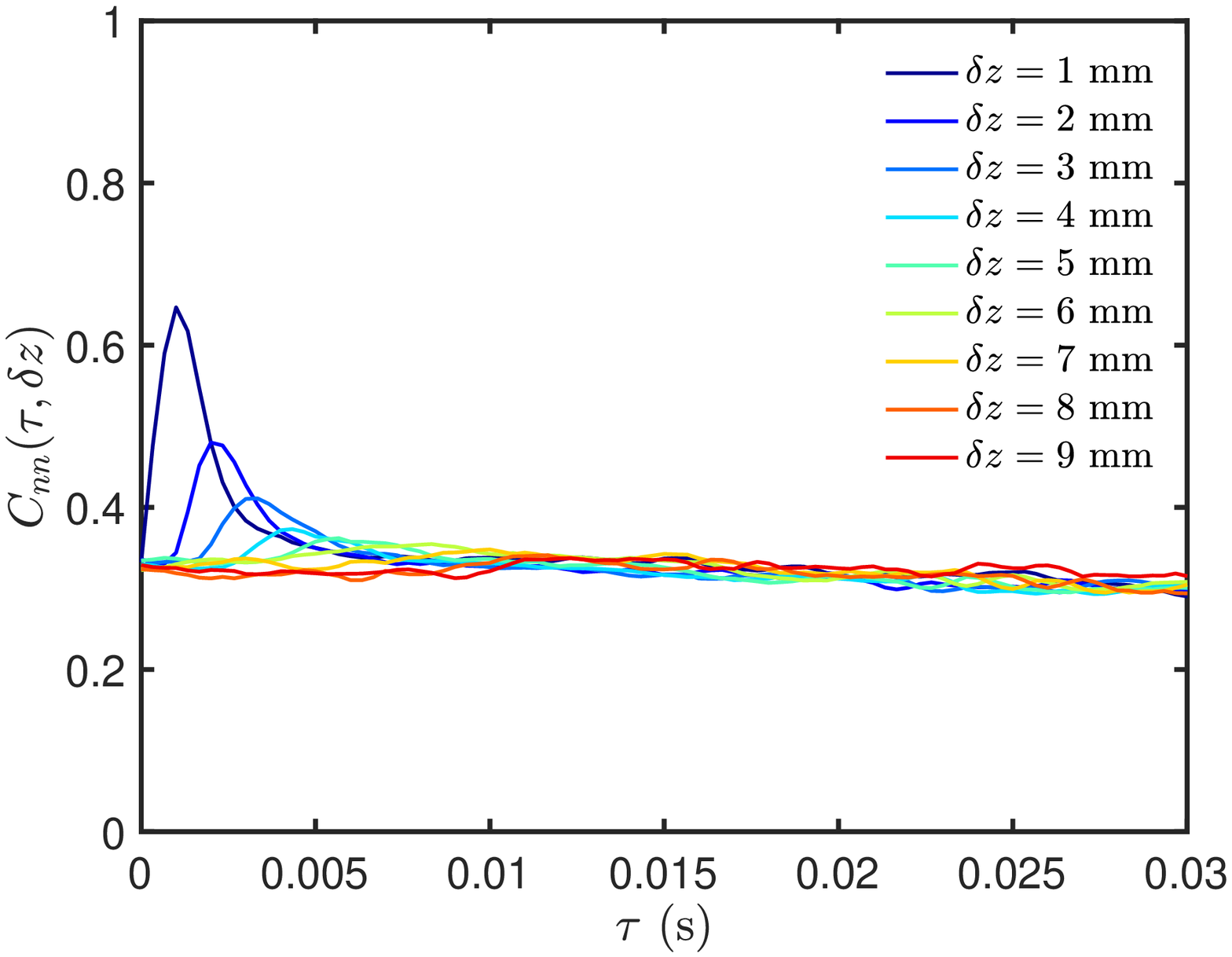}
\caption{\label{cross_corr} Temporal variation of the cross-correlation of particle number densities at different heights in the measurement volume. Top panel: the case of kn1\_3mb [DS2]; bottom panel: kng1\_8mb[DS3]. The shift of the peaks with time and the broadening of the shape indicate that the concentration evolves like an advection-diffusion process.}
\end{figure}

In Figure~\ref{cross_corr}, we plot the cross-correlation of particle number densities at two different locations as defined by Equation~\eqref{crossc}. For both the kn1\_3mb [DS2] (top) and kng1\_8mb [DS3] (bottom) data, $C_{nn}(\tau,\delta z)$ decreases in magnitude and broadens as the distance between the two locations increases, which is consistent with the behaviour of an advection-diffusion process\footnote{The larger value of $C_{nn}(\tau, \delta z)$ at long time-lag in the bottom panel is simply due to the higher average particle number density of the kng1\_8mb [DS3] run.}. The time at which the cross-correlation peaks corresponds to the time the particles take to move through the distance $dz$ at the mean velocity. The broadening of the cross-correlation curves is due to the velocity fluctuation of the particles. For the kng1\_8mb [DS3] (bottom) case, the rms velocity is much larger than the kn1\_3mb [DS2] (top) case, as shown in Table \ref{rms_table}, which explains the faster de-correlation of the particle number density observed in the bottom panel.

This broadening of the cross-correlation curve implies that the clustering of particles that we observed in the measurement volume is a dynamic effect rather than some initial concentration variations introduced at the bottom of the apparatus; our measurement volume is $\sim 50$ cm above the bottom where the particles are entrained in the air stream, which converts to $\sim 1.25$ seconds of travel time at a mean particle velocity of $\sim 0.4$ m/s. Consequently, any initial inhomogeneities in particle concentration would have been smeared out, judging from the de-correlation time shown in Figure~\ref{cross_corr}.

\subsection{Evidence for collective continuum-drag reduction}\label{cont-drag}

\begin{figure}[ht]
\includegraphics[scale=0.5]{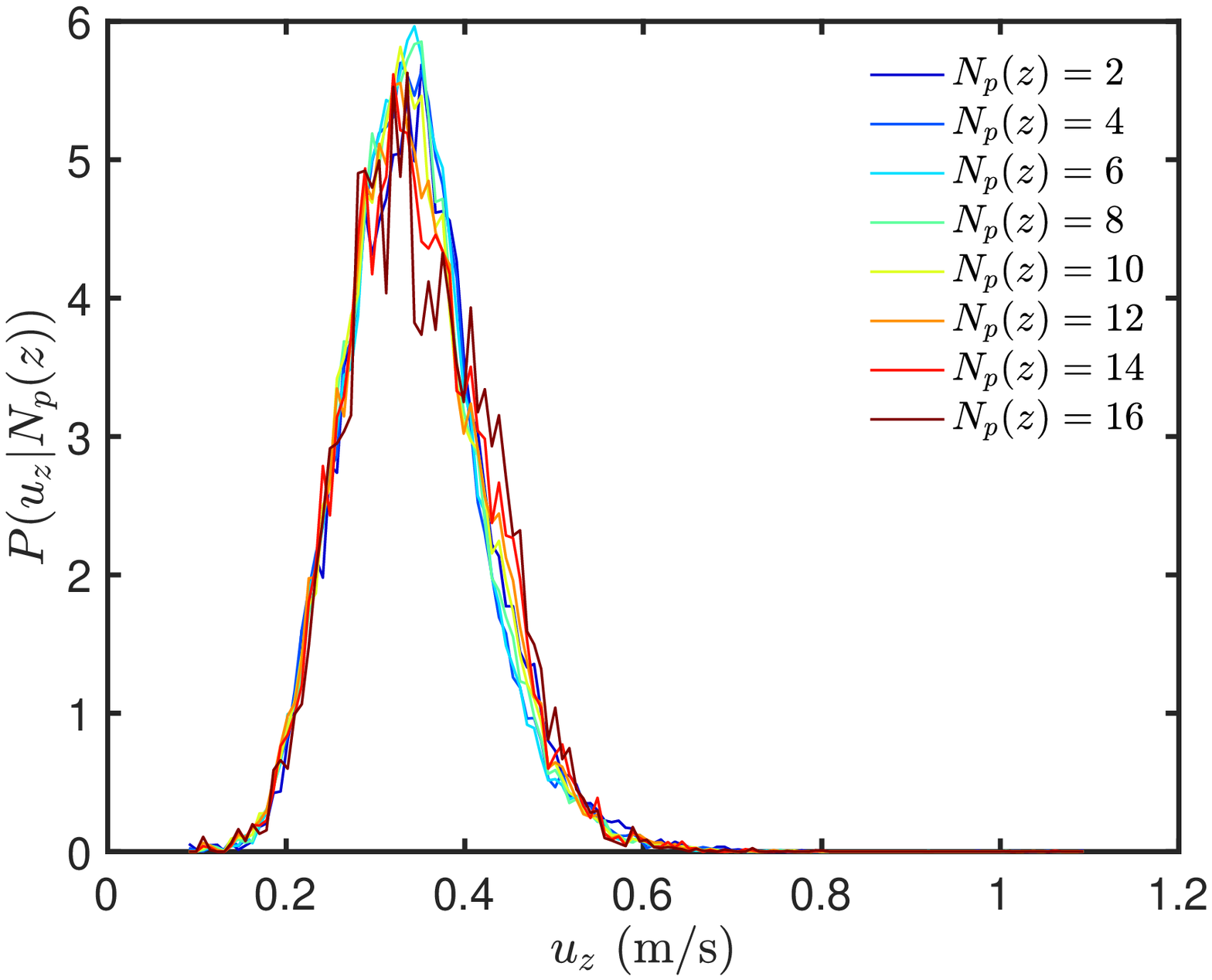}\\
\includegraphics[scale=0.5]{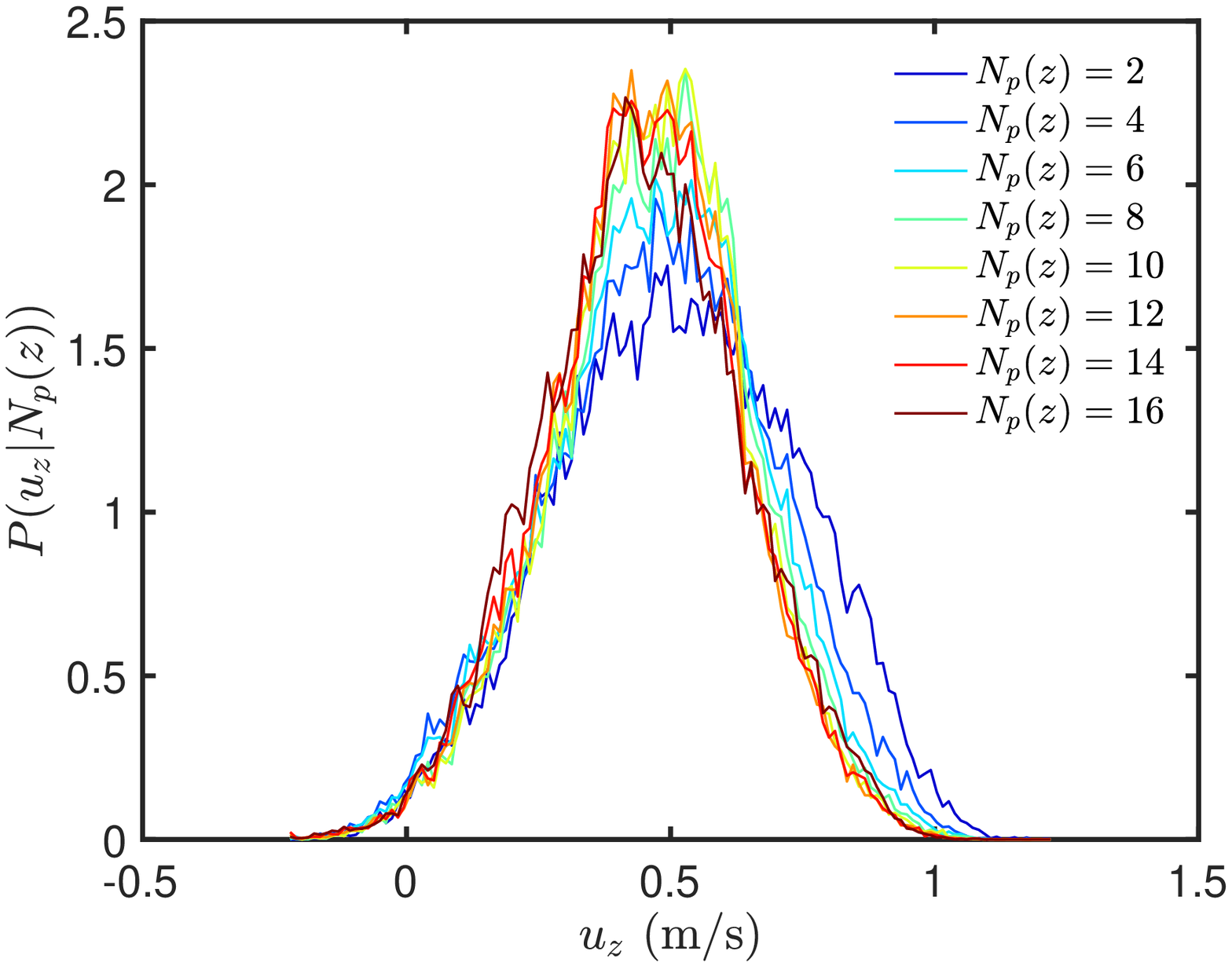}\\
\caption{\label{smallscale_pdf} PDFs of $z$-direction velocity, conditional upon the number of particles in a 1-mm slab, for kn1\_3mb [DS2] (top panel) and
  kng1\_8mb [DS3] (bottom panel). Curves shift towards lower velocities with increasing number density in the lower panel. }
\end{figure}

Having demonstrated particle aerodynamic focussing and spatial clustering, let us now consider how the vertical particle velocities depend upon the local particle density. 
To do so, we measure the PDF of the vertical component of the instantaneous particle velocities $u_{z}$, conditioned upon the number of particles in the 1 mm high slab which the particle under investigation belongs to. The resultant PDFs, $P(u_{z} \vert N_{p}(z))$, are shown in Figure~\ref{smallscale_pdf}, where the two panels are for the two data sets with gas pressures in ascending order. 

The sets of PDFs corresponding to the two $Kn<1$ data sets (middle and bottom panels) show that, as $N_{\rm p}$ increases, the peak of the distribution both narrows and shifts to lower positive velocity. For the $Kn \sim 1$ data, there are some differences in the PDFs depending upon $N_{\rm p}$, but with no clear trend; perhaps there is some broadening of the distribution with increasing number density but no obvious shift in the mean. 

\begin{figure}[!h]
\includegraphics[scale=0.5]{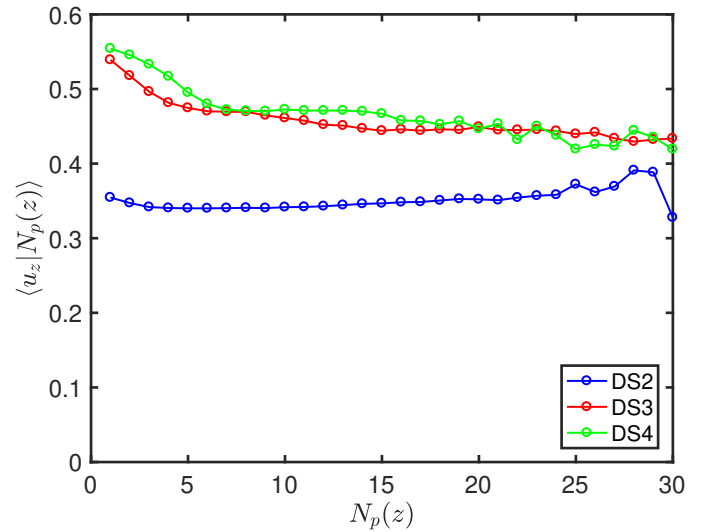}\\
\caption{\label{drag_reduction} Mean settling velocity in m~s$^{-1}$ for kn1\_3mb [DS2] in
  blue, kng1\_8mb[DS3] in red, and kng1\_5mb[DS4] in green, conditional upon the number of
  particles in a 1-mm cylindrical slab. Drag dissipation is reduced for high density regions in the $Kn<1$ data, yet not in the $Kn\sim1$ data.}
\end{figure}

These observations are summarised and clarified in Figure~\ref{drag_reduction}, which shows the mean of all instantaneous vertical particle velocities as a function of the number of particles found in a 1-mm slab. The two curves representing data with $Kn <1$ decrease as $N_{\rm p}$ increases, in contrast to the $Kn\sim 1$ data set. 

While the net velocity of the particles is always positive, a decreasing positive value represents a tendency for the particles to resist being pushed upwards by drag forces. The dependence on number density for the two $Kn<1$ data sets strongly suggests collective particle-drag reduction. That we do not observe this behaviour for the lower gas-density data set could be attributable to the considerable slip one expects on the surface of the particles. Indeed, how the mean $u_{\rm p}(z)$ is always slower for this data set than for the other two is consistent with the fact that, for the same gas velocity, $u_{\rm t}$ is higher for particles in the Epstein drag regime and so one expects greater $\delta u$ between the carrier and disperse phases. Similarly, the momentum diffusivity, expressed by the dynamic viscosity, $\mu=\eta/\rho_{g}$, is higher in a low density medium and so perhaps the local gas accelerations caused by the particles decay more readily than when the momentum is allowed to diffuse more slowly. 

Figures~\ref{smallscale_pdf} and \ref{drag_reduction}, which show different outcomes with regard to collective drag reduction for each of the two representative data sets, provide more insight into the slightly dissimilar behaviour observed between the two data sets in Sections~\ref{sec:relvel} and \ref{clustering}. This will be discussed in Section~\ref{discussion:drag_kn}.

\section{Discussion}\label{discussion}

\subsection{Inter-particle interactions and the consequences of low $\phi$}\label{discussion:interactions}

The experimental apparatus we use for this study bears some resemblance to a fluidised bed, which is a type of system known to show various signs of collective particle phenomena. An important distinction between this and previous studies is the drastic contrast in material density of the two phases, which is within $10^{7}<\frac{\rho_{\rm{p,m}}}{\rho_{gas}}<10^{10}$, depending on the gas pressure of our experiment. Therefore, we can achieve relatively high average solid-to-gas ratio, $\epsilon \approx 1$, expressed by Equation~\eqref{loading}, even with extremely low average filling factors, $10^{-7} <\phi<10^{-6} $, expressed by Equation~\eqref{filling}. The parameter range of the system in question can be compared directly to the situation in a PPD. Assuming a MMSN disc, at 1 AU, the material-density contrast between rock, with $\rho_{\rm{rock,m}}\sim3$ g cm$^{-3}$, and gas is $\frac{\rho_{\rm{rock,m}}}{\rho_{gas}}=2\times 10^{9}$. 
For a dust-to-gas ratio of unity in the midplane, we obtain a volume filling factor $\phi \approx 4.3 \times 10^{-10}$. 
Such extremely diffuse mass distribution is in some sense the essence of why rapid particle concentration mechanisms such as SI are required to facilitate self-gravity. 

It might be imprudent to try to study this extremely diffuse limit, either of the carrier or disperse phase, in a laboratory flow, since the total volume under consideration can never approach interplanetary length scales and so continuum fluid assumptions would not be met. Still, this study mimics the extremely diffuse regime, due to the relatively large values of $Kn$ on the particle scale and particularly in the sense that the system meets the criterion for the solid phase to be considered a \emph{pressureless} fluid. In a PPD, the solid component is pressureless because the particles do not -- in their laminar equilibrium drift configuration -- directly interact, either collisionally or viscously. The dusty fluid therefore has no effective `temperature' in the kinetic sense. The laboratory flow we produced is the same, which is quantified by the criterion $T_{f}>t_{d}$, demonstrated to be true in table \ref{exp_table}. This criterion is demonstrated to be true in the MMSN by \cite{lambrechts}.

Collisions were not observed in our experimental system. We estimate the collision frequency in the measurement volume between any two particles, per unit time, Z, as is done in the kinetic theory of gases: 

\begin{equation}
Z=\frac{1}{2}V_{\rm{rel}}\pi d_{\rm{p}}^{2}n^{2}/V_{\rm meas}.
\end{equation}
 Using the values of $d_{\rm p}$ and $n$ reported in table \ref{exp_table}. Figure~\ref{struct_a} shows that the square root of $D_{ll}$ for particles with separations $\vec{l}\lesssim 1$~mm is approximately 0.05~ms$^{-1}$, consistent with the findings in Figures~\ref{cyllinder_a} and \ref{cyllinder_b}, that the relative velocity at separations $\delta r <1$mm and $\delta z <1$mm approaches 0.05 ms$^{-1}$, which we adopt as $V_{\rm{rel}}$. We obtain $Z \approx 10^{-2}$ collisions s$^{-1}$ in the measurement volume, and therefore we could observe no more than one collision in each of our experimental data sets, which are each, cumulatively, $\sim 30$~seconds. We note that the collision frequency calculated this way is an over-estimate, because the relative velocity is dominated by the vertical component, and so a direct collision will also depend upon two particles having a favorable vertical alignment.  

\subsection{On the negligibility of electrostatic forces}\label{electrostatic}

The relative velocity statistics in Section~\ref{sec:relvel} 
 bear direct witness to particles in the act of spontaneously aggregating. Recall that the particles are always on average approaching one another. However, notably, they do so only in the direction coincident with the mean direction of drag force. We stated in Section~\ref{apparatus} that we ignore electrostatic forces between our particles. We reinforce this point to explain why the clustering phenomemon we observe is fluid-dynamical in nature and unrelated to electrostatic attraction.

We use stainless-steel particles, which cannot become charged by friction. Recall the fundamental properties of conductive metals: all excess charge resides on the surface, with charge density $\sigma_{\rm e}$; electrons freely flow within, to, and from them -- because there is no stable equilibrium point at which the potential energy has a minimum. Furthermore, in a uniform electrical field, the total force on an uncharged conductor is identically zero (see, e.g. \citealt{landau_lifshitz}, for formal proofs). 

There is no external energy source in this system to induce or sustain a net charge on the particles\footnote{The energy of our light source is too low for the photo-electric effect to occur on the metallic surfaces, and even if it were not, the result would be net repulsion, not attraction.}. Still, supposing that the particles are not completely neutral, the field around each particle of surface area A and charge density $\sigma_{e}$ would have everywhere a potential energy, in the direction n normal to the surface $E_{n}=4\pi \sigma_{e}=-\partial {\Phi}/\partial{n}$, where $\Phi$ is the potential. The total charge Q on the sphere, inscribed within a Gaussian surface, is $Q=\sigma_{e} A$. It would require a very high surface charge density to overcome the potential energy of our small, heavy particles, $E_{\phi}=\rho_{\rm{m,p}}(gh_{\rm meas}-\frac{1}{12} \pi d_{\rm{p}}^{3} u_{\rm pz} )$, where $h_{\rm{meas}}$ is the height at which measurements are made which, along with all other variables, is defined in Section~\ref{techniques} of this paper. 

The attractive force given by Coulomb's law $F \propto \frac{Q_{i}Q_{j}}{d^2}$ between particles i and j is directed along the separation vector $d$ between particles. By stark contrast, we find that the direction of the attractive phenomenon we observe in Figures~\ref{cyllinder_a} and \ref{cyllinder_b} is not along the separation vector, but strictly along the vertical component of it. There is indeed a quickening of the attractive behaviour in the vertical direction with decreasing vertical distance, however over rather great distances of up to 4~mm $\sim80-90 d_{\rm{p}}$, where the Coulomb force should be extremely weak, and we do not find any particle pairs at separations close to $d_{\rm{p}}$ (even though the positional accuracy of the measurements is much smaller than this scale, see Section~\ref{techniques}), where the Coulomb forces actually, in general, operate. 

\subsection{Drafting and the consequence of Re$_{\rm p}$} \label{discussion:drafting_re}

Particles that systematically drift in a PPD correspond to sizes of pebbles or boulders (mm-m), depending upon location or disc model assumptions. Such particles are very close to the transition given by Equation~\eqref{draglaws} and in the MMSN, this transition is given by $9/2 \lambda =6.4$ cm $\left(\frac{r}{\rm AU}\right)^{2.75}$ i.e. 6 cm particles at 1 AU. In Figure~\ref{mfp}, we demonstrated that our experiments bracket this transition. 

In either the Stokes or Epstein drag regime, an important property of flow around individual spheres is that it is symmetric, both in space and time \citep{suspensionbook}. As a pertinent example, \cite{You_good2005} exploited this property to linearise the Navier-Stokes equations and derive an analytical dispersion relation for the SI, and refer to this simplification as a ``terminal velocity approximation". When removing the time dependence of the particle velocity, it is sometimes customary to also retract Navier's name from the momentum balance equations and refer to them simply as `Stokes equations' or `Stoksian' flow, and in practical terms this mainly means that the drag law (be it `Stokes' or `Epstein') is linear. A flow can be considered Stoksian when the shearing rate is small compared to the viscous stress and Re$_{\rm p}<<1$; in Section~\ref{techniques} we demonstrated that this is true in all of our experiments.

There are several highly-studied regimes in particle-pair and collective-particle sedimentation phenomena that are due to flow symmetry breaking by increasing Re$_{\rm p}$ \emph{much higher than 1}. A notable example is `wake attraction', which occurs when the flow separates from the downstream side of a sphere to generate a wake, where the low-pressure conditions essentially create a vacuum and pull a closely aligned and relatively nearby (never more than 10 $d_{\rm{p}}$ downstream) particle into the wake. While various studies of this effect differ regarding to precisely which Reynolds number is required for this effect can result in particle clustering\footnote{This variation depends upon whether it is a 2D or 3D system, whether the particles are tightly confined by walls, how high is $\phi$, and whether the wake sheds vortexes or not.}, all studies (see e.g. \citealt{fortes_joseph_lundgren_1987, feng_hu_joseph_1994, KAJISHIMA:2002,uhlmann_doychev_2014}) are unanimous in stating that it requires moderate-to-high Reynolds numbers, meaning $50<Re_p<850$, which are decidedly non-Stoksian values. Such relatively high values of Re$_{\rm{p}}$ are required because the mechanism quite literally involves a well-separated or even turbulent wake, e.g. of the types beautifully visualised in \citealt{tietjens1957applied,vanDyke:1982}.  

In addition to our experiments being in the incorrect regime to expect wake attraction, the phenomenology as it is formally studied is inconsistent with the details of our data. Wake attraction requires a very favorable alignment, and in Figure~\ref{cyllinder_b} the aggregation effect occurs for both vertically aligned \emph{and misaligned} particles -- as expressed by how all of the curves appear the same for every value of horizontal separation. While wake attraction is a relatively long-range effect, it is not so long as what is found here; as mentioned above in Section~\ref{electrostatic}, the separation at which the particle layers aggregate in this system is much greater than 10 $d_{\rm{p}}$ downstream. 

Due to all of these considerations, we resist any speculation that the aggregation in this system is simply a manifestation of this well-known result, even though it may remind one of classical studies of strong flow-symmetry breaking.

\subsection{Drag dissipation and consequence of drag regime ($Kn$)}\label{discussion:drag_kn}

We remind the reader that the two `representative' data sets we have presented are those with values of $\epsilon$ for which inertial coupling between the carrier and disperse phases is non-negligible. This criterion is considered important for reaching critical growth rates both in the formulation of a parallel shear flow with Keplerian gravitational potential \citep{You_good2005}, and as in the present case of sedimentation with constant gravitational acceleration \citep{lambrechts}.  We achieve this criterion in two different ways: in kn1\_3mb[ds2] ($Kn\sim1$), we use fewer particles and lower gas pressure; in kng1\_8mb[ds3] ($Kn<1$), we use more and slightly larger particles, with the gas pressure over twice the gas pressure of kn1\_3mb[ds2].

In Section~\ref{cont-drag}, we find that even for the same local number density conditions, drag is reduced only for the Kn<1 data. We do not equate the aggregation behaviour we find in Figures~\ref{cyllinder_a} and \ref{cyllinder_b} with the lowered drag dissipation rate we report in Figure~\ref{drag_reduction}. Our data unambiguously show that spontaneous mass concentration occurs even in the absence of a correlation between particle number density and mean settling rate. We believe this effect, when it happens, plays an auxiliary role, i.e. by increasing the magnitude of the baseline value of $\delta u_{z}$ in the bottom panels in Figures~\ref{cyllinder_a} and \ref{cyllinder_b} and by shortening the timescale for the cross-correlation of number density to decay in the bottom panel of Figure~\ref{cross_corr}. Because of this contribution, particles find it easier to `catch up' to one another and are exchanged more rapidly between high and low-density regions.

As noted above, due to the high contrast in material densities between the two phases, high $\epsilon$ does not imply high $\phi$. So, while increases in the local particle number density by several orders of magnitude proportionately augment $\epsilon$, the volume filling factor $\phi$ remains very low. Therefore, the collective drag reduction we observe (only for Kn<1, not for Kn$\sim$1) does not come from augmenting the total surface area presented to the flow. 

It is much more likely that here, as in \cite{You_jo2007j} and \cite{Jo_you2007j}, the collective momentum of the particles is slowing the gas locally, which in turn reduces the drag the particles feel. Unfortunately, we cannot visualise or measure the local gas flow simultaneously with the inertial particles. However, the dataset for which collective drag reduction does \emph{not} occur, while active aggregation \emph{does}, serves as proxy and supports this hypothesis. 

\section{The state of the art}\label{soa}

We interpret the spontaneous aggregation behaviour we observe to fulfill the predictions of the `mass-loaded rain' simulations in \cite{lambrechts}. 
We emphasise, however, that the interpretation of the \cite{lambrechts} simulations--along with attempts to grasp the physical meaning of the mathematical and numerical calculations that produce SI--is being constantly revisited by theory (see e.g. \citealt{Squire_Hopkins:2018b}). 

While the interpretation of our results is guided by fundamental principles, there is at the same time very little pre-existing framework for understanding the detailed physics across scales: to our knowledge, multi-particle physics in the transition drag region, additionally with such low particle packing fractions ($\phi$), has never been studied before. 

We believe that, by virtue of the very low values of $\phi$, we have removed the primary cause of inter-particle perturbations, namely long-range hydrodynamic interactions, which are deemed responsible for strong velocity and density fluctuations in fluidised beds or, equivalently, particle suspensions \citep{,Batchelor_jfm,Guazzelli_2011,Kalthoff, tee_mucha_brenner_weitz_2008}. Yet, the current state of the art in understanding the decay of the perturbation field around particles comes from investigations of low $Kn$ systems where pure continuum fluid assumptions apply. We therefore highlight the question of how such perturbations decay in rarefied gas of $Kn \approx 1$ or greater as a potentially impactful future direction for investigation. 

Since the measurements of \cite{cunningham} were made, there has been significant progress in developing an experimentally verified unified theory for drag on individual objects in transitional flow (\citealt{Karniadakis_etal:2005,Fissell2011}). However, the results are quite dependent on the boundary conditions both at the particle-fluid interface and at relatively long range. The boundary-condition dependence of fluid drag presents notorious challenges in studying multi-particle systems, because the presence of neighbouring particles in the flow alters the shearing rate at particle-fluid interfaces. There is a storied history of attempts to understand how viscous drag is modified due to this fact, building primarily on the seminal works of \cite{darcy_early, einstein_viscosity:1905,CARMAN}, and \cite{Brinkman}. A drag force correction can be derived, and many authors have proposed variations on this theme (see, e.g. \citealt{Hamdan:2016,Lasseux:2017}). However, all such proposed corrections universally take the form $F_{d}=F_{\rm{d,Stokes}}(1+f(\phi))$, where $F_{\rm{d,Stokes}}$ and $\phi$ are defined the same as in this work.\footnote{Although often written in terms of the porosity, which is $P=1-\phi$.} This is necessarily the case, because drag by definition depends on the total surface area presented to the flow. Due to the vanishing values of $\phi$ in these experiments, such theories cannot explain the apparent drag reduction shown in the top panel of Figure~\ref{smallscale_pdf}. 

Whether collective drag reduction occurs, or whether it does not, depends upon $Kn$. The experimental result is nearly in agreement with the simulations presented in \cite{lambrechts}, who only considered particles in the Epstein regime and find a very weak effect of enhanced particle settling by particle swarms formed in a non-linear instability. There is perhaps a seed of insight here regarding the long-standing question of why strong clumping in the non-linear phase of SI is particle-size dependent, and so future work to derive a precise explanation for the $Kn$ dependence should be worthwhile. We suggest that an important factor to take into consideration when pursuing this line of inquiry is that the kinematic viscosity is equivalent to a momentum diffusivity, which increases inversely with decreasing gas density. See also the semi-analytical analysis in \cite{lambrechts}, in which it is shown that increasing viscosity has a damping effect on the toy model that is studied there. \cite{Squire_Hopkins:2018b} also address how differences in drag regime have an impact on the growth of resonant drag instability, which is a family of instabilities to which, they argue, SI belongs. 

We point out that wake formation and associated attractive forces in \emph{rarefied gas} is not a topic of previous investigation. Therefore, the possibility that we are the first to observe an effect related to wake attraction in a completely new regime cannot be completely dismissed. However, even if some form of wake-related behavior is operating in this system, it has a very different characteristic than traditional sedimentation. Drafting events in dilute 3D sedimentation generally do not lead to clustering, but to the renowned "draft, kiss, tumble" (DKT) effect which, as indicated by the name, involves a momentary but fleeting contact between particles \citep{fornari_picano_brandt_2016}. As expressed above, we do not observe contact. Moreover, simulations of sedimentation show that isolated DTK events do not dominate the velocity statistics, but can be identified as a bump in the tail in the negative end of the settling velocity PDF; they do not dominate the mean because a very unique alignment is required. We would also be able to detect this bump on the left-hand side of the $z$-direction velocity PDFs in Appendix \ref{appendix:pdfs}, and do not.\footnote{As a caveat, we do not have enough data here to study extremely rare events in the statistical tails.}  

We urge caution in making direct analogies to wake attraction as it has previously been studied as primarily as a particle-pair process. We find tentative evidence in Section~\ref{sec:relvel} (see also Appendix \ref{appendix:moments}) for a developing pressure phase lag with heuristic similarity to the mechanism proposed in \cite{youdin_singlefluid}. It is well known that flow through densely packed porous beds results in a pressure drop \footnote{This is the operational principle in the right hand panel of Figure~\ref{seeding}, in which the packing fraction is maximum in order to create a steady low-pressure flow in the experimental test chamber.}. However, in our experimental test chamber, low $\phi$ implies that there is no flow impedance to induce the local pressure gradient, in the classical sense of \cite{darcy_early, CARMAN}.

We find more grounds for interpreting our results in the physical principles inherent to SI, than we do from classical sedimentation experiments and theory of suspension dynamics. We cannot find any major contradiction in the predictions of the SI and the results of the present study. However, reconciling volume-averaged, two-fluid models with discrete systems is in general an outstanding field of inquiry in sedimentation \citep{brennen_2005}. In a real system, one must grapple with the fact that particles are discrete entities that also interact individually with free-flowing molecules.  
Therefore, our ongoing collaborative efforts include code development starting from gas and solid kinetics \citep{Herminghaus_Mazza:2017}, in order to test the constitutive equations of the macroscopic scale.
\normalfont
\section{Summary}\label{summary}
We constructed an experimental facility uniquely suited for the study of dust-grain aerodynamics under conditions similar to those in a PPD. The apparatus contains a suspension of solid spherical particles (of size 15--65~$\mu$m) settling under constant gravitational acceleration against a low-pressure gas stream. We calibrated the fluid state parameters and gas-velocity field in the absence of solid particles to confirm that the bulk flow is laminar and incompressible and that the conditions are steady, as discussed in Sections~\ref{techniques} and \ref{character}. 

We repeated particle-sedimentation experiments at four fixed pressures in the mbar range and featured the results from two experimental data sets, each of which representative of being on either side of the Stokes--Epstein drag law transition. The typical volumetrically averaged solid-to-gas mass density ratio of these two data sets was close to one.

Using a particle-tracking technique, we measured individual particle motions in three dimensions, within a cm from the centre of the chamber, within which distance the gas-flow velocity varies by only a few per cent. The particle-tracking statistics reveals a complex particle-velocity structure that varies as a function of scale.  

The statistics of relative velocity between particle pairs shows that, on average, particles approach one another when their vertical separation is small. While this effect depends strongly on vertical separation, it is independent of horizontal separation, at least within the range of the measurement volume. It is interesting to note that the vertical direction is the direction of the mean relative motion between the gas and the solid phases.  This dynamic phenomenon operates in concert with variations in the mean particle density, measured at fixed position while particles advect through the measurement window, and therefore corresponds to the mean particle density variation in the vertical direction. Such a variation thus can be interpreted as particle-density waves.

In addition to the observation of convergent motions leading to particle-dense layers, we find enhanced mean settling velocities that depend upon the local particle number density.  This effect is suggestive of collective particle-drag reduction. However, it is only clearly seen when the Knudsen number corresponds to nearly continuum flow conditions around the particles, and does not occur for the data set with $Kn$ approaching free molecular flow. Since it is only present in the $Kn<1$ experiments, yet both the $Kn\sim1$ and $Kn<1$ data exhibit complex convergent particle motions, the collective drag reduction might be neither exactly the same as, nor the cause of the particle aggregation behaviour; rather, it may be a secondary effect that occurs in systems where particle clustering is already emerging, but this secondary effect helps accelerate the clustering. 
See also \cite{You_jo2007j} for an explanation of the role of energy dissipation by collective-drag forces in SI (Section~5 of that article).

The two observations -- on-average approaching relative velocities and long-lasting particle density variations -- indicate that the disperse phase of the system in question is unstable. The particle density variations are however extremely localized in scale and also illustrate a diffusive tendency, and so we find no indication that they are static structures.
We interpret the strong directional dependence of the observed inhomogeneities to be a wavenumber selection process, such that the flow is sensitive to disturbances in the $K_{\rm{z}}$ direction, and either $K_{\rm{x}}$ and $K_{\rm{y}}$ are $0$, or slightly larger than $2\pi/w_{\rm{meas}}$, with $w_{\rm{meas}}$ the width of the measurement volume $V_{\rm{meas}}$. Because we measured the gas flow to be laminar in the absence of particles, we have strong reason to suspect that the observed highly localized behaviour arises from the dynamic interaction between the two phases. We explore an appropriate parameter range to represent the diffuse particle distribution in PPDs, and thus our experiments provide evidence for the presence of particle-gas drag instabilities in an astrophysical context.

The most appropriate theoretical framework for interpreting the observed instability comes from astrophysical flows where so-called `streaming' or `drafting' instabilities (or `resonant drag', more broadly) are theoretically predicted to occur and are considered a promising route to planetesimal formation. Such theories have not previously been directly tested. This work is the first to generate a real physical system that meets the essential physical assumptions of these models, most closely that presented in \cite{lambrechts} of particle sedimentation in a non-rotating system. We subsequently find a spontaneous concentration phenomenon to self-organize. Specifically, in a pressure-driven flow with order-1 mass loading of particles in which coupled inertial particles experience linear drag forces, trailing particles tend to draw near to leading particles, localized dust-to-gas density enhancements occur, and--for a certain drag regime--so does collective particle drag reduction. 

Since the dilute-gas flow vessel used here is the first of its kind, we expect its existence to enable further novel laboratory-based investigations of particle aerodynamics in astrophysical flow. We have demonstrated the ability to achieve steady-state conditions for fixed dust-to-gas density ratio and for fixed Knudsen number; further experiments to systematically vary these two parameters independently will be an important future direction within the existing capabilities of the facility. Additional development of a moving measurement system will be required to assess how and when the instability saturates and if the growth rate depends upon varying the system parameters. Developing the capability to visualize the gas response to the inertial particles would also be a useful innovation.

\begin{acknowledgements}
    We thank Hubert Klahr and Philip Armitage for helpful input and also acknowledge inciteful conversations with Greg Voth and Andrew Youdin. This work was supported by the German \emph{Deut\-sche For\-schungs\-ge\-mein\-schaft, DFG\/} collaborative research centre 963 \emph{Astrophysical Flow Instabilities and Turbulence}, and by the grant ”Bottlenecks for particle growth in turbulent aerosols” from the Knut and Alice Wallenberg Foundation (Dnr. KAW 2014.0048). HLC also received support within the framework of the NCCR PlanetS supported by the Swiss National Science Foundation. The anonymous reviewer's constructive critique improved the manuscript. The mechanical drawings of the experimental apparatus are credited to Dr. Artur Kubitzek.  
\end{acknowledgements}

\bibliographystyle{aa} 
\bibliography{biblio.bib} 

\begin{appendix}

\section{Intermittent statistical distributions}\label{appendix:pdfs}

\begin{figure*}[!h]
\begin{overpic}[scale=0.45]{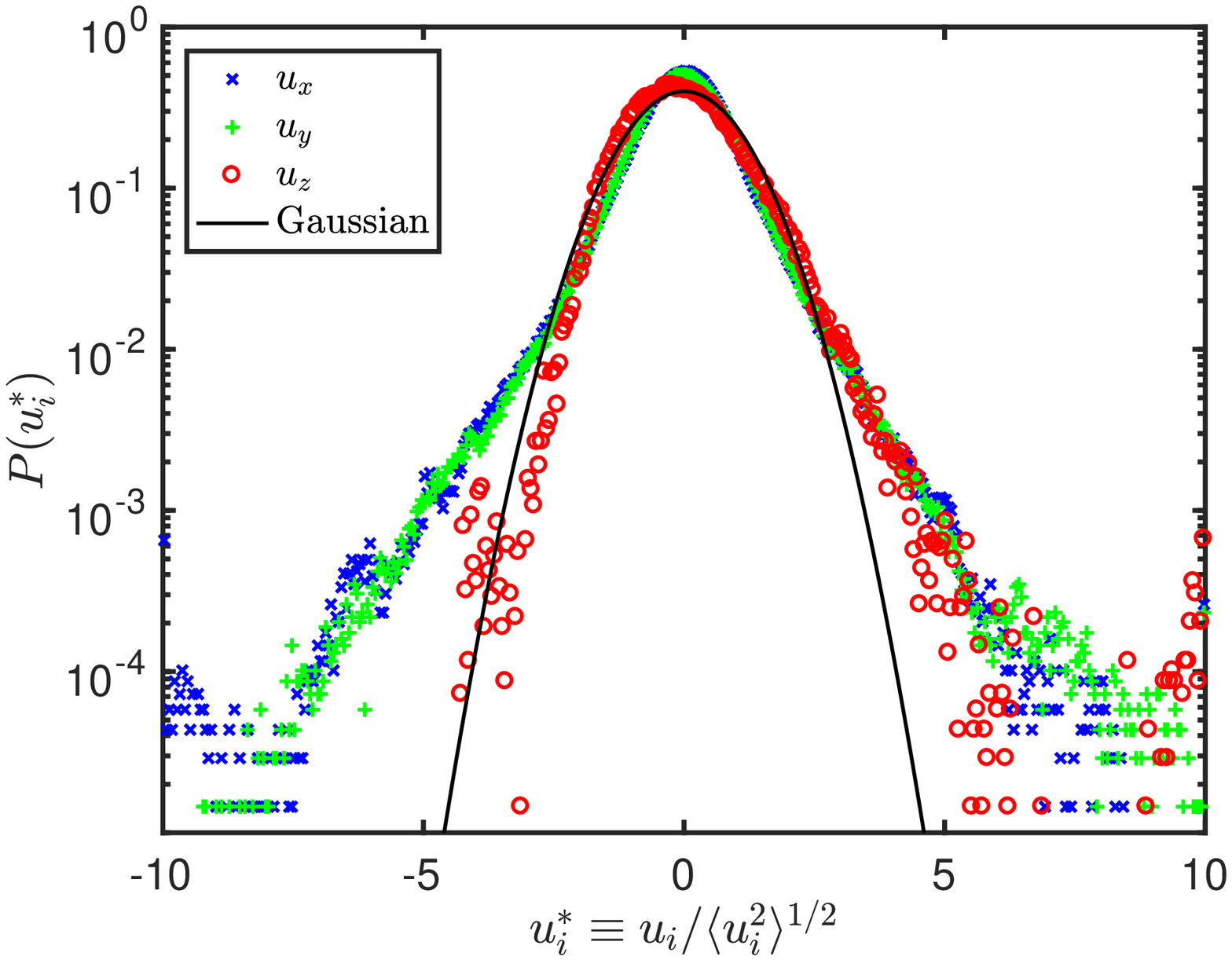}
\end{overpic}
\begin{overpic}[scale=0.45]{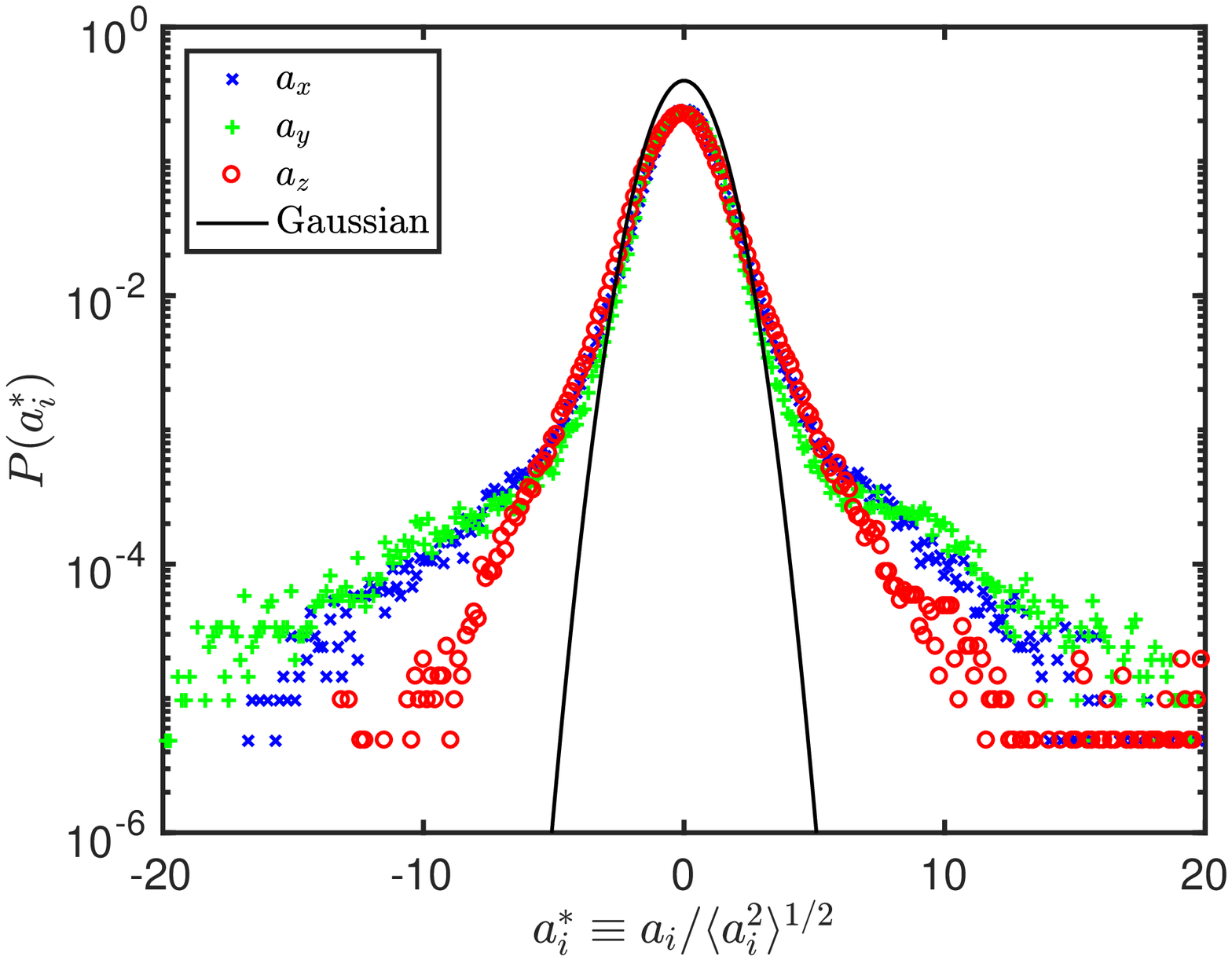}
\end{overpic}
\begin{overpic}[scale=0.45]{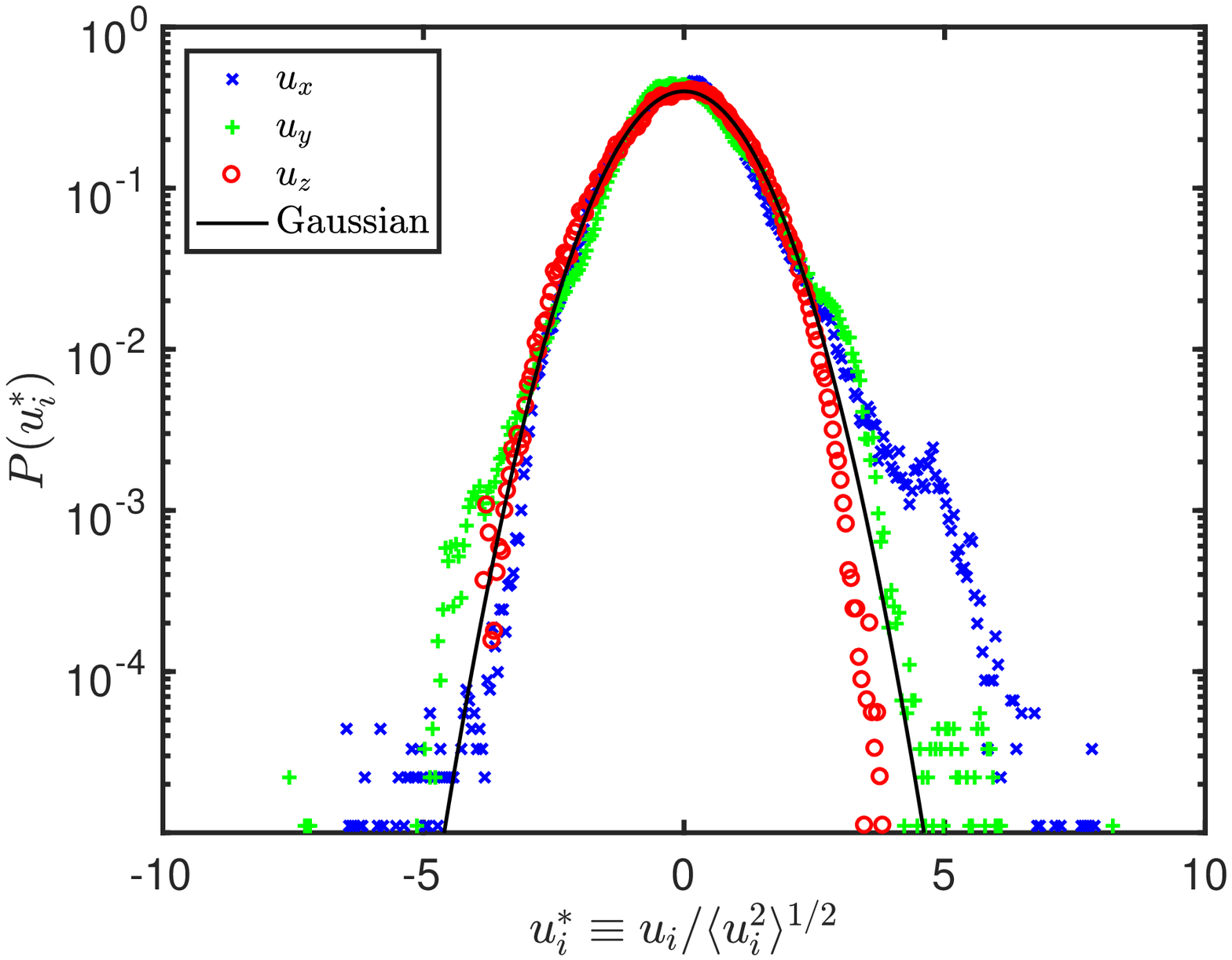}
\end{overpic}
\begin{overpic}[scale=0.45]{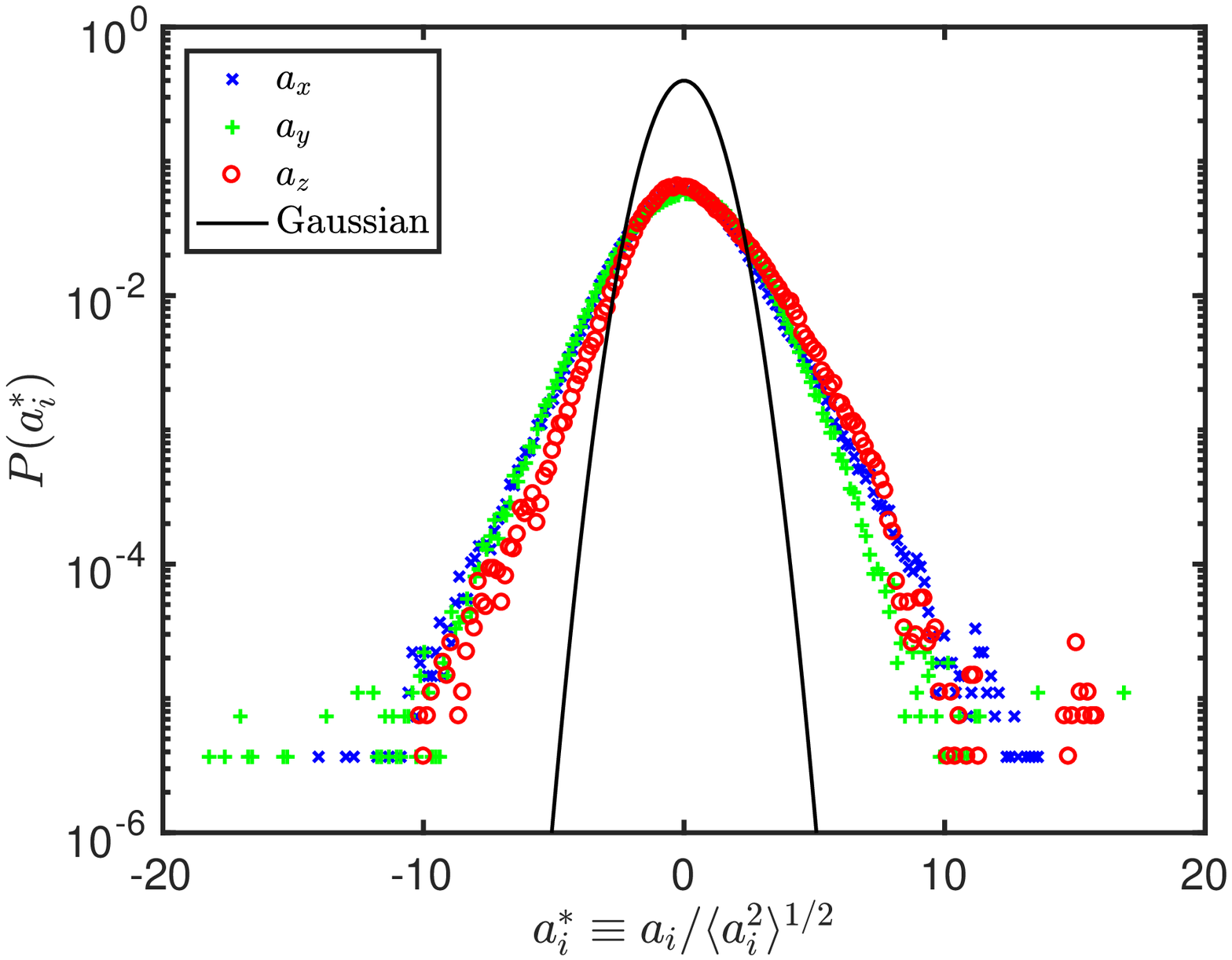}
\end{overpic}
\caption{\label{appendix:velocity_pdfs} Velocity
  (left-hand panels) and acceleration (right-hand panels) for two representative data sets. Blue x's, green plus marks, and red circles correspond to the $x$, $y$, and $z$ components, respectively. All
PDFs are normalised by their rms value. A Gaussian curve is overplotted for comparison.}
\end{figure*}

We present an overview of the particle behaviour by considering the particle velocity and acceleration PDFs. For each of the two representative data sets, Figure~\ref{appendix:velocity_pdfs} shows in the left-hand panels the PDF of instantaneous velocity normalised by the rms velocity for the distribution, $u^{\star}_{i} \equiv u_{i}/<u_{i}^{2}>^{1/2}$, for each of the $i$ components. Similarly, the PDF of instantaneous acceleration, normalized by the rms acceleration, $a^{\star}_{i} \equiv a_{i}/<a_{i}^{2}>^{1/2}$, is shown in the right-hand panels of the same figure.  The colour and symbol coding in Figure~\ref{appendix:velocity_pdfs} refers to the three different components, where $z$ is vertical and $x$ and $y$ are horizontal. An equivalent Gaussian curve is over-plotted for comparison.
We notice in the $Kn\sim1$ case, that both the velocity and acceleration PDFs deviate from a normal distribution. The same is not true of the $Kn<1$ case, since the velocities are nearly Gaussian, but the accelerations are clearly not. 
There are a number of possible reasons for the different appearance of these sets of distributions. In the $Kn<1$ case,
The number density is relatively high, and so it is possible that coordinated motions of the particles serve to reign in the excursions, even while the accelerations remain intermittent. The rms velocity, which can be read off of Table \ref{rms_table}, is much higher, and so excursions may not be as apparent due to the normalisation, whereas the opposite is true in the horizontal component PDFs of velocity.  

\section{Higher-order moments of the relative velocity increment in axisymmetric form}\label{appendix:moments}

\begin{figure}
\begin{overpic}[scale=0.45]{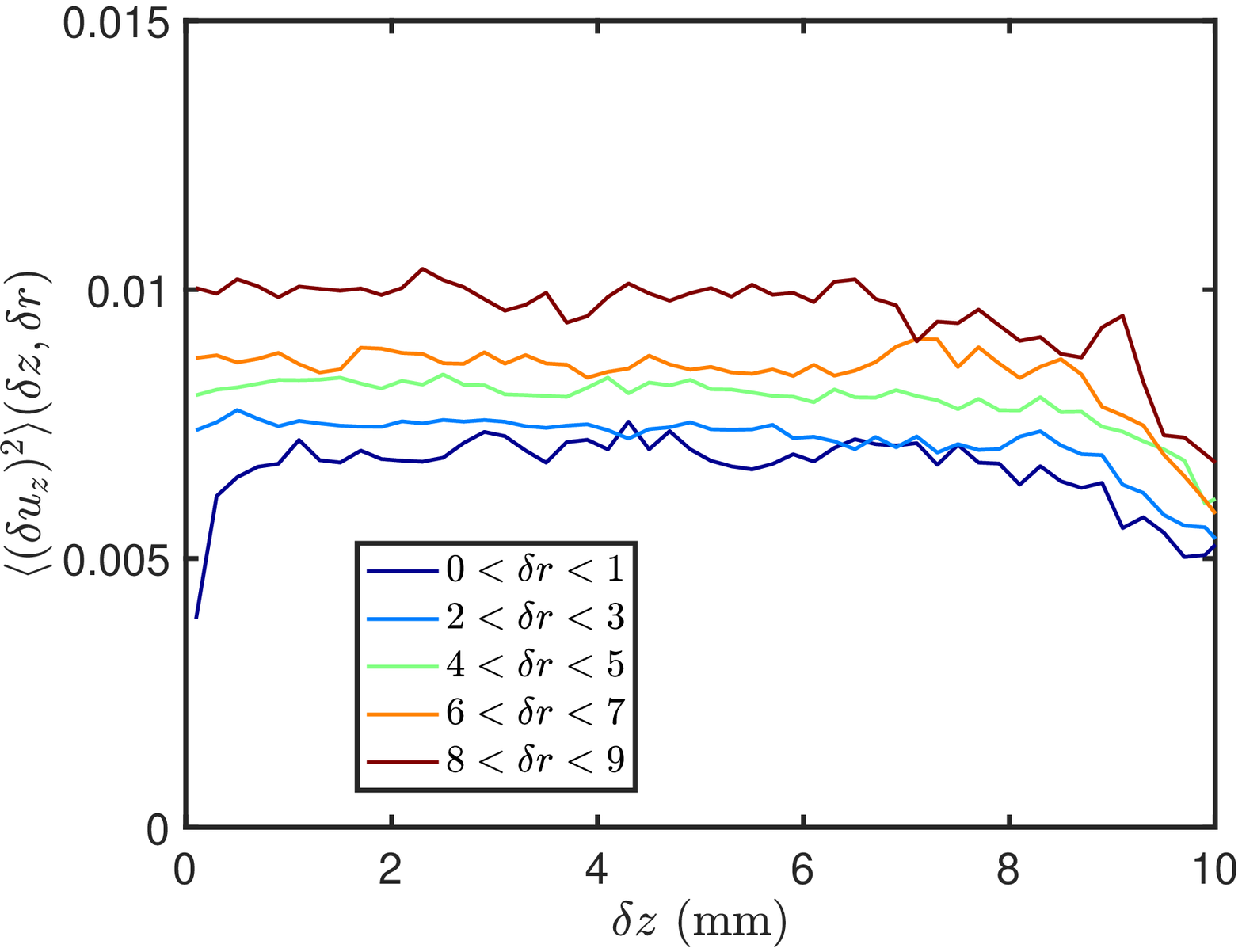}
\end{overpic}
\begin{overpic}[scale=0.45]{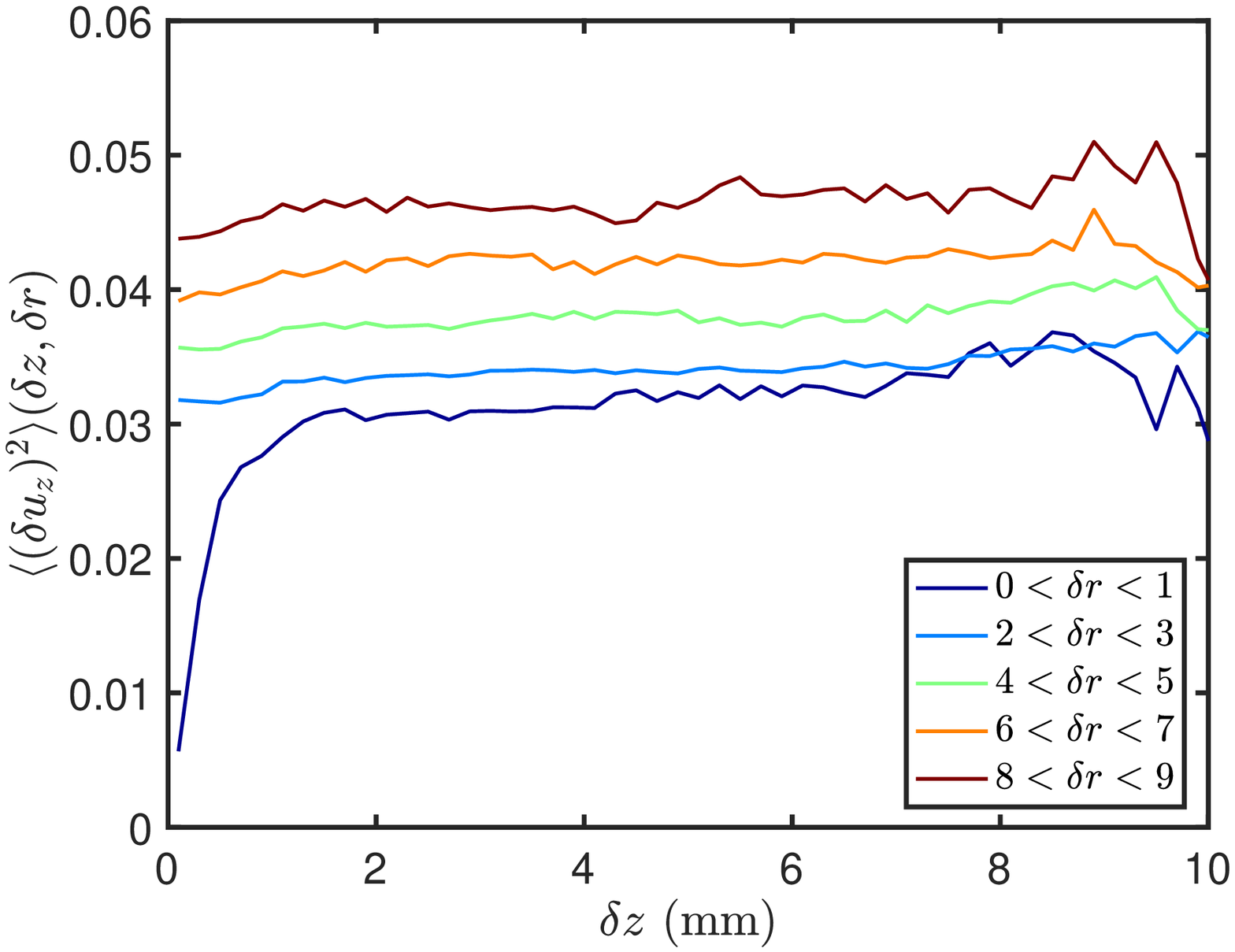}
\end{overpic}
\caption{\label{appendix:2ndmoment} Second-order moment of $\delta u_{z}$ vs. $\delta z$, conditioned upon horizontal separation $\delta r$. Top (bottom) panel: for $Kn \sim$1  ($Kn<1$) data. Conditioning and coloring same as for Figure~\ref{cyllinder_b}.}
\end{figure}

\begin{figure}
\begin{overpic}[scale=0.45]{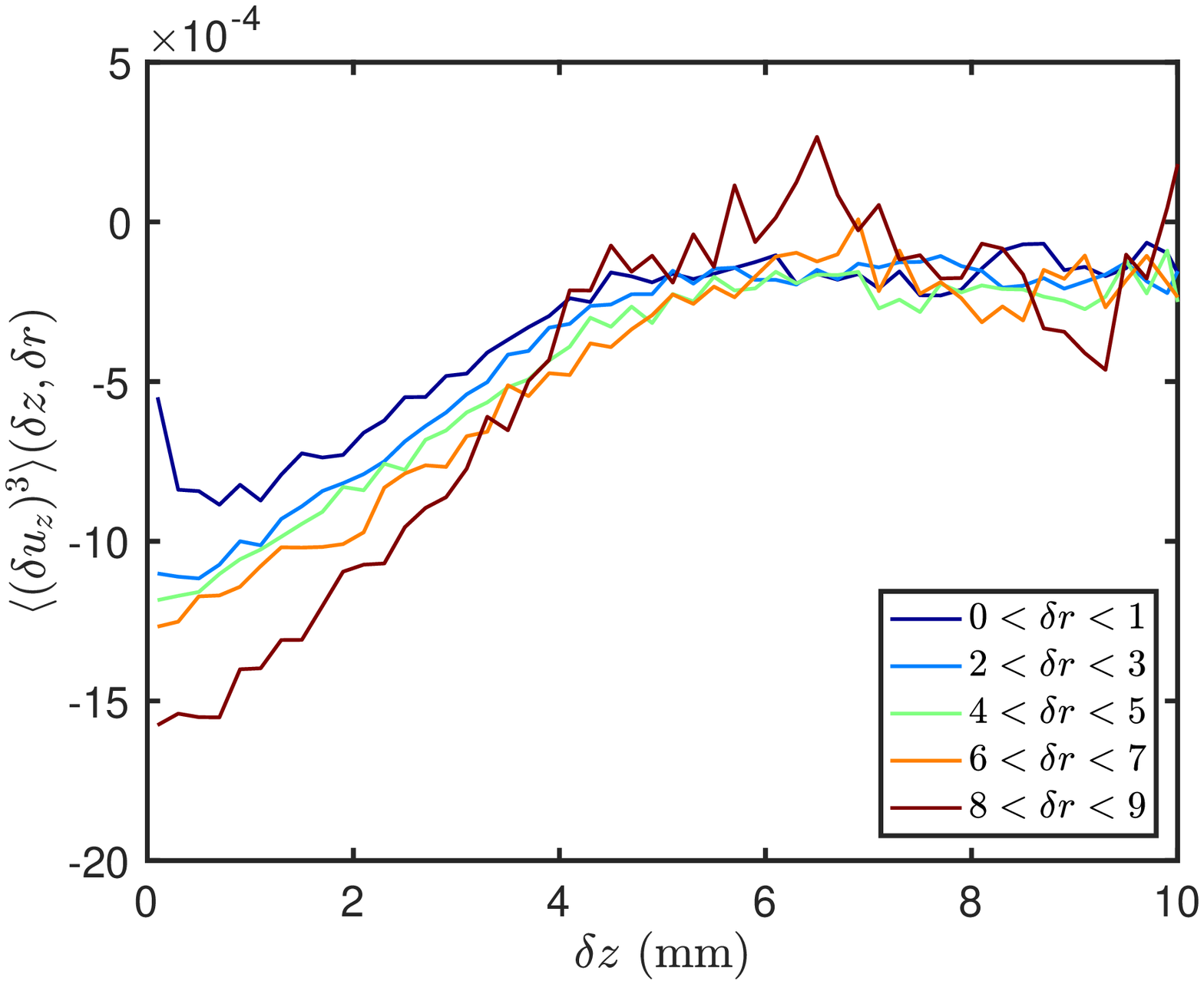}
\end{overpic}
\begin{overpic}[scale=0.45]{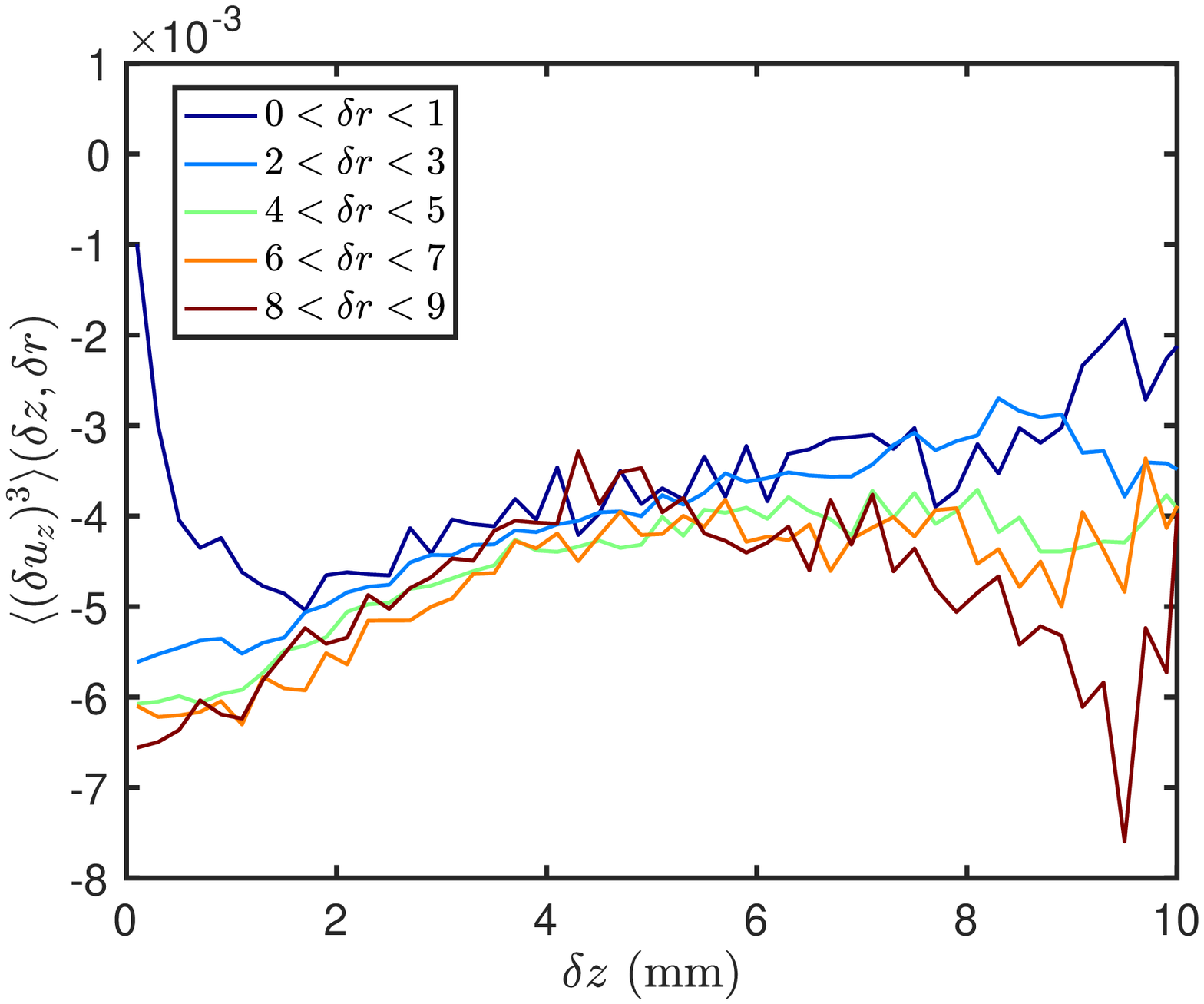}
\end{overpic}
\caption{\label{appendix:3rdmoment} Third-order moment of $\delta u_{z}$ vs. $\delta z$, conditioned upon horizontal separation $\delta r$. Top (bottom) panel: for $Kn \sim$1  ($Kn<1$) data. Conditioning and coloring same as for Figure~\ref{cyllinder_b}.}
\end{figure}

In Section~\ref{sec:relvel}, the mean relative velocity is presented in component form and in an axisymmetric cylindrical coordinate system, with $r=\sqrt{x^{2}+y^{2}}$, and $z$ the vertical component. Here we present higher-order moments of the relative velocity, using the same geometry and conditioning. Figures~\ref{appendix:2ndmoment} and \ref{appendix:3rdmoment} show the second-order and third-order moments of the relative vertical velocity respectively, conditioned upon separation $\delta r$, plotted against vertical separation, $\delta z$, for each of the two representative data sets.   

The second-order moment reveals a dependence of $\delta u_{z}^{2}$ on radial separation, $\delta r$, for both data sets: the placement of the curves on the $y$-axis decreases with descending radial separation. The spread in values of $\delta u_{z}^{2}$ is much larger than the flow-profile velocity variation shown in Figure~\ref{struct_a}. Therefore, we take the trend in both panels of Figure~\ref{appendix:2ndmoment} to indicate that the correlation in particle motion decays with radial separation. This result may also be consistent with a developing pressure phase lag, e.g. particles at small radial separation are close to the lagging pressure-wave peak and those at larger separation respond to the pressure-wave trough. 

There is a notable difference between the two data sets: for the $Kn\sim1$ data shown in the top panel, the second-order velocity difference in the $z$--direction turns towards smaller values for increasing $\delta z$. This indicates that, for separations larger than the wavelength of the unstable wave mode, the relative velocity will eventually return to zero. The same does not occur in the $Kn<1$ data shown in the bottom panel.

The major difference between the two data sets is illustrated in Section~\ref{cont-drag}: the $Kn<1$ data shows a dependence of settling velocity on particle concentration but the $Kn\sim$1 data does not. Only when this effect is present can it serve to increase the velocity dispersion and hence prevent the particle relative velocity to return to zero for large distance. 

The second-order moment, being even, is unsigned and so it does not indicate direction. The apparent influence of the collective drag effect can therefore be better seen in the third-order moment, $\delta u_{z}^{3}$ vs $\delta z$, shown in Figure~\ref{appendix:3rdmoment}. The difference between the two data sets is striking. In the top panel, which shows the result for the $Kn\sim1$ data, the third-order vertical velocity difference approaches zero relative velocity. In the bottom panel, which shows the result for the $Kn<1$ data, there is an apparent turnover that depends upon $\delta r$ and occurs around $\delta z$ of 4-5 mm. The turnover obviously cannot be due to the gas flow profile because it would then be apparent in both data sets. Instead, it seems that we slightly pick up a non-zero horizontal wave mode. The implication is that the collective drag reduction effect is likely to increase the wave numbers k$_{\rm x}$ and k$_{\rm y}$, and supports our interpretation of the wave-like nature of the clustering we report on in Sections~\ref{sec:relvel} and \ref{clustering}.

\begin{figure}
\begin{overpic}[scale=0.45]{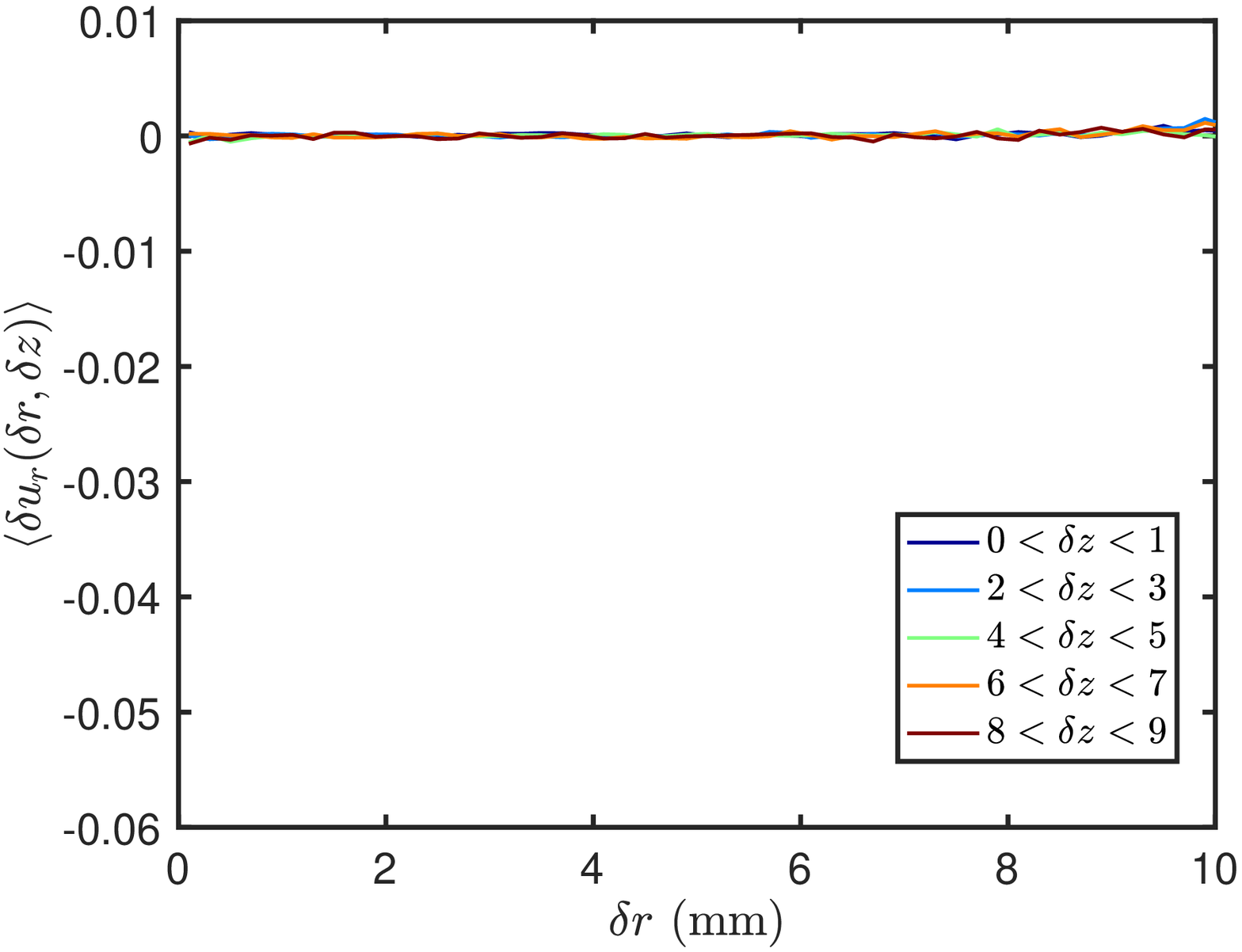}
\end{overpic}
\begin{overpic}[scale=0.45]{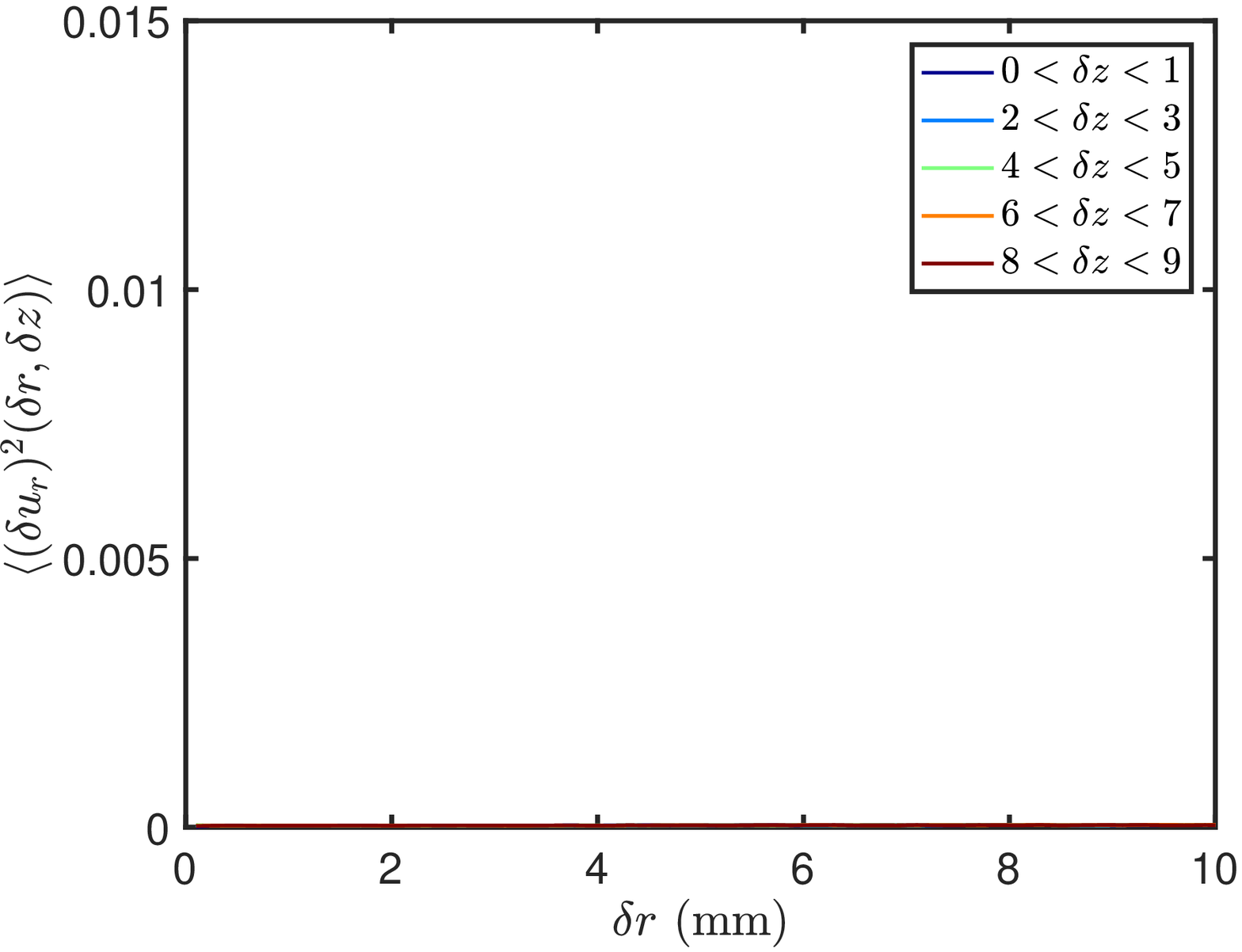}
\end{overpic}
\begin{overpic}[scale=0.45]{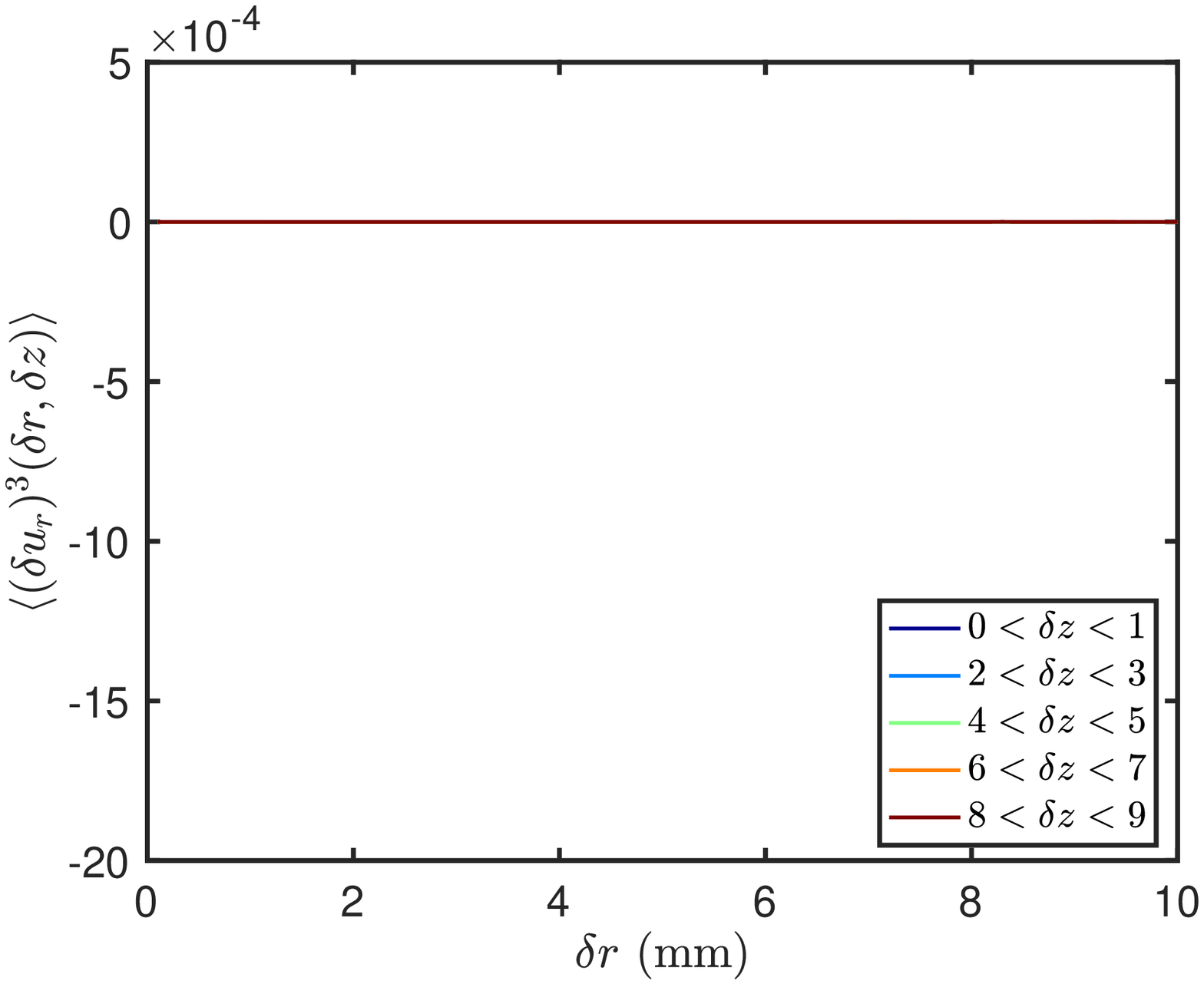}
\end{overpic}
\caption{\label{grad_r_ds2} Shown for $kn\sim1$ data; Top, middle, bottom: first-, second-, third-order moments of the velocity difference in the radial direction, conditioned upon vertical separation, as a function of radius. }
\end{figure}

\begin{figure}
\begin{overpic}[scale=0.45]{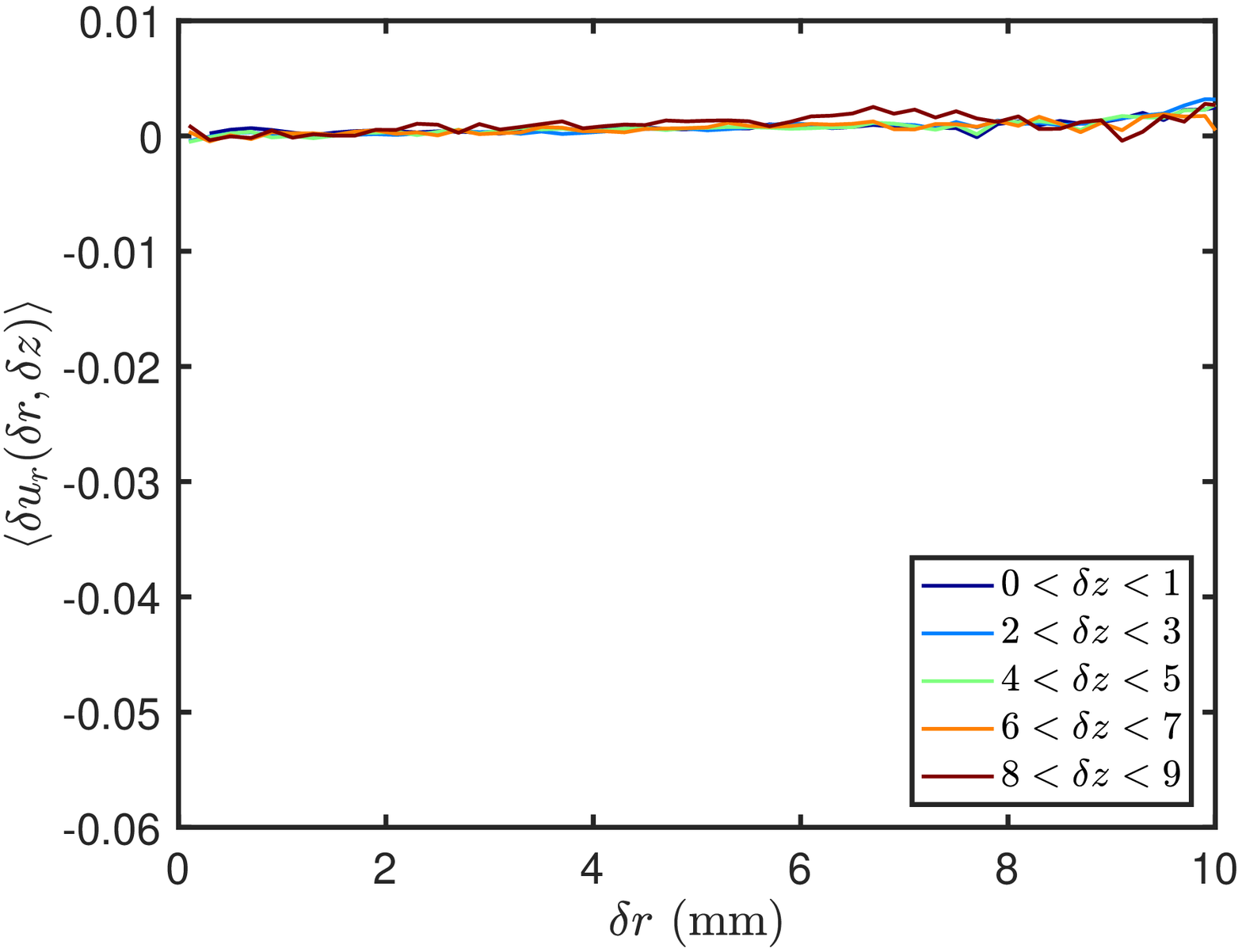}
\end{overpic}
\begin{overpic}[scale=0.45]{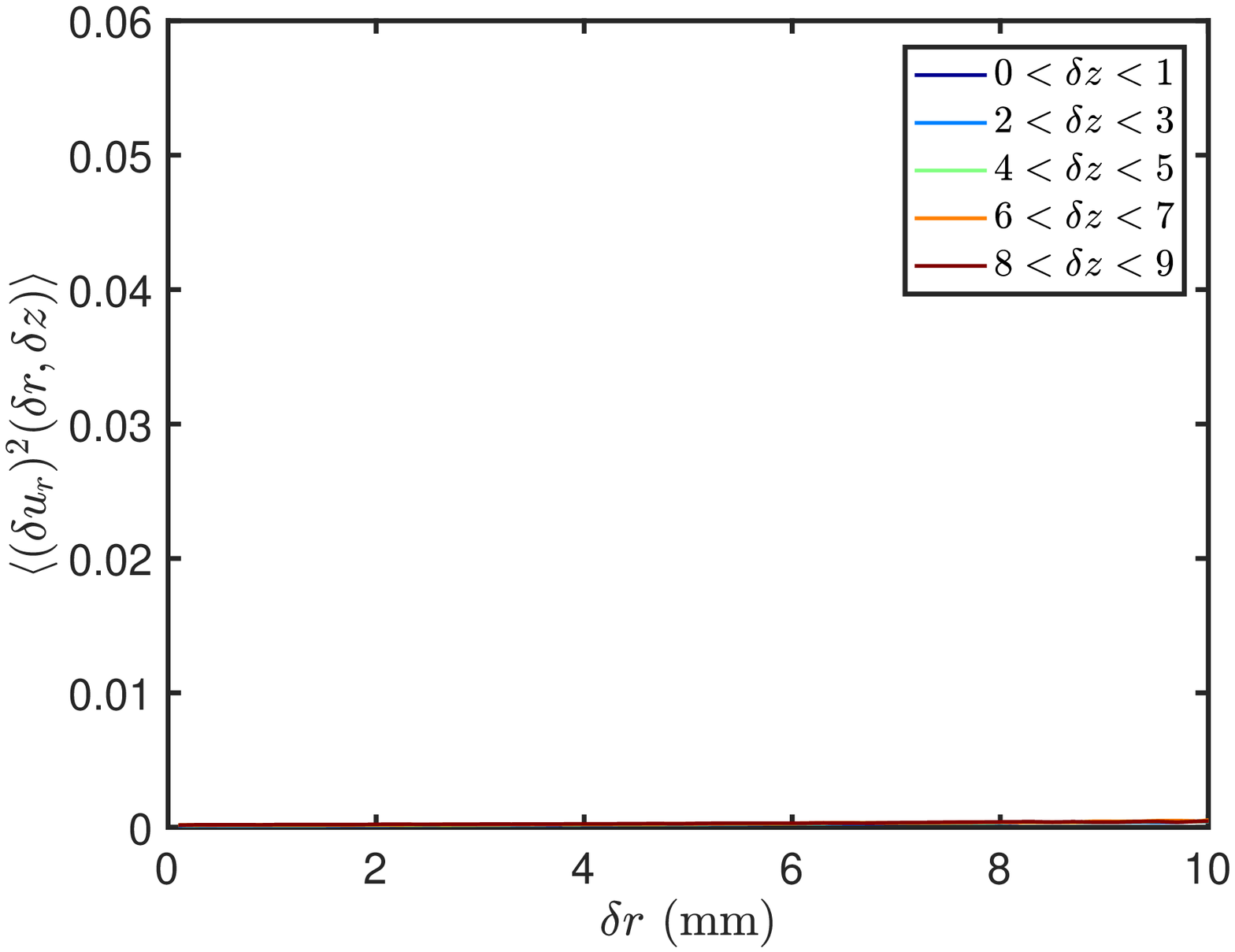}
\end{overpic}
\begin{overpic}[scale=0.45]{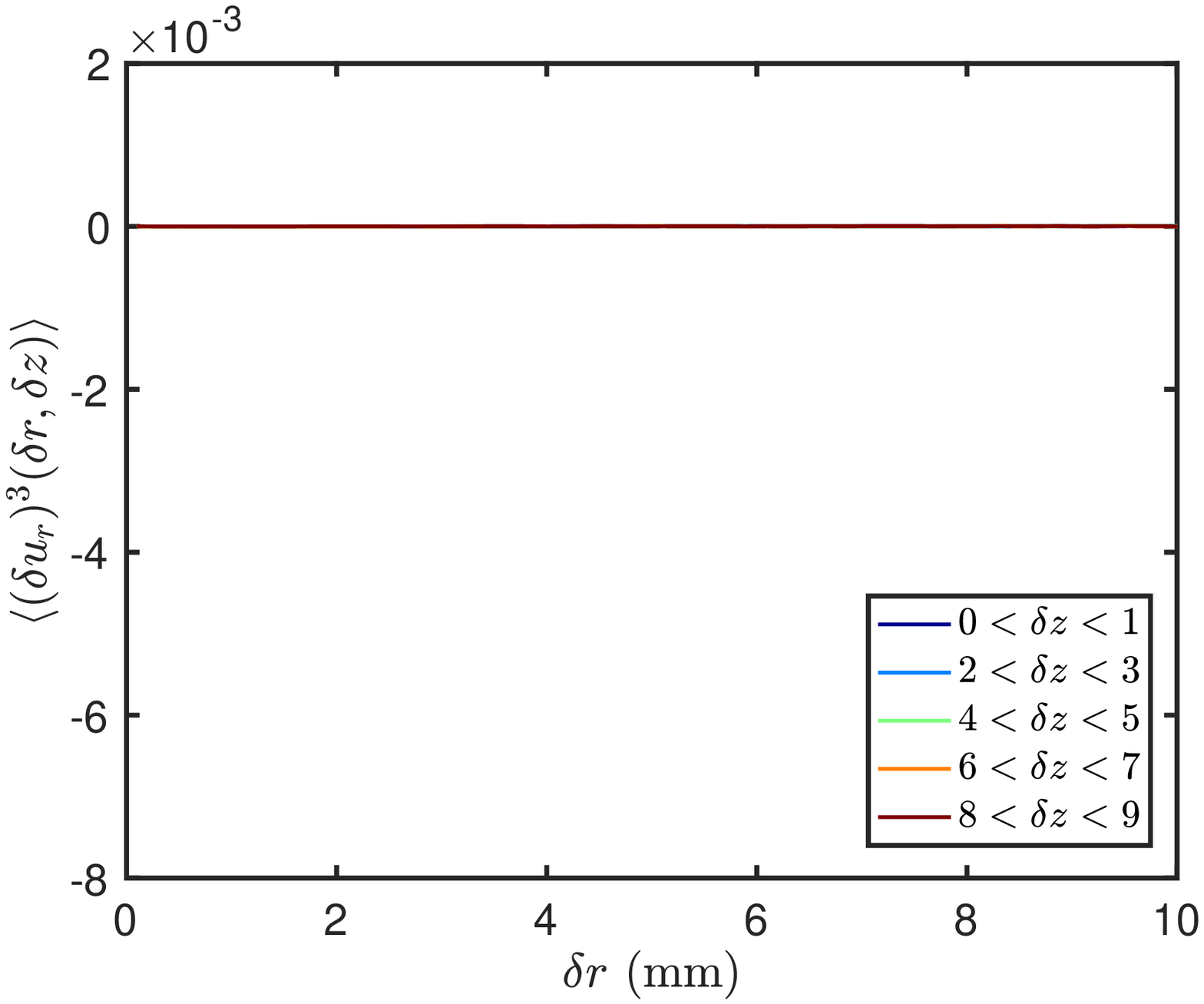}
\end{overpic}
\caption{\label{grad_r_ds3} Shown for $kn<1$ data; Top, middle, bottom: first-, second-, third-order moments of the velocity difference in the radial direction, conditioned upon vertical separation, as a function of radius. }
\end{figure}

For completeness, we also show the moments of the velocity difference in the horizontal direction in Figures~\ref{grad_r_ds2} and \ref{grad_r_ds3}. We see that, despite the complex velocity field we observe, the relative velocity remains homogeneous in the horizontal spatial directions, up to third order. Therefore, this system meets the essential criteria for fluid instability: the base state must be statistically homogeneous in at least one spatial direction and the governing equations in equilibrium must be deterministic, and not time dependent; given enough time to develop, and a free energy source (here the differential gas-dust settling motion of $\approx 1$m/s), the system should approach a new steady state, characterized by non-linear waves \citep{cross_greenside}. We illustrated in Figure~\ref{terminalvelocity} that the equilibrium vertical velocities are not time dependent. The right-hand panel of Figure~\ref{terminalvelocity} shows that the fluid instability predicted by \cite{lambrechts} to occur specifically in this experimental system, has had time to nearly fully develop. According to figure~2 of that publication, the growth reaches its saturation state in no more than 5-10 T$_{\rm{f}}$, which is comparable to the travel time of particles in this system before reaching the measurement location. Figure~\ref{cyllinder_b} of this paper shows that the particles organise into density waves. Figures~\ref{autocorrb_ds2} and \ref{autocorrb_ds3} show that, while structures (and voids) survive long enough for the particle density to remain correlated during their passage through the measurement window, the number density eventually shows the same correlation as average seeding density, implying that small-scale heterogeneities form due to growth of perturbations upon a steady base state. Finally, in this appendix, Figures~\ref{grad_r_ds2} and \ref{grad_r_ds3} illustrate that the deterministic equations of the new, unstable state remain homogeneous in two spatial directions.

Figures~\ref{grad_r_ds2} and \ref{grad_r_ds3} also reinforce why the aggregation effect cannot be due to Coulomb forces, which would not cause zero radial component. Furthermore, it is abundantly clear from these figures that the particle flux is not controlled by any hypothetical anomaly outside the view of the measurement window. 

\end{appendix}

\end{document}